Remote Sensing Image Enhancement through Spatiotemporal Filtering

THESIS

**Presented in Partial Fulfillment of the Requirements for the Degree Masters of Science in the Graduate School of The Ohio State University**

By

Hessah Albanwan, B.A.

Graduate Program in Civil Engineering

The Ohio State University

2017

Thesis Committee

Rongjun Qin, Advisor

Alper Yilmaz

Charles Toth

# Abstract

The analysis of time-sequence satellite images is a powerful tool in remote sensing; it is used to explore the statics and dynamics of the surface of the earth. Usually, the quality of multitemporal images is influenced by metrological conditions, high reflectance of surfaces, illumination, and satellite sensor conditions. These negative influences may produce noises and different radiances and appearances between the images, which can affect the applications that process them. Thus, a spatiotemporal bilateral filter has been adopted in this research to enhance the quality of an image before using it in any application. The filter takes advantage of the temporal information provided by multi-temporal images and attempts to reduce the differences between them to improve transfer learning used in classification. The classification method used here is support vector machine (SVM). Three experiments were conducted in this research, two were on Landsat 8 images with low-medium resolution, and the third on high-resolution images of Planet satellite. The newly developed filter proved that it can enhance the accuracy of classification using transfer learning by about 5%,15% and 2% for the three experiments respectively.

Acknowledgements

I would like to express my sincere gratitude to my advisor, Dr. Rongjun Qin, for his continuous support of my Master's thesis, his patience, motivation and guidance throughout the research.

Secondly, I would like to extend my appreciations to my dearest family and friends for all their encouragement and support. My deepest gratitude to my parents for all their love and unconditional support.

I also would like to thank Kuwait University and The Ohio state university for giving me this opportunity to pursue my dreams and goals.

Finally, I would like to thank Planet Labs Inc. for providing us with satellite images to assist and complete this research.

To everyone who made this possible in their way, the work would not have been completed without you, thank you.

Vita

January 2013 .......................................B.S. Civil Engineering, Kuwait University

July 2015............................................Ministry of Public Works, Kuwait

Publications

Alsulaili, A., Albanwan, H., Alsager, B., Almeer, A., & AlEssa, L. (2014). An Integrated Solid Waste Management System in Kuwait. *2014 5th International Conference on Environmental Science and Technology*, (pp. 54-59).

Fields of Study

Major Field:  Civil Engineering

Table of Contents

List of Tables

List of Figures

Chapter 1: Introduction

     Assessing land use and land cover (LULC) is crucial to many applications such as urban planning, agriculture, climate monitoring and change detection ( (Kiptala, Mohamed, Mul, & Cheema, 2013). Lands are classified into several categories based on their usages and natures such as urban areas, agricultural areas, barren lands, water bodies, soils, and rocks. Investigating LULC maps and classes spatially can be beneficial, not only to distinguish between types of lands, but also useful in decision making, land management, evaluation and conservation of natural resources such as water used for irrigation (Kiptala et al., 2013), soil resources and forests ( (Butt, Shabbir, Ahmad, & Aziz, 2015).

     Remote sensing image analysis techniques have strong capability of aiding in planning and exploring lands. Moreover, remote sensing tools are often used to examine LULC over time and allow monitoring and detecting terrestrial changes of lands, objects, and phenomena, in addition to keeping track of the updates on land use and cover maps (Kiptala et al., 2013). One of the tools and methods to explore LULC is classification using available satellite images, which can provide information related to the land composition and areas of different types of land classes. These satellites vary in their specifications, manufacturers, spatial and temporal resolution. One of these available satellites with low-medium resolution is Landsat; it has multiple bands that help to explore a variety of phenomena. However, this satellite imagery is known to have low

accuracies and resolutions. Therefore, most studies agreed that in order to get accurate and reliable results, additional ground truth data should be conducted and compared to the data obtained spatially from the satellite imagery (Kiptala et al., 2013; Butt et al., 2015). Alternatively, high-resolution satellite imagery can provide highly detailed maps and images of areas, and can be helpful in exploring urban regions on a wide scale. Planet satellite imagery has a variety of high resolutions ranging from 3-7 meters and a high temporal resolution up to daily availability. Processing such time-series images faces the problems of radiometric discrepancies, very often highly non-linear, such as atmospheric effects or high reflectance of surfaces. Traditional remote sensing atmospheric correction techniques using radiative transferred models have demonstrated success in reflectance recovery, while to utilize such methods accurately requires many in-situ information such as AOD (aerosol optical depth), looking angle, weather, etc. It is therefore, important to find out simple and effective ways for correcting non-linear spectral reflectance before performing classification and change detection.

In this research, a new spatiotemporal bilateral filter is proposed to enhance the quality of images that are used in classification. The filter operates by measuring pixels' closeness and resemblance in three dimensions spectrally, spatially and temporally. We evaluated the performance of the spatiotemporal filter through transfer learning using SVM (Support Vector Machine) on a sequence of multi-temporal satellite imageries. The transfer learning provides automatic guidance to classify all images based on labels from a single source (Ma, Xu, Wu, Wang, & Chen, 2016). SVM is a well-known supervised classification method that operates in higher dimensions and solves an optimization problem to look for the optimal hyperplane that best separates two classes ( Xiong,

Zhang, & Chen, 2010). These methods were applied on medium and high-resolution images of Landsat 8 and planet satellites. This work analyzes and compares images before and after classification, as well as, before and after image filtering. It also reports the performance of spatiotemporal bilateral filter based on the accuracy of the classification on filtered and original images.

## 1.1 Objectives

The primary objective of this research is to investigate the effect of using the spatiotemporal bilateral filter on the classification method. The second aim is to investigate the enhancement of such methods on the transferability of learned samples. Enhancing the quality of images provides a better distinction between different features within an image such as roads, buildings, trees, ground, shadows, etc. Therefore, facilitate and improve the digitizing process for applications like GIS, in addition to updating and interpreting maps and images' contents. Furthermore, it can be useful to have a rough estimate on what is on the ground and allows the exact or close enough capture and extraction of areas, surfaces, and shapes. Finally, it allows monitoring any variations or changes over time that can be used in the change detection.

## 1.2 Following sections

The next chapters cover the following; first, we will introduce the relevant works of classification, transfer learning, and spatiotemporal filtering. In chapter three, we will present the proposed spatiotemporal filtering method for enhancing the radiometric qualities of multi-temporal datasets. The fourth chapter discusses the experimental results. Finally, we draw the conclusions, limitations and potential direction of future works.

Chapter 2: Literature Review

**2.1 Time series analysis of remotely sensed data**

Processing time-sequence satellite images is an important and common research area in remote sensing. Analyzing several multi-temporal images is more valuable and informative than using a single image. Processing multi-temporal satellite data have been used widely in change detection (Zhou, Li, & Chen, 2011), image registration, radiometric correction, environmental monitoring, and so on (Jianya, Haigang, Guorui, & Qiming, 2008). The key element in the analysis of multitemporal satellite images is the radiometric conditions, as since these images will be correlated and compared. Usually, the acquired images experience differences in their appearances, either due to different lighting conditions such as the sunlight intensity and direction, or different metrological conditions such as the existence of clouds, snow, rain, etc. Another influence on images' appearances is the seasons; natural elements on the ground such as vegetation and barren lands change over seasons, and so does their intensities in the images (Paolini, Grings, Sobrino, Jime´Nez Mun˜ Oz, & Karszenbaum, 2006). All of these mentioned factors in addition to the satellite sensor conditions variation due to aging and operating environments will cause the images to have different looks, intensities, and in some situations, affect their qualities. Therefore, radiometric correction should be applied to remove any possible noises and normalize the images to achieve radiometric homogeneity (Paolini, Grings, Sobrino, Jime´Nez Mun˜ Oz, & Karszenbaum, 2006). For this reason, a newly developed spatiotemporal bilateral filter is adopted to adjust the radiances and spectral conditions of the images, therefore, improving the classification using transfer learning.

## 2.2 Classification

One of the major tasks in remote sensing is classification, where images are broken down into categories and subcategories based on their contents. Classification is crucial since it distinguishes different land classes such as water surfaces, buildings, trees, grass, shadows, etc. This step facilitate the digitizing process to produce categorical and thematic maps to be used later on for other applications. Accuracy in classification is very important since it can be used to capture the precise areas, spaces, dimensions, and locations.

### 2.1.1 Classification using Support vector machines (SVM)

In general, there are two types of classification in statistical learning approaches either the supervised or unsupervised. The prior involves providing the program with labels for which the classes are known, which will be used in the training step. The latter, on the other hand, does not need inputs labels or data for the training, where it groups and clusters the dataset based on the closeness and resemblance in color. Support Vector Machine (SVM) is one example of supervised classification methods, where image pixels are being trained to create an organized and classified image. Several Studies on SVM have shown that due to its simplicity and practicality it has outperformed other methods for classification such as Decision Tree (DT), Neural Networks (NN) (Hussain M., Wajid, Elzaart, & Berbar, 2011) and Random Forest (RF).

Classification using SVM have been used in different sorts of applications such as in the medical sector (Hussain et al., 2011; Srivastava & Bhambhu, 2010), land classification, object and pattern recognition, satellite imagery (Srivastava & Bhambhu, 2010), etc. SVM require minimum inputs and reference data for the training, which makes it convenient for classification (Matykiewicz & Pestian, 2012). Furthermore, for remotely sensed data, SVM has proven its efficiency in many of the studies on multispectral or hyperspectral satellite images (Mankar, Khobragade, & Raghuwanshi, 2016; Srivastava & Bhambhu, 2010). Like any other approaches, SVM has its own limitations, like the computational complexity due to the size of input data, which affects the time and speed of the process (Prajapati & Patle, 2010). Other possible shortcomings to this approach are the right selection of kernel function and having an appropriate amount of labeling data for training, testing, and validation.

SVM operates by finding the best separator between different classes. There are three main components of SVM; the first component is the hyperplane that separates different features. The second component is the maximum margins that define the location of the hyperplane, and finally, kernel function that helps SVM to perform in higher dimensions. The size of the training sample and the choice of the kernel function are two factors that influence the accuracy in classification Training input dataset is very important during classifying phase and that is because it describes the labels, feature, and attributes. SVM can work in a linear and nonlinear way by different choices of kernel function. For the linearly separable datasets, simple linear kernel often used, whereas, for the non-separable situation, a variety of options are available, including polynomial, radial basis function (RBF) or sigmoid. Many studies have shown that using RBF as kernel function is preferable since it returns the highest accuracies in classification, which makes an ideal option for the SVM algorithm (Prajapati & Patle, 2010). Moreover, there are two options for training, the first one is one-against-one where discrimination between each pair of labels takes a place; it is simply the basis of binary classification (Hsu & Lin, 2002). The second alternative for the training method is one-against-all where each class is being discriminated and compared with the rest of classes at once (Hsu & Lin, 2002; Prajapati & Patle, 2010). The choice of either one of the two options depends on the purpose and nature of the task. For example, one-against-one is known for its efficiency and higher accuracy, but in many circumstances, training one-against-all is preferred since it compares a single class with all other classes not just a pair of classes, thus it is time efficient (Milgram, Cheriet, & Sab, 2006).

### 2.1.2 Transfer learning

Transfer learning is another important field in machine learning and is highly effective in the classification process. It has been used widely in the classification of images, looking for texts, objects or features through the internet, etc. (Ma et al., 2016). To explore the way transfer learning works, let us say there are two datasets that are available image 1 and 2, and say image 1 has already been trained, classified and the labels already available. In this case, it is easier to use what was learned in the learning process of the first dataset to classify the new dataset of image 2. Thus, instead of creating new labels from scratch and training all over again, transfer learning speeds this process by using the known information from previously trained and labeled dataset to classify the new data. Transfer learning is valuable not only for saving time but also for when labels and training inputs are not accessible and available (Ma et al., 2016).

However, the data in the source (reference classifier) and target do not always match (due to differences in feature space and appearances of the images). Thus, a transformation between the two features spaces is often performed to match the data in both domains. Some studies acquire as much as possible of the training data to cover up all possible features that might be encountered in transfer learning (J. Lim, Salakhutdinov, & Torralba, 2011), but increasing the training data will affect the computational complexity. Other studies found different algorithms to transform the data between multiple source images, but that would also raise the work load and time. Therefore, to reduce the work load and enhance the classification results using transfer learning, the spatiotemporal bilateral filter is proposed (it relates heterogeneous data from multisource images).

Despite the usefulness of transfer learning, it still faces many challenges. One of the issues for investigation is the accuracy drop after transfer learning. This drop-in accuracy can be due to differences in the inputs like the intensities or that some of the new data were not presented in the previous training, thus, can be misled. Moreover, transfer learning can be used with any method of classification such as Decision Tree (DT), SVM, Neural Networks (NN), Naïve Bayesian and Random Forest (RF). Each of these methods varies in their algorithms and applications. For instance, SVM is mostly used for a high dimension of features, whereas (NN) is a more complicated method to classify images by using connected neurons and mimicking human brain in the way it processes the information (Xiong et al., 2010). Other methods like DT has a hierarchal structure for processing data, thus not well suited to handle large feature spaces (computational complexity is affected by the size of data) (Otukei & Blaschke, 2010); Naïve Bayesian, on the other hand, is used for simple tasks.

Enhancing classification accuracy has always been the motive, and one option to raise the accuracy is to use an additional source of information (other than pixel values). (Qin, Tian, & Reinartz, 2015) used height information of digital surface models (DSM) along with corresponding orthoimages to extract features and enhance classification and this approach had proven its efficiency in raising the accuracy. Other methods use vegetation indices such as NDVI to locate vegetation or eliminate them for a specific feature detection (Dejan, Kosmatin Fras, & Petrovič, 2011). The combination of both methods is implemented in some situations to get more accurate results as was performed in (L., S., W., & Li, 2012), where both the height and vegetation indices were used to get information about the whereabouts of a particular feature like buildings. To sum up, external sources of information can be highly useful for validation and raising the accuracy of classification.

**2.3 Spatiotemporal Bilateral Filter**

In image processing, a digital image is represented in the form of square grids; each cell of the grid has a digital value or number, and these squares are known as pixels. Hence, any image has two domains, one is the spatial domain and the other is the range domain. The space domain is represented by the positions of pixels in the x and y dimensions of the image. On the other hand, the range domain is the pixels values and these values together with the bands in the image carryout the colors of the image. Any enhancement or filtration of an image will have to modify either or both domains.

A conventional bilateral filter is an example of a nonlinear filter, which operates on two domains: spatial and spectral. This filter is mostly used when edges are to be maintained in the image, while smoothing of the colors is demanded to get a better appearance (Tomasi & Manduchi, 1998). To understand the bilateral filter, we must go over the Gaussian filter first. The Gaussian filter is a simple linear, low pass, non-preserving technique that smooths the entire image. Moreover, it only processes the edge positions and locations of the pixels in the image and filters them within a pre-specified window size based on their distances from the central pixel in the window. Likewise, the bilateral filter also considers the space domain as in the Gaussian filter, but at the same time, it uses the intensity information to filter and preserve the edges. Therefore, the bilateral filter measures the closeness of the pixels in their locations, as well as, the similarities in their intensities in the range domain. Of course, a number of parameters such as the window size and the sigma values (band-width) denoting standard deviation control the amount of filtering in each domain. The computed weights together with the original pixels values are multiplied together to get the new enhanced image.

  It is worth mentioning that one of the advantages of the bilateral filter is its ability to detect and fix inliers and outliers by replacing them with the modified pixel values ( Durand & Dorsey, 2002). This happens when the measured difference of intensities is large, which means that these pixels have no relations between them, therefore, should be assigned less weight. Furthermore, it has been used for edge detection and preserving, and is considered in some cases a better replacement to anisotropic diffusion method (Durand & Dorsey, 2002). Another advantage to this filter is that it allows removing certain effects due to bad lighting conditions such as blurriness, darkness, under exposed images ( Eisemann & Durand, 2004), over sharpness, shadows and red eye effect due to flash (Bennett & McMillan, 2005).

In many cases, researchers prefer dealing with a sequence of images taken at different dates or times. These set of images are considered valuable since they contain more information about a scene than a single image does. For this reason, most of the remote sensing analyses use multi-temporal images or data and process them simultaneously to take advantage of the temporal information in each of them. Studies have been trying to modify bilateral filter to be able to use it in the temporal domain, however, most of these attempts have shown their success in the video enhancement area (Bennett & McMillan, 2005) or pair of images taken from a camera ( Durand & Dorsey, 2002; Eisemann & Durand, 2004). Very few studies have incorporated the spatiotemporal filter in complex satellite images.  (Bennett & McMillan, 2005), used a temporal-spatial bilateral filter to enhance low dynamic range (LDR) video scenes. This research's technique is based on monitoring the pixels values and observing the changes within a small window. If the pixel value remains constant, it uses the temporal bilateral filter with dissimilarity value to discern between the values. On the contrary, if there is a great change of the central pixel within the window only spatial-median bilateral filter is applied. Finally, tone mapping was applied to fix the low exposure of light and LDR in the same manner as in ( Durand & Dorsey, 2002), which also uses the bilateral filter. Another research by ( Eisemann & Durand, 2004) used modified bilateral filter and shadow correction on two consecutive images to correct the dark and over sharp images under different lighting conditions. They used the shadow and color information from two images: one without a flash and the other with flash. They also used an approach called cross-bilateral filter and modified it for the image with no flash to capture the edges. The use of two images with different environmental lightening conditions allowed the research to use the information of both images and fix them accordingly. According to (Shreyamsha, 2015), an image that is out of focus can be enhanced and the details can be retrieved or filled by using a new image fusion technique. This technique is based on an approach that is called cross-bilateral filtering. It uses multisensory, multi-focus and multisource images, find their weights with respect to each other, and finally, uses the weights and values of original images to get the new fused image. Another approach adopted by (Qin et al., 2015), performs a 3D spatiotemporal bilateral filter on a single satellite image along with multiple stereo satellite images that include the digital surface model DSM. This method creates probability maps using random forest classification, followed by a 3D bilateral filter to enhance the image in three dimensions spectral, spatial and temporal. The temporal dimension used in (Qin et al., 2015) processes the height information provided from the stereo satellite images taken at different dates.

To summarize, bilateral filter has been used widely often for noise reduction (Bennett & McMillan, 2005; Liebold & Maas, 2016), tone mapping ( Durand & Dorsey, 2002; Eisemann & Durand, 2004), successful replacement to anisotropic diffusion for edge detection ( Durand & Dorsey, 2002), image enhancement (Shreyamsha, 2015) and so on. With all these application and modifications of the bilateral filter, not so many studies have used the temporal domain in the bilateral filter on the multitemporal satellite images. Thus, it would be ideal to test the newly developed spatiotemporal bilateral filter in three dimensions in the spectral, spatial and temporal domains on several satellite images.

**2.4 Limitations**

The major challenge of this project is the classification is accuracy. Several factors will contribute to the accuracy assessment such as the type of image filtered or original, and learning method for classification whether it is individually classified or classified through transfer learning. Thus, to maintain a certain level of accuracy, the operation or system must go under many constraints and trails. Another issue affecting the accuracy of classification is the image quality; an example of these effects is the spatial resolution, which plays a major role in showing the precise details in the image. Usually low to medium resolution images hide details like small houses and minor roads, these small features can be obvious in the high-resolution images. Another inevitable factor that influences satellite image quality is the atmospheric conditions such as cloud cover, haziness, reflectance of sunlight on surfaces, season and time of the captured image. All these factors may degrade the quality of the dataset and produce less informative images. Furthermore, the filter parameters also play a significant role in the accuracy of the results, setting any of the three bandwidths (sigma) in the filter to a high value will cause the image to blur, and may lose important information. On the contrary, the smaller the values assigned to the bandwidths (sigma), the less filtering effects will take place. Therefore, testing several parameters before starting the experiment is important to reach optimal outcomes.

Chapter 3: Methodology and Theories

This chapter covers theoretical aspects of this research. There are mainly three steps for performing a spatial-temporally enhanced classification: first, the image sequence will be registered and the radiance will be coarsely corrected using a linear model, secondly, we will apply the proposed spatial-temporal bilateral filter to enhance the sequence of images, finally, we will perform classic and transferred classification to the processed images (Figure 1). Three experiments were conducted in this research, experiment II and I were performed on different datasets of Landsat 8 images, whereas experiment III was made on Planet satellite images. All experiments follow the same steps.

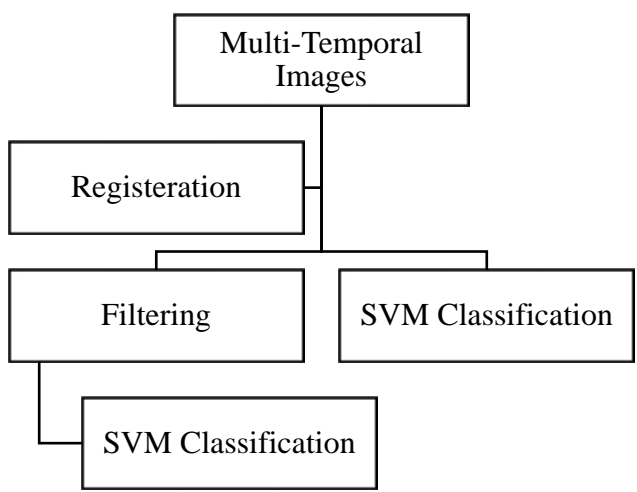

*Figure 1. Research Steps*

### 3.1 Study Area

There are three study areas in this work; all are in Columbus, Ohio, including five temporal images for each study. The first two datasets (used in experiment I and II) involves images taken from Landsat 8 (low to medium resolution images) with dimensions of 2000 × 2000 pixels (see Figure 2, Figure 3)These images from both experiments were taken in the periods of winter, summer, and spring (see Table 1). The data used in experiment I (Landsat 8 images) is in the countryside, and it mostly involves natural areas like barren lands and vegetation (see Figure 2). On the other hand, the data used in experiment II is mostly urban region (houses, buildings, and roads) (see Figure 3). The high-resolution images taken from planet satellite have dimensions of 1534 × 1534 pixels (see Figure 4), and the resolution of 3.0 meters; the details on the seasons are shown in Table 1. The main components of the images are water surfaces, barren land, and vegetation, building and roads; the details on classes are shown in Table *2*. A number of classes is often governed by the purpose of the work and how many details can be extracted from the images. To demonstrate, buildings and roads can be distinguished and discerned only through high-resolution images. However, for low-resolution images as in the case of Landsat 8, it is hard to discriminate between the two classes, thus, these two classes were merged into one class known as impervious surfaces. The input images must undergo several preprocessing steps before using them in the experiments. The first step is resolution merge (pan-sharpening) for Landsat 8 images, by taking advantage of the fact that all the bands have 30-meter resolution with band number 8 of 15-meter resolution (panchromatic band). Hence, a software named ERDAS Imagine (Earth Resource Data Analysis System) was used to do the pan sharpening and move the bands to a higher resolution of 15-meters. The second preprocessing step is registration, which will be discussed in the following section.

It can be seen that the temporal images in each dataset vary greatly in their radiometric appearances, highly non-linear even for the same class. This is mainly due to the difference in radiometric conditions in each image. The dates and seasons are one factor that causes the images to have significant differences in appearances. The sunlight effect, the sunlight angle, and its reflectance on urban and nonurban features on the ground may also influence images appearances and radiometric properties. For example, the river and lake in image 4 in experiment I and II have some parts (mid and upper sections in experiment I, and the upper section in experiment II) where the water surfaces appear in white due to high reflectance of sunlight on that surface, while in other images it is simply blue. Another example is the high urban reflectance of buildings and roads in image 2 and 5 in experiment III. These influences on the appearances of the images affect the quality and the analysis of these images. Therefore, processing multi-temporal satellite images require applying radiometric corrections to reduce the differences between the images for comparison purposes.

Table 1. Input Images Details

| | | | | | |
|---|---|---|---|---|---|
| **Experiment I Dataset: Landsat 8 Images** | | | | | |
| **Image** | 1 | 2 | 3 | 4 | 5 |
| **Date** | 2016-05-27 | 2016-06-12 | 2016-09-16 | 2016-02-21 | 2015-09-14 |
| **Season** | Spring | Spring | Summer | Winter | Summer |
| **Experiment II Dataset: Landsat 8 Images** | | | | | |
| **Image** | 1 | 2 | 3 | 4 | 5 |
| **Date** | 2016-05-27 | 2016-06-12 | 2016-09-16 | 2016-02-21 | 2015-09-14 |
| **Season** | Spring | Spring | Summer | Winter | Summer |
| **Experiment III Dataset: Planet Images** | | | | | |
| **Date** | 2017-02-26 | 2016-09-02 | 2016-10-14 | 2016-10-05 | 2015-08-28 |
| **Season** | Winter | Summer | Autumn | Autumn | Summer |

Table 2. Classes and Their Descriptions

| No. | Classes | Description | Miscellaneous |
|---|---|---|---|
| 1 | Water Surfaces | Lakes, rivers, ponds | |
| 2 | Barren Land | Sand, rock, areas with leafless trees | |
| 3 | Vegetation | Trees, grass | |
| 4 | Impervious Surfaces | Roads, buildings, infrastructures | For Landsat images |
| 5 | Roads | Highways, minor roads | For Planet images |
| 6 | Buildings | Houses, commercial buildings | For Planet images |

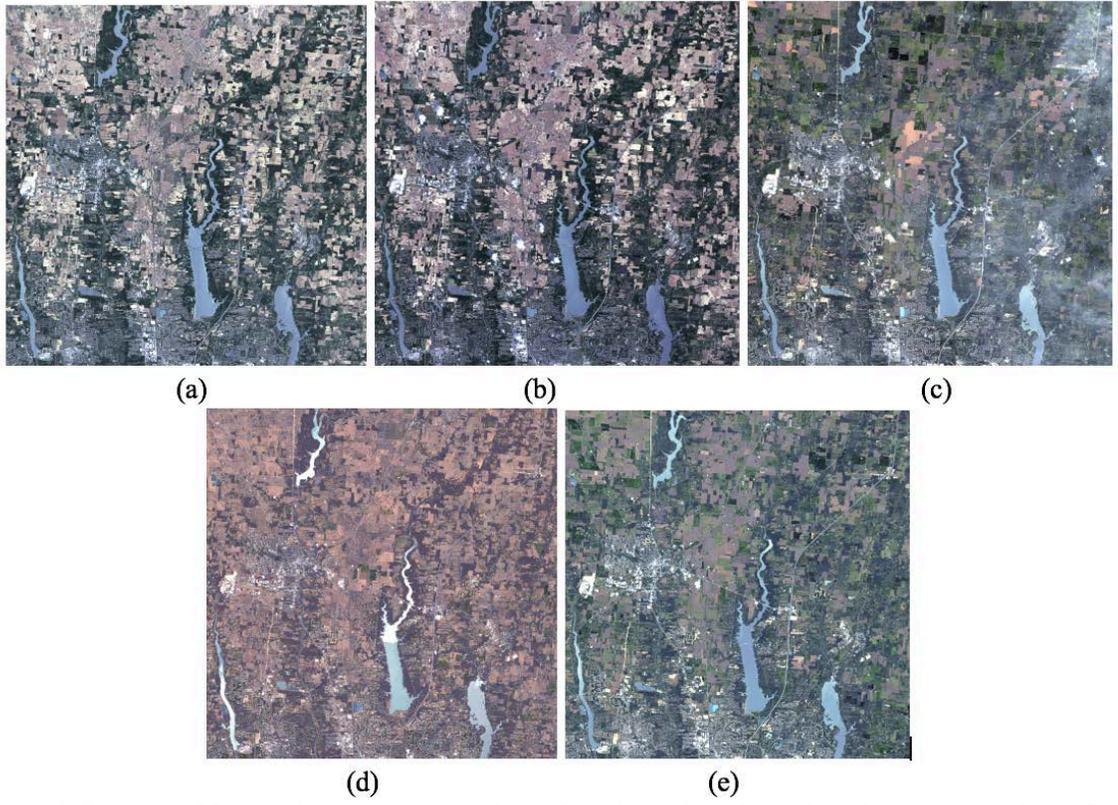

*Figure 2. Experiment I Dataset. Landsat images, (a) Image 1, (b) Image 2, (c) Image 3, (d) Image 4 and (e) Image 5.*

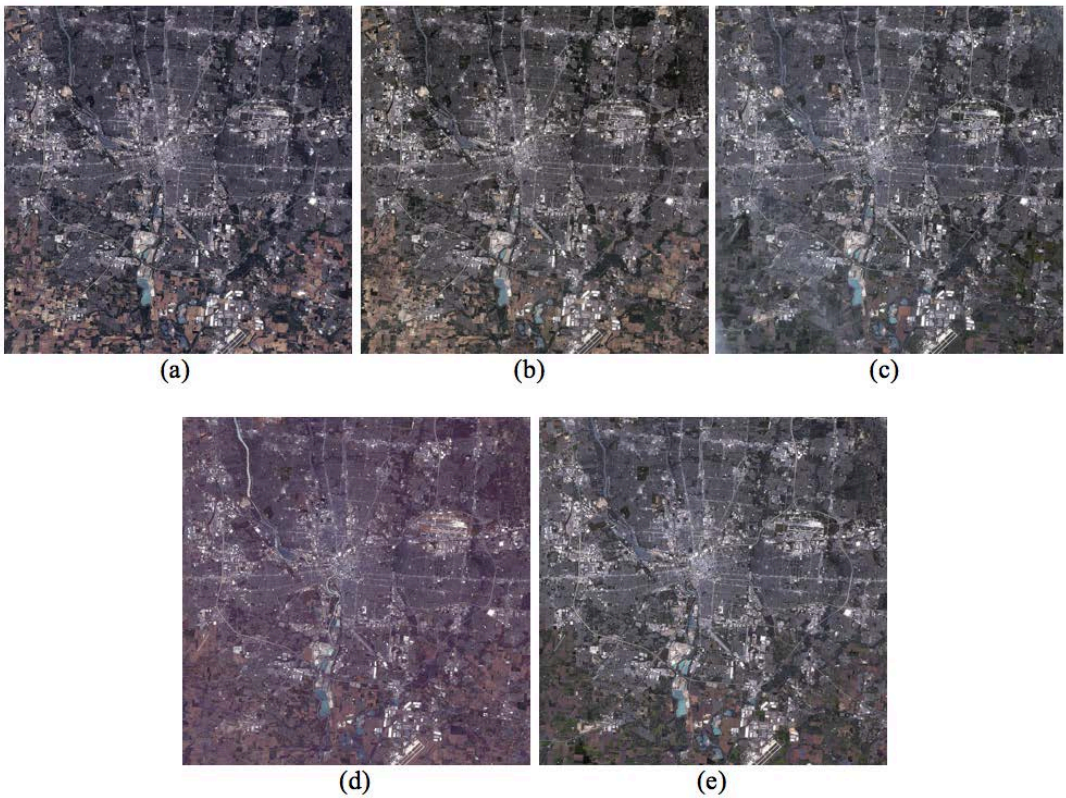

*Figure 3. Experiment II Dataset. Landsat images, (a) Image 1, (b) Image 2, (c) Image 3, (d) Image 4 and (e) Image 5.*

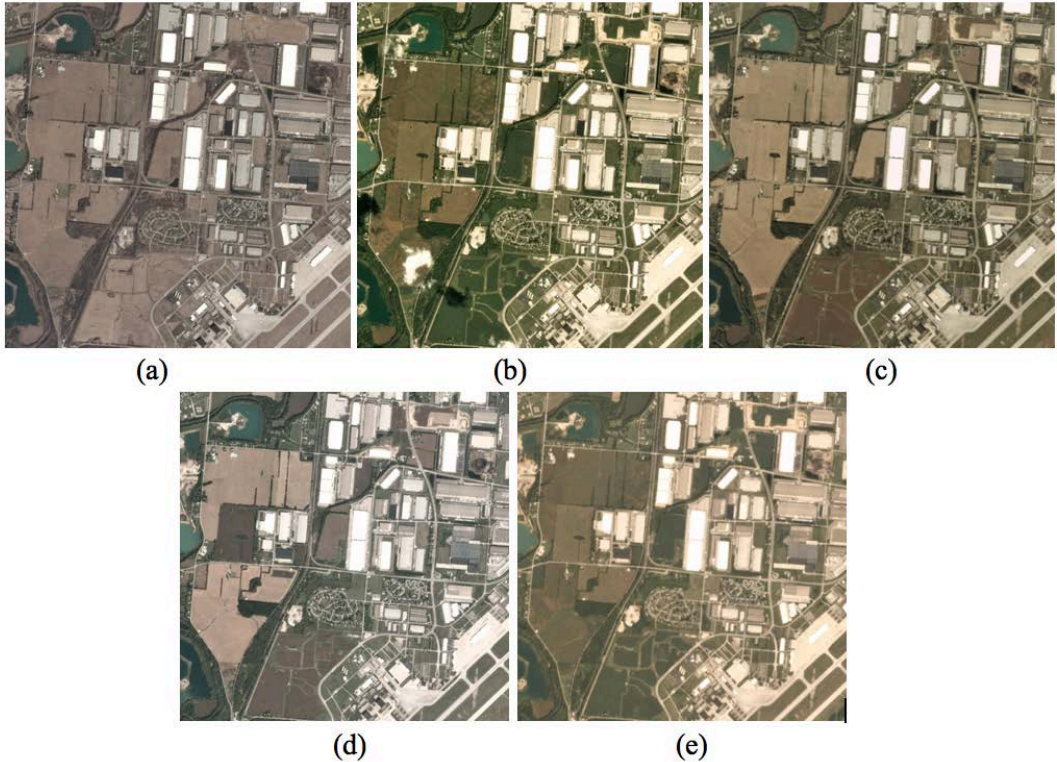

*Figure 4. Experiment III Dataset. Planet images, (a) Image 1, (b) Image 2, (c) Image 3, (d) Image 4 and (e) Image 5*

**3.2 Image Registration**

Although these datasets have been geo-referenced, precise alignments are still needed to facilitate per-pixel operation. Since this research is using five multi-temporal images, they must be overlaid correctly pixel-wise to be able to process them at a later step. Registration is important to align all the images and fix any possible distortions in positions. These distortions can be due to scenes taken from different sensors, different dates, times or even angles of view (Zitová & Flusser, 2003), or sometimes due to image cropping. These distortions between the images need to be eliminated via precise registration. There are many methods to do image registration; it can either be feature based or intensity based (Zitová & Flusser, 2003). The feature-based method works by finding the corresponding matching features such as lines, corners, points, etc., and this can be achieved by using control points (CP). Locating CP in not always accurate and in some methods it is performed manually, thus, it will have high potential for misalignment due to human errors or incorrect matching. Whereas, the intensity based approach uses the color values to look for the related color patterns and register them automatically on that basis. Generally, there are several methods that use and modify the feature and intensity based methods to do image registration or finding specific objects. Among these methods is template matching (Jia, Cao, Song, Tang, & Zhu, 2016), least squares adjustment, cross correlation, Fourier transform (Guizar-Sicairos, Thurman, & Fienup, 2008), and so on.

We use the intensity-based method to achieve high accuracy registration. This method is more accurate than using the feature point based approach. In this research, the intensity-based method is used, in addition to an affine transformation algorithm, where images are being registered based on the resemblance of the pixels values. Registration includes a number of steps, which can be summarized as the following: first, a reference image for which all other images are being referenced to is chosen. The second step is to find similar corresponding points or patterns between the reference image and the target image. After that, affine transformation parameters such as rotation, translation, shearing and scaling are measured. Finally, these transformation parameters are used to transform the remaining pixels to the new coordinate system that matches the reference image. MATLAB (Matrix Laboratory) has ready available functions and tools to do intensity-based-registration and compute transformation parameters. The function (imregtform) is used to measure the parameters of the affine transformation by providing it with the reference and target images. The following step will be using (imwarp) to register the target images using the conducted transformation parameters. Furthermore, it allows using an optimizer and metric configurations to optimize and find the right and exact parameters for registration by performing several iterations. Eventually, the five output datasets will end up having the same coordinate system and conforming locations of all pixels.

### 3.3 Classification

Selecting the appropriate classification method can be difficult and requires trying number of methods to reach the optimum technique. Machine learning provides a variety of options to perform classification and categorization of data by processing the it in a statistical manner. Among these methods are nearest neighborhood, linear and nonlinear SVM, Naïve Bayes, Neural Network (NN) and Random Forest (RF). In general, classification in machine learning is divided into two types: supervised and unsupervised. The former approach relies on providing the program with known labels for each class for the training, while; the latter has no labels or prior information on the data labels, thus, the program on its own clusters and finds patterns and relations among the inputs. Furthermore, classification of images can be accomplished by using only single-source information such as color denoted by the pixel values (used by this work), or the color values along with other external source of information such as ground truth data or height (Qin et al., 2015).

### 3.1.1  Support Vector Machine (SVM)

SVM operates in three stages: the first where it is provided with the classes and labels for the training; the second is the testing where it uses the function that was created in stage one to classify the remaining data; finally, the validation stage to measure the accuracy of the technique. The kernel function used in this project is radial basis function (RBF); it is used since it can transform the data into a higher dimensional space to get better classification outcomes. Furthermore, the training is performed using one-against-all where each class is being compared with the rest of the classes, and it can result better outcomes in terms of computational speed especially with large size of labels and training data.

The major work and the most challenging part in designing the SVM classifier is finding maximum margins that will define the separating hyperplane. SVM operates by finding the optimal hyperplane with maximum margins that separate classes the two main components that define the hyperplane are: the bias b and the normal w, and it is represented by the inequalities of equation (1). These inequalities are known as the decision functions, where, every new class is being tested whether it belongs to a certain class or the other by knowing if it is $\geq 1$ or $\leq -1$.

$$f(x_i) = y_i\,(w^T.x_i + b)\ \geq,\leq\ \pm 1 \qquad (1)$$

- $(w^T.x_i+b) \geq 1$        Feature belong to either of the two classes
- $(w^T.x_i+b) = 0$        On the hyperplane
- $(w^T.x_i+b) \leq -1$        Feature belong to either of the two classes

Where

- $y_i = \pm 1$
- i {1, …., N} number of a training set,
- $x_i$ training input set,
- b, bias

Since the project compares all classes at once using one-against-all classifier, a feature $x_i$ is said to belong to a certain class if the value of the decision function is maximized as follows:

$$f(x_i) = \arg max_{i=1...N} \ (w^T.x_i + b) \tag{2}$$

During classification, the two closest points or features of each class are located, and the maximum furthest distance between them ($\frac{2}{||w||}$) is measured, this is called maximum margin and is defined by the support vectors. The greater the margin between the classes the better optimization of the hyperplane, hence better separation. The mid-point between the support vectors is the hyperplane that is represented by a line; the two vectors are away from the hyperplane by a distance of ($\frac{1}{||w||}$) from each side. Figure 5 demonstrates the main components of SVM.

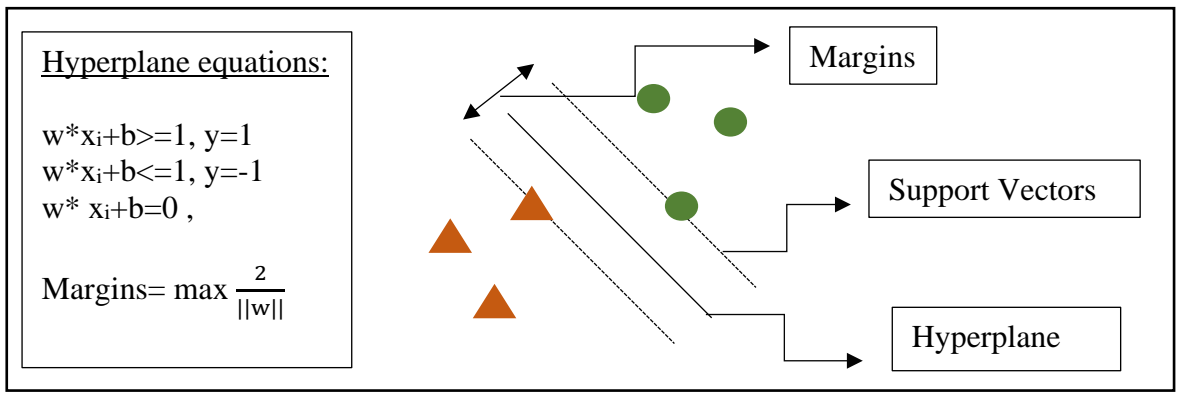

*Figure 6. Support Vector Machine (SVM) Components*

SVM maximizes ($\frac{2}{||w||}$) and minimizes $||w||$ in order to find the optimal hyperplane and to find the optimum distances of the support vectors one of the following options of optimizers are used including Lagrange series, Legendre series or Laplace. However, the best option seems to be using Lagrange series (see equation (3)) since it maximizes and minimizes without considering any constraints that reduce the unknowns and aids in finding the optimal hyperplane. Knowing these margins will help to reach the optimal decision separation surface.

$$L_D = \sum_{i=1}^{n} \alpha_i - \frac{1}{2}\sum_{i=1}^{n}\sum_{j=1}^{n} y_i y_j \alpha_i \alpha_j x_i^T x_j \qquad (3)$$

Where

- $\alpha_i \alpha_j \sim$ Lagrange multipliers
- Max($L_D$)
- $0 \leq \alpha_i \leq C$
- $\sum_{i=1}^{n} \alpha_i y_i = 0$
- RBF Kernel function: $K(x_i, x_j) = \exp\{-\|x_i - x_j\|^2 / (2\sigma^2)\}$

To sum up, SVM works as a statistical tool to separate data by providing it with the proper labels and inputs to train the algorithm. It also creates a hyperplane that can clearly and optimally distinguish between different features and labels, in addition to minimizing any errors due to misclassified data. Since it is a learning technique, it is based on iteration to find an optimal classifier that can clearly separate trained and non-trained data.

### 3.1.2 Transfer Learning

Transfer learning is a powerful concept for effort and time saving when it comes to training the input data. It is also valuable in the case where several images exist, and input labels were not available, thus, it works by assigning a single classifier or training set to classify the rest of the dataset. In this project, five images were conducted for each experiment, and for each image, we had to create 4-5 masks for all classes (these masks are used in the training of classifier). Creating several masks for each image is time-consuming; therefore, having only one reference to training for each experiment to classify the rest of the images is highly recommended to reduce time and effort. However, the most difficult part of the learning process is finding the right source or reference for training input (Ma et al., 2016). Thus, it is important to first visually inspect the datasets and find the one that has a variety of colors and information, in some situations a few iterations must be performed to decide the optimum source and reference for the training. Moreover, in many situations transfer learning is accompanied by lower accuracies due to the difference of input and testing data, thus having an abundance and variety of training data would be helpful and could increase the accuracy of classification. Another factor, which can boost the accuracy of transfer learning, is reducing the differences between the satellite images (as was discussed in chapter 2 in the time series analysis). If the images have nothing in common and their spectral appearances differ extensively, then a lot of features would be classified incorrectly. However, if this difference between the images is reduced by applying a filter or any other radiometric correction approach, then there is a good chance to fix possible misclassifications. Therefore, due to the filter, there will be no need to perform any transformation (of feature spaces) between the source and target datasets in transfer learning.

### 3.4 Spatiotemporal Bilateral Filter

Bilateral filter is a nonlinear filter that smooths the image while preserving the edges. It is best used in de-noising and homogenizing the colors in the image. Since the goal of this project is to reduce noises and differences between the images, to improve classification using transfer learning, a modified bilateral filter is implemented and tested as a primary step to classification. The main purpose of the test is to decide whether using this spatiotemporal bilateral filter as a prior step to SVM enhances the classification method or not.

According to (Liebold & Maas, 2016), the best option to reduce noise in an image is to use time domain in any filter used. On this study made by (Liebold & Maas, 2016), they have used the temporal domain with several filters such as bilateral, moving average and Gaussian to monitor cracks in beams. The research used 1D convolution for the temporal approach, to reduce any noise like white noise. The results showed that the best filter for noise reduction was bilateral combined with saw tooth filter for the time domain. As stated by ( Bennett & McMillan, 2005), for a static camera taking images at the same location with no moving objects, summing and weighting (in another word simple averaging) the temporal pixel values would provide an optimal solution to result pixels' true values. Despite the fact that this method is suitable to get a rough estimate of the images' contents, this averaging will produce similar outcomes at the end, where images will not explore any variations between them. To maintain the individuality of each image, the differences and Euclidian distances among images and their bands must be considered. Previous studies that used bilateral filter combined with time domain, showed an enhancement to the results and proved that it is beneficial to incorporate the filter with time. Thus, there are three domains in this filter, the first one is in the spatial domain, and since it is concerned with the positions of pixels, it is also a Gaussian filter. This filter manipulates and processes image contents based on neighboring pixels within a pre-specified window:

$$\text{Spatial:} \quad w_s = e^{-\frac{((x-m)-(y-n))^2}{2\sigma s^2}} \quad (4)$$

Where

    x, y: are the positions of the pixels in the window

    m, n: positions of the central pixel in the window

The next part is the spectral domain; it is concerned with the intensities and pixel values. Filtering in this domain is responsible for preserving the edges. The bilateral filter can operate on grayscale or color images (Tomasi & Manduchi, 1998); it is perfect for edge maintaining since it does not generate any phantom colors along the boundaries and it reduces the phantom effects if it exists. The bilateral filter operates on each band individually; thus, it preserves the individuality of each band. As mentioned before, spatial filtering smooths the image, whereas, range filtering preserve the edges. (Tomasi & Manduchi, 1998) suggested using CIE-LAB color space to minimize the effect of getting the false colors, phantoms, and shadings along the edges. The color information is used as a reference to homogenize the pixels colors within the pre- specified window. The equation of this filter is as follows:

Spectral: $$w_r = e^{-\frac{(I(x,y,t)-I(m,n,t))^2}{2\sigma r^2}} \quad (5)$$

Where

I(x,y,t)= pixel value within the window for image at date t

I(m,n,t)= pixel value of the central pixel within the window for image at date t

The last weight computed in the filter is in the temporal domain, which is the new addition in this project. It works by comparing pixel values of a single image with the same pixel in different images of different dates. The comparison comes in the form of differencing the pixels of the single image with another image by using all bands (measuring the Euclidian distance) to capture the differences between them. The following represents the equation used for the temporal filter:

Temporal: $$w_t = e^{-\frac{L_t}{2\sigma t^2}} \quad (6)$$

$$L_t = \sum_1^i \sqrt{\frac{\sum_1^b (I_{p,b} - I_{i,b})^2}{\sum b}} \quad (7)$$

Where

$I_{p,b}$ : Ip image being filtered

$I_{i,b}$ : Ii image being differenced from image Ip

i : Number of images

b : Number of bands

To sum up, the algorithm used by the three filters combined and normalized in a single equation to form the spatiotemporal bilateral is shown in equation (8). Filtering requires normalizing the values eventually to ensure all values sums up to one, thus dividing by the total weight, w.

$$\text{Filter, F} = \sum_{i=1}^{size}\sum_{j=1}^{size}\sum_{k=1}^{\#imgs} \frac{ws*wr*wt*I}{w} \qquad (8)$$

$$w(m,n,k) = ws \times wr \times wt$$

Where, w: is the total weight

There are many factors controlling this spatiotemporal bilateral filter. One of these factors is the window size; the greater it is the more blurring occurs. Since it depends on the Gaussian filter to blur based on the location of pixels, the furthest pixels to the centric pixel will be assigned less weight than those nearby. The other factors are the three sigmas, representing the amount of filtering on the image and the standard deviation of which the weights and pixels are to be distributed. For filtering in the spatial domain, increasing $\sigma_s$ will increase smoothing and may lose a lot of information. On the other hand, lower $\sigma_s$ will lead to less blurred image and may not be able to capture or reduce all noises in the image. Additionally, since blurring in the spatial domain is the same as Gaussian filter it will not preserve the edges as $\sigma_s$ increases, so care must be taken when deciding on a value of $\sigma_s$. Similarly, for $\sigma_r$ filtering in the range domain, the higher it is the more blurring will be on the boundaries, and vice versa. Finally, since $\sigma_t$ is the main investigation area this research, several parameters must be tested in different trials to decide the optimal parameter. For this work in particular, the window size, $\sigma_s$, and $\sigma_r$ were fixed into moderate values that will not hide a lot of details, but filters at the same time and $\sigma_t$ was tested for several parameters to attain solid conclusions and for comparison purposes.

Chapter 4: Experimental Results

In chapter 4, the results of the three experiments are mentioned in detail, the spatiotemporal bilateral filter results will be presented in the beginning. After using the filter as a preprocessing step to improve the images, the classification results will be demonstrated. Finally, evaluating the technique used required accuracy assessment and a performance measurement, which will be discussed and analyzed in the last section.

**4.1 Preprocessing steps**

As mentioned in chapter three, there are ten multi-temporal images with different dates conducted from Landsat 8 and other five from Planet satellite. The main preprocessing steps made in the research is pan sharpening applied on low-medium resolution images of Landsat 8. The next step is registration for the precise alignment of images. Lastly, creating masks to be used as labels and inputs for the training in the classification phase.

**4.2 Spatiotemporal Bilateral Filter**

Since transfer learning is used in classification, a lot of data is prone to misclassifications due to differences of the training source from the target data. Thus, it is very important to find a filter that settles and reduces the radiometric differences between the images. This filter will be used as preprocessing operation to result less variant images and better accuracies in transfer learning outcomes, in addition, to provide a visually enhanced classification outcomes. The filter has been implemented on five images, where each image is being filtered based on the information it has (intensities and geometrical distances), in addition to the contribution from the rest of the images. The three domains contributing in this filter are spectral, spatial and temporal as shown in Figure 7. The spatial and spectral domains form the regular bilateral filter, and are responsible for noise reduction and homogenizing the colors within a region while maintaining the edges. Of course, the more weight applied in both domains, the more filtering and blurring are introduced. Therefore, if the details are to be maintained the bandwidths parameters ($\sigma_s$ and $\sigma_r$) must be set to moderate and reasonable values. The new addition to this filter is in the temporal domain, where the filtered image uses information from the rest of images by finding the differences between bands and assigning weights accordingly (the algorithm was discussed in chapter 3). Since differencing a single band is not sufficient to measure the variations between images, measuring the differences between the bands in the form of Euclidian distance is more useful. These differences allow capturing the major variations between images, therefore, maintaining or preserving the structure and color components of the image. In this work, the parameters $\sigma_s$ and $\sigma_r$ have been fixed to (7 and 50) with window size (5) for all of the experiments except for experiment II the window size used is (11). The temporal parameter $\sigma_t$ has been tested on all images with values ranging from (0 to 0.9) with increment 0.1.

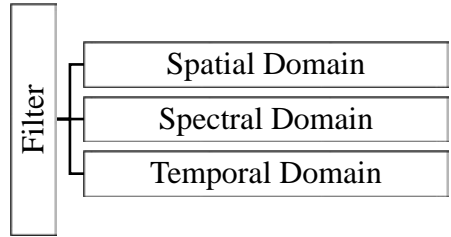

*Figure 7. Filtering Steps*

### 4.2.1 Experiment I

The filtering results of experiment I are depicted in Figure 7 through Figure 11. There are apparent differences before and after applying a spatiotemporal bilateral filter; first, it can be observed that images will start to lose their original appearances over $\sigma_t$. At low values of $\sigma_t$ (around 0 to 0.3), the filter will preserve the original structures and appearances of the images (see a, b, c, & d in Figure 7 to 11), however, as $\sigma_t$ increases it will be hard to distinguish between the images. Higher values of $\sigma_t$ (larger than 0.4.) assign more weights and result less variant images (see e, f, g, k, i, j in Figure 7 to 11); the images are forced to have similar pixel values, thus all images will look almost the alike with minor variations between them.

The main benefits of this spatiotemporal filter are that it reduces the noises, normalizes the intensities and facilitate the transformation of transfer learning. Moreover, it reduces any undesired atmospheric effects such as clouds as in image 1 (see Figure 7), haziness as in image 3 (see Figure 9), shadows covering barren lands that might be falsely interpreted in the classification as buildings due to the close pixel values (see Figure 7 and Figure 8). Another benefit to this technique is that it reduces the high reflectance of water surfaces due to sunlight effect, which turns the water color into white and causes misclassification (see Figure 10). The filter fills the gap by measuring the differences between pixels' values in the images; the higher weights applied the more filtering will happen, and the less natural and atmospheric effects will be encountered. One more observation is that the filter operates based on the majority of features in a specific location, and transforms the image on that basis. To explain this more, notice that most of the images have barren lands in the mid-upper part of the images, thus, as $\sigma_t$ increases, all the filtered images will have barren land in that region (similarly to other features and classes). Generally, applying the filter to these images can present a good estimate on the locations of natural landscapes and cities on the ground.

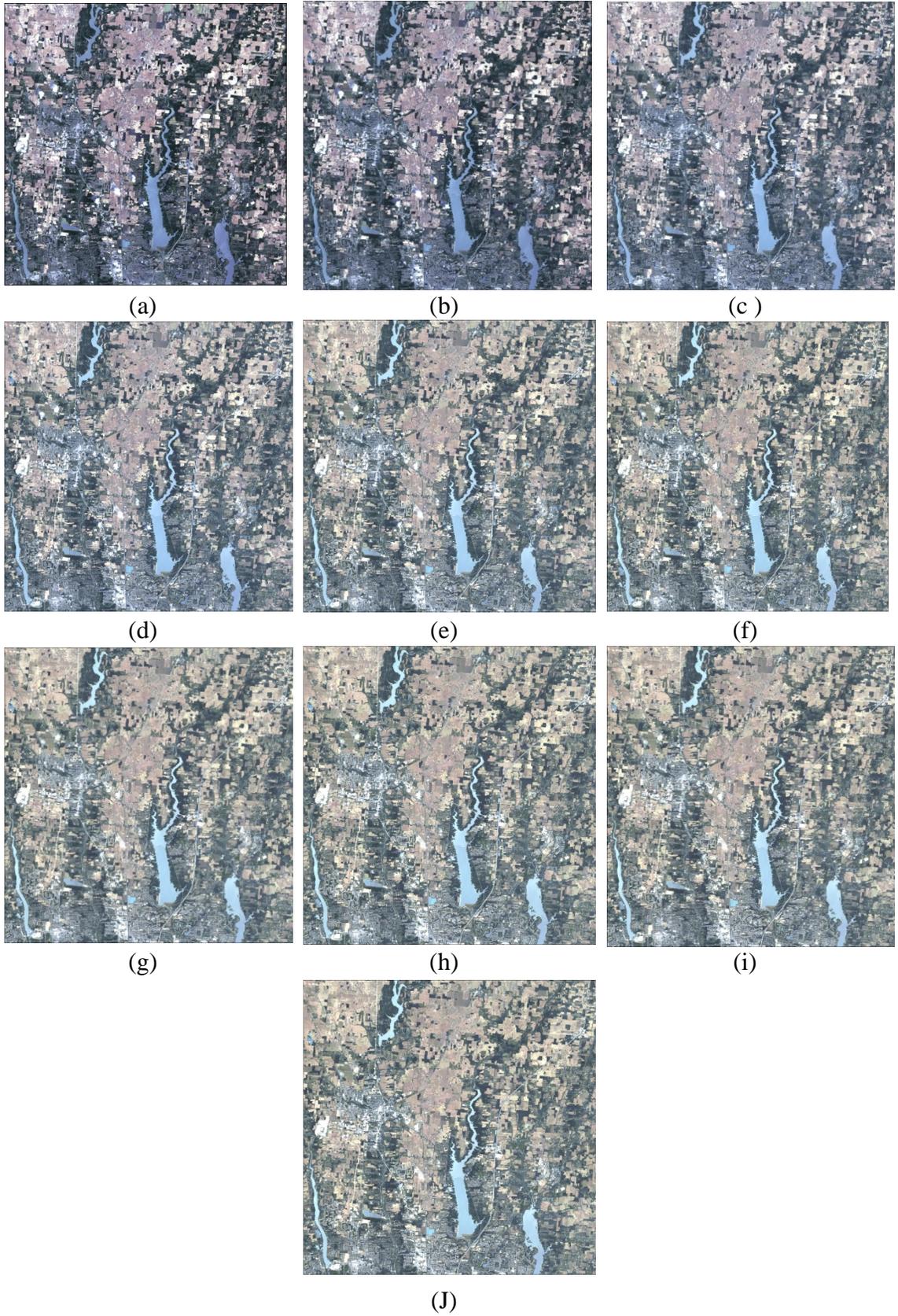

*Figure 8. Experiment I Filtering Results of Image 1. Filtering Results of Parameters σ$_t$ from 0 to 0.9 marked from a to j*

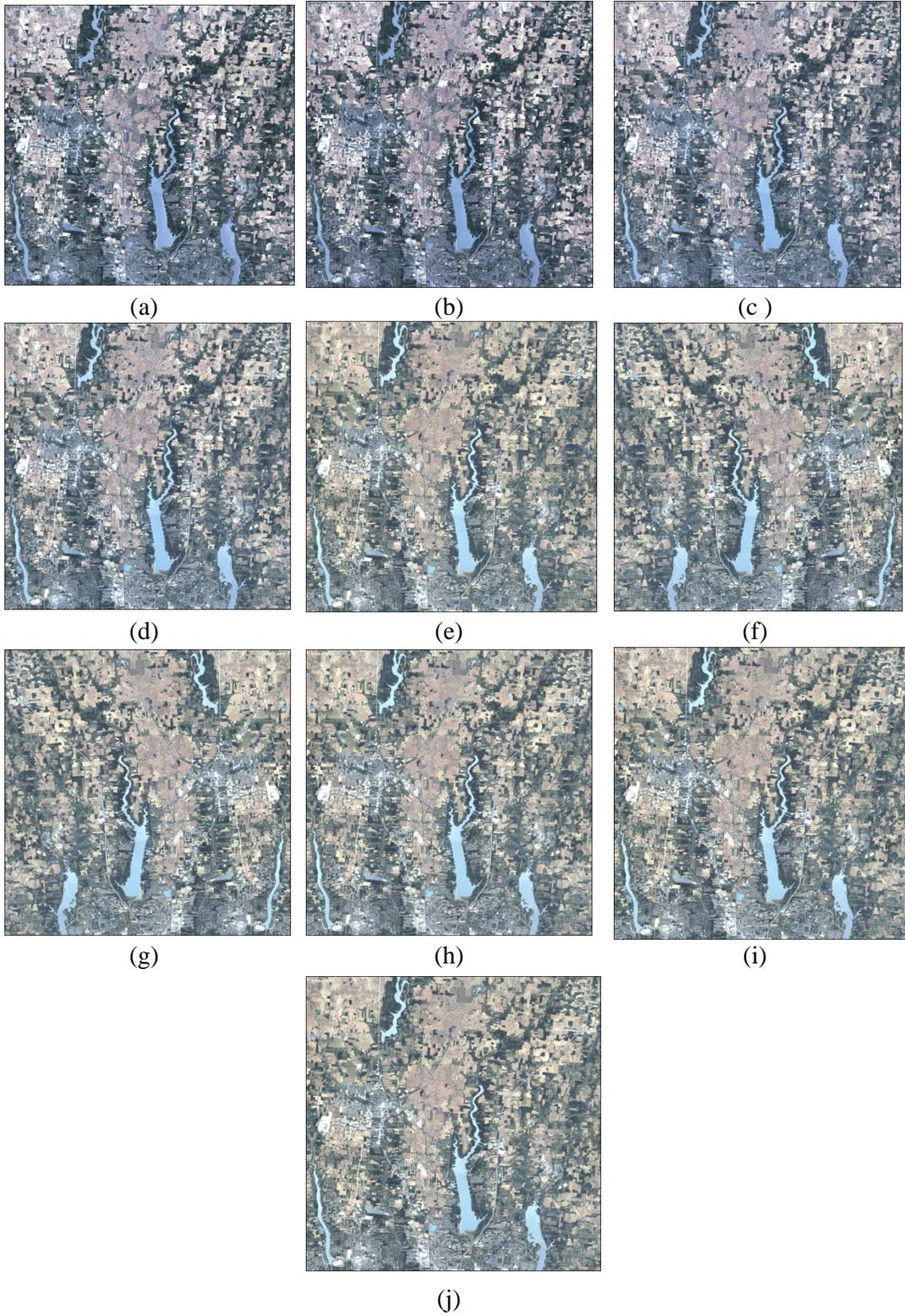

*Figure 9. Experiment I Filtering Results of Image 2. Filtering Results of Parameters $σ_t$ from 0 to 0.9 marked from a to j*

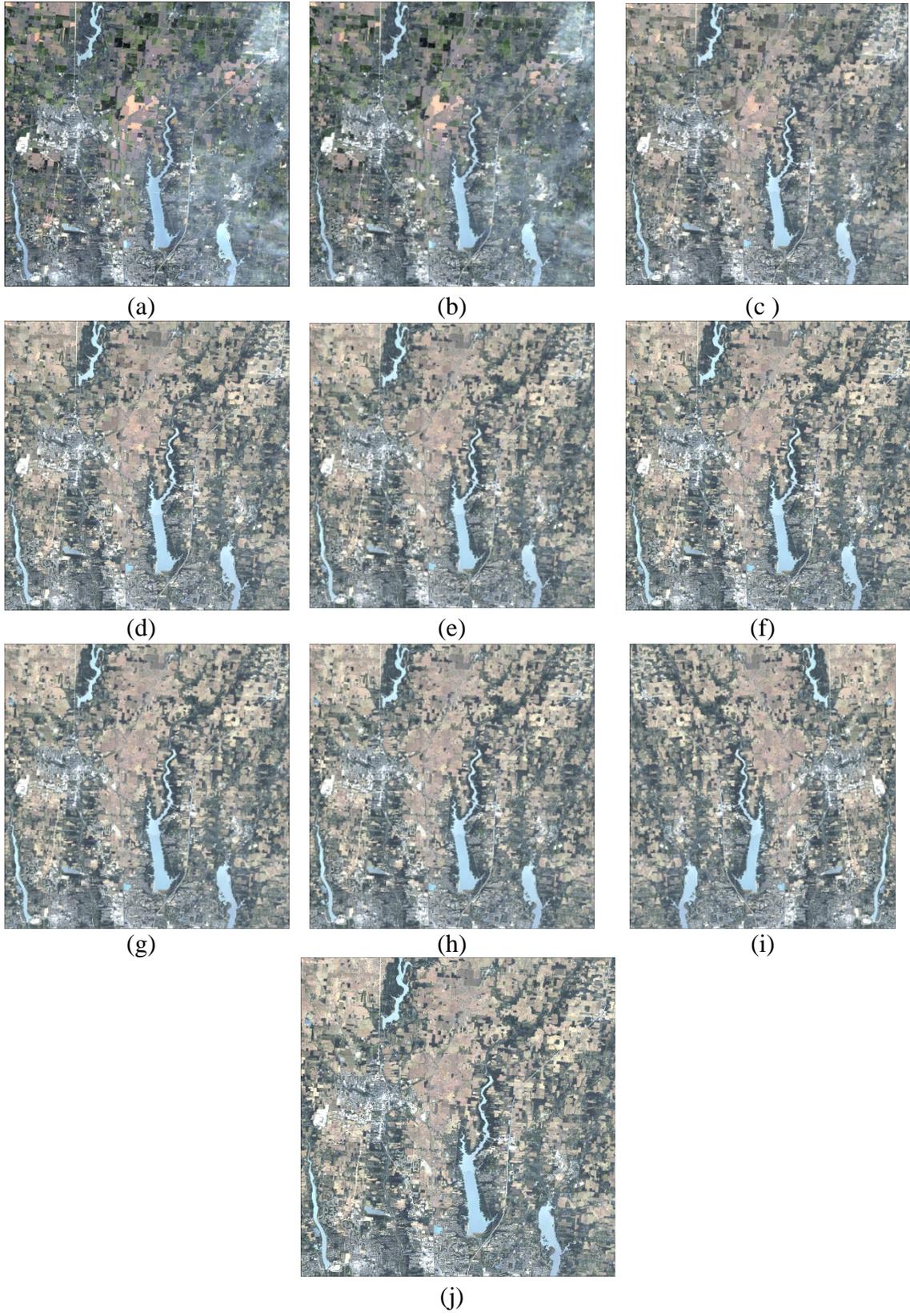

*Figure 10. Experiment I Filtering Results of Image 3. Filtering Results of Parameters σ$_t$ from 0 to 0.9 marked from a to j*

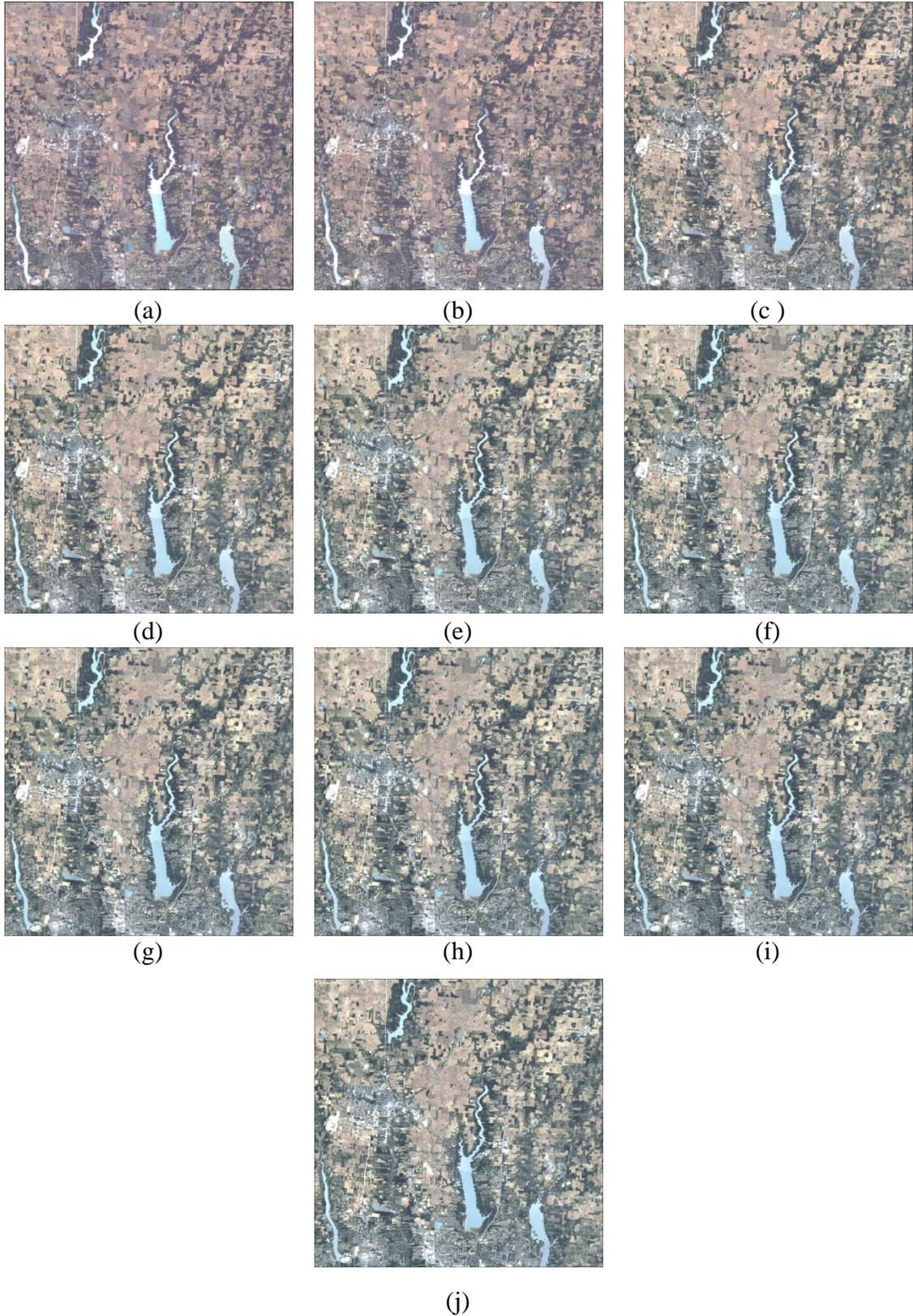

Figure 11. Experiment I Filtering Results of Image 4. Filtering Results of Parameters σt from 0 to 0.9 marked from a to j

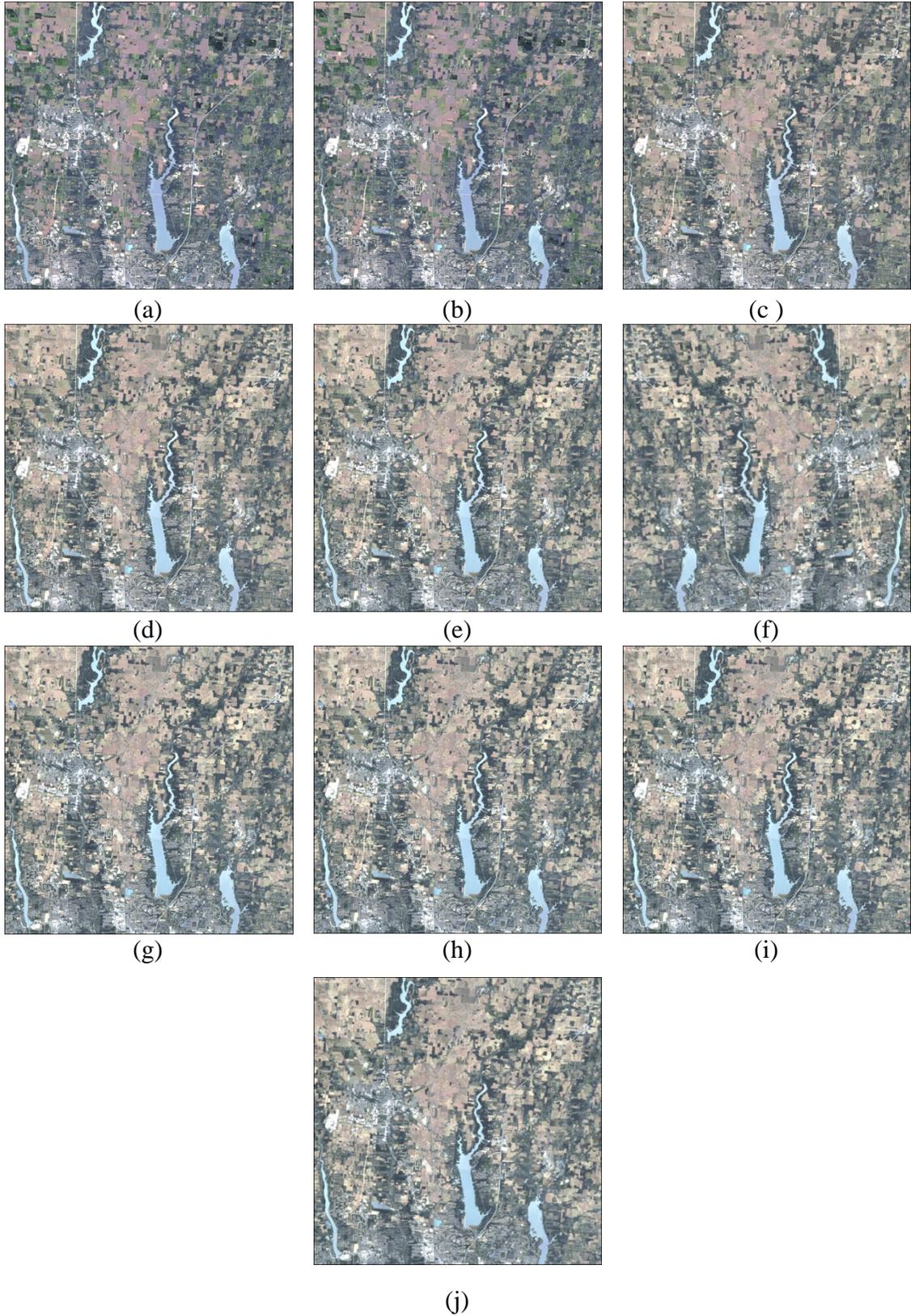

*Figure 12. . Experiment I Filtering Results of Image 5. Filtering Results of Parameters σt from 0 to 0.9 marked from a to j*

### 4.2.2 Experiment II

The filtering results of experiment II are shown in Figure 13 to Figure 17, and similar conclusions to experiment I can be drawn in experiment II. The differences between the images will be reduced and the images will start looking alike with minor variations as $\sigma_t$ rises. Moreover, the five images will maintain their original looks within the bandwidths $\sigma_t$ from 0 to 0.3; beyond these values (after $\sigma_t \geq 0.4$) all images will start to look alike with slight disparities among them. It can also be seen that most of the images have natural features at the lower part of them and manmade structures at the upper part, and since the filter works by giving weights to those of pixels with similar values, as $\sigma_t$ increases in value, all images will have these same components at the same location. Moreover, the filter will improve any negative effects on the quality of the images that would cause misclassifications. For example, haziness as in the lower left corner of image 3 (see Figure 15) will be reduced, and sunlight effect on the river in image 4 will also be corrected at $\sigma_t=0.2$ (see Figure 16). In addition to these corrections, any small clouds, shadows, or any blockage to the scene will be either eliminated or reduced by the filter. However, because the data in experiment II is mostly urban region with residential areas, buildings, and roads, filtering will cause some features with similar pixel values to be combined and blended together such as roads and small houses. As a result, it would be hard to distinguish between these two features, thus it would be more convenient to create a single class for these two features.

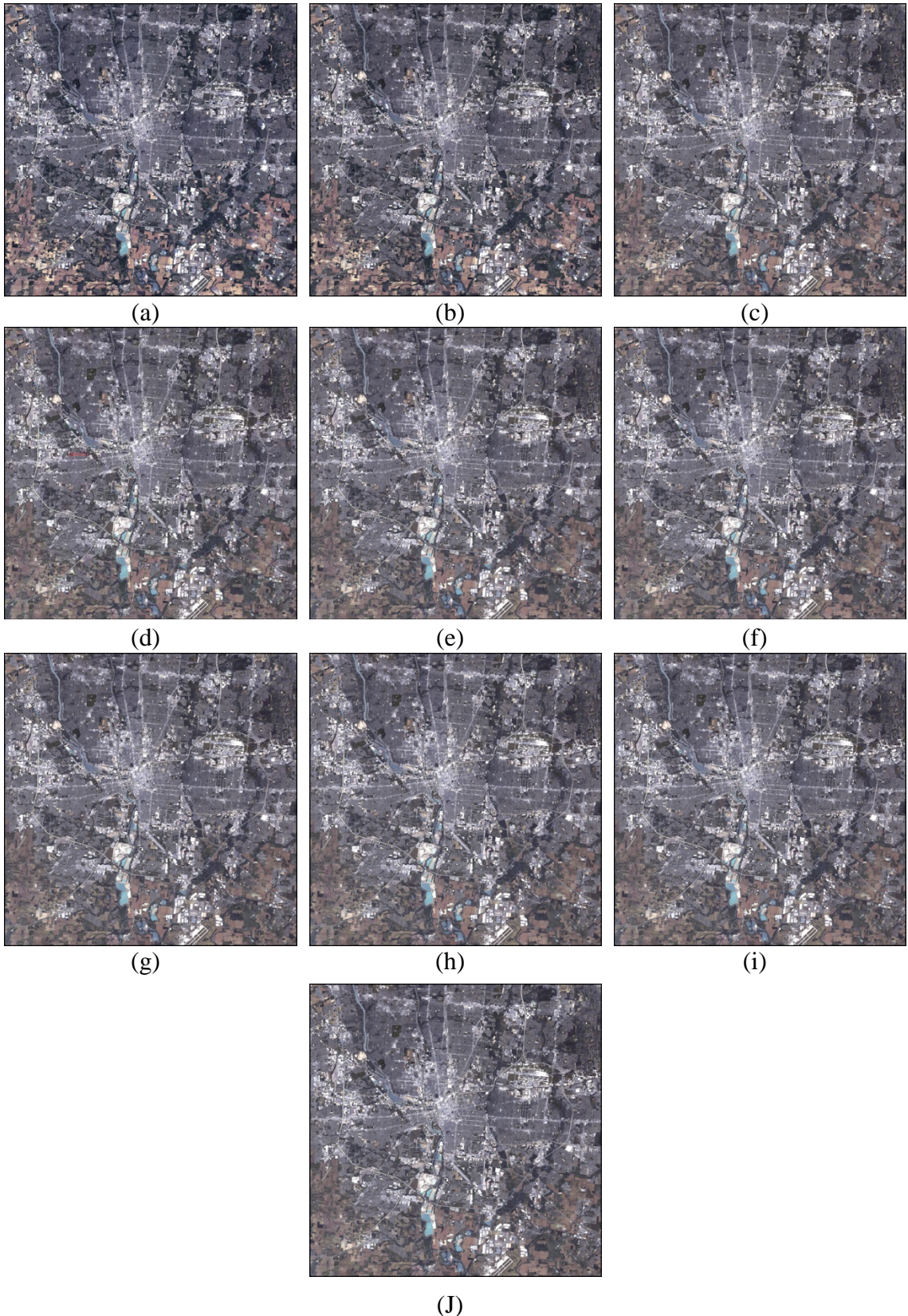

*Figure 13. . Experiment II Filtering Results of Image 1. Filtering Results of Parameters σt from 0 to 0.9 marked from a to j*

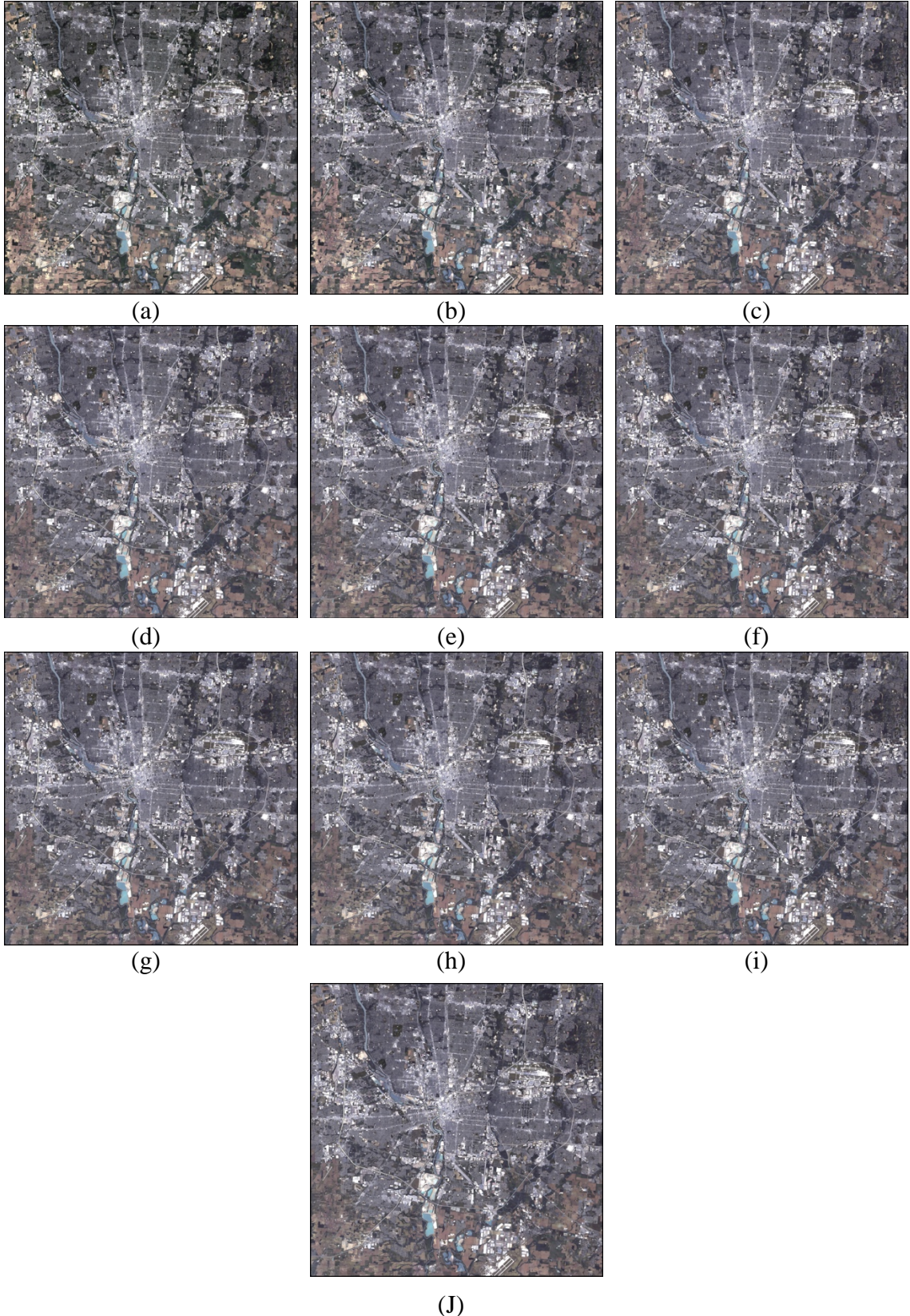

*Figure 14. Experiment II Filtering Results of Image 2. Filtering Results of Parameters σt from 0 to 0.9 marked from a to j*

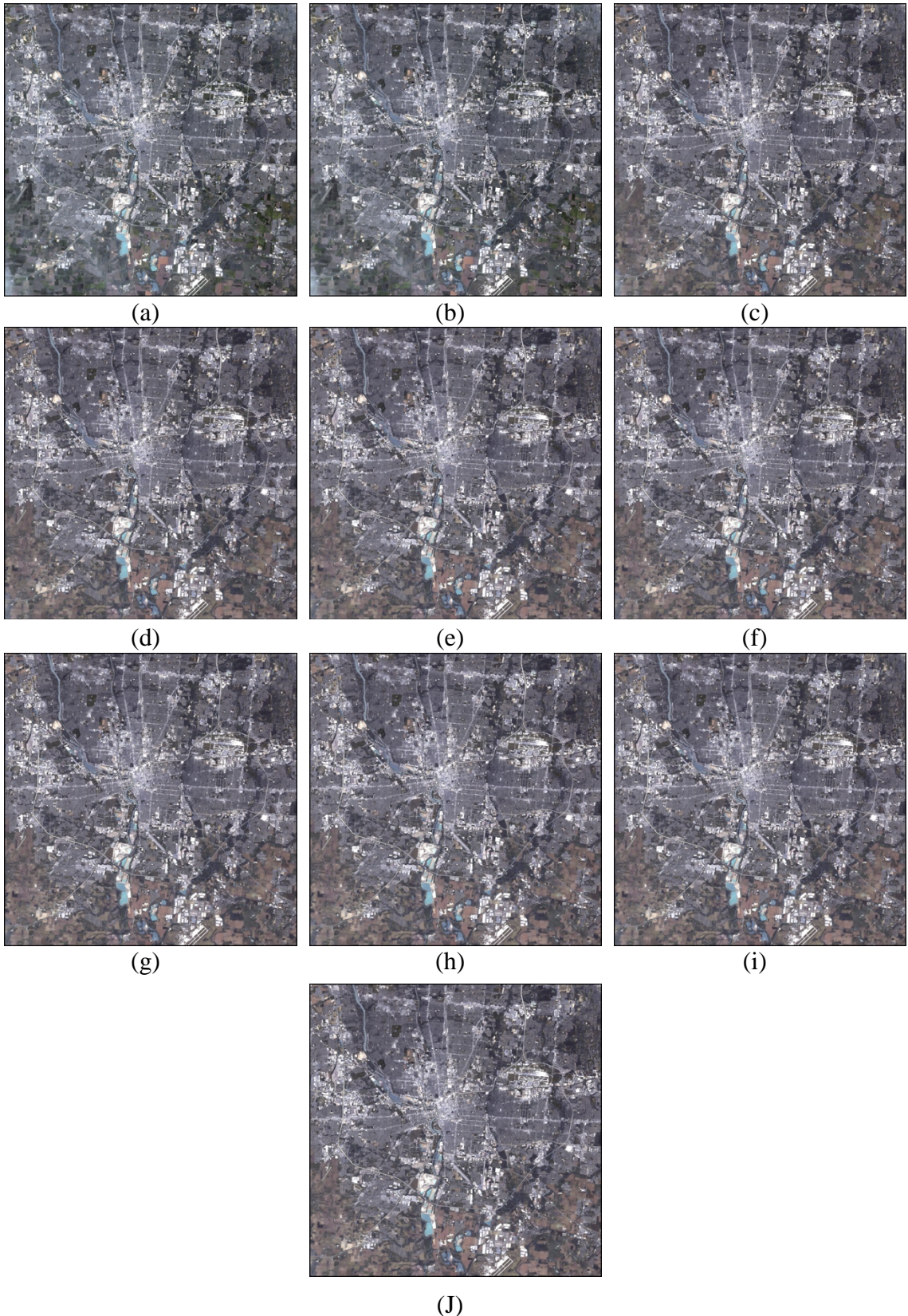

*Figure 15. Experiment II Filtering Results of Image 3. Filtering Results of Parameters σt from 0 to 0.9 marked from a to j*

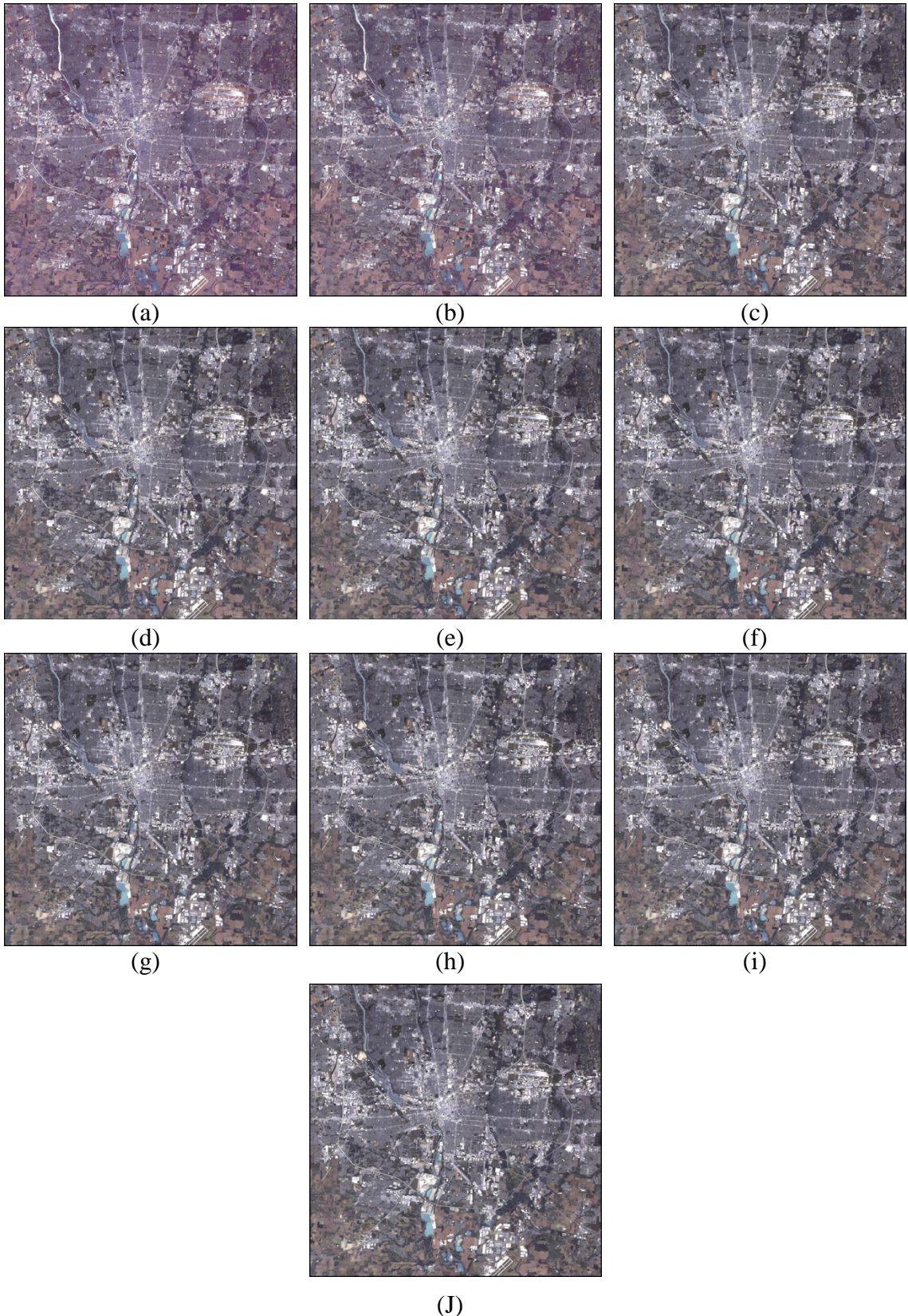

*Figure 16. Experiment II Filtering Results of Image 4. Filtering Results of Parameters σt from 0 to 0.9 marked from a to j*

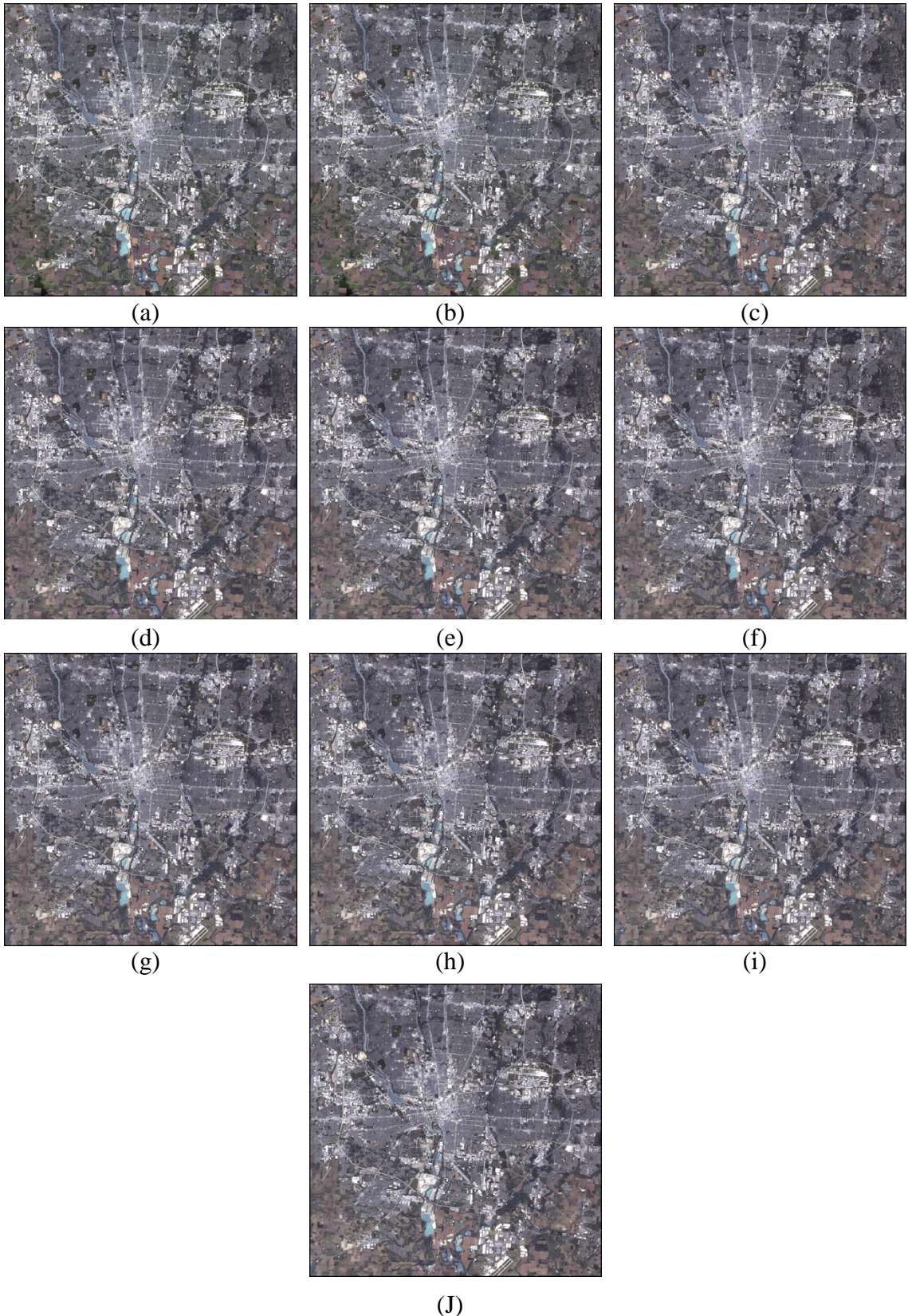

*Figure 17. Experiment II Filtering Results of Image 5. Filtering Results of Parameters σt from 0 to 0.9 marked from a to j*

### 4.2.3 Experiment III

High-resolution images do not differ in the process of the filter from medium resolution images, and so is the findings, the results of filtering are shown from Figure 18 to Figure 22. High-resolution images involve studying small geographical areas, thus, if the quality of the image is adversely affected, it will be highly noticeable since it covers large space of the image. The spatiotemporal bilateral filter in its turn will carry this noise to the rest of the images. To demonstrate more, the cloud cover in image 2 (see Figure 19) was passed to the rest of the images starting at the bandwidth $\sigma_t$ around 0.4 to 0.5. Moreover, the urban reflectance of buildings and roads is considered another source of noise in the image and can be a reason for incorrect classifications. Therefore, images that experienced high reflectance like in, image 2 and 5, soon after parameter $\sigma_t$ is around 0.1 to 0.2 will be less reflective (see Figure 19 and Figure 22). Some edges will be strengthened by this filter this can be noticed in the filtered results, and at the boundaries between roads and other features in particular. Finally, scenes with a specific feature at the same location will be filtered to that precise feature as $\sigma_t$ increases, for instance, the five images have common barren land in the lower left corner, thus more filtering will turn this area into barren land.

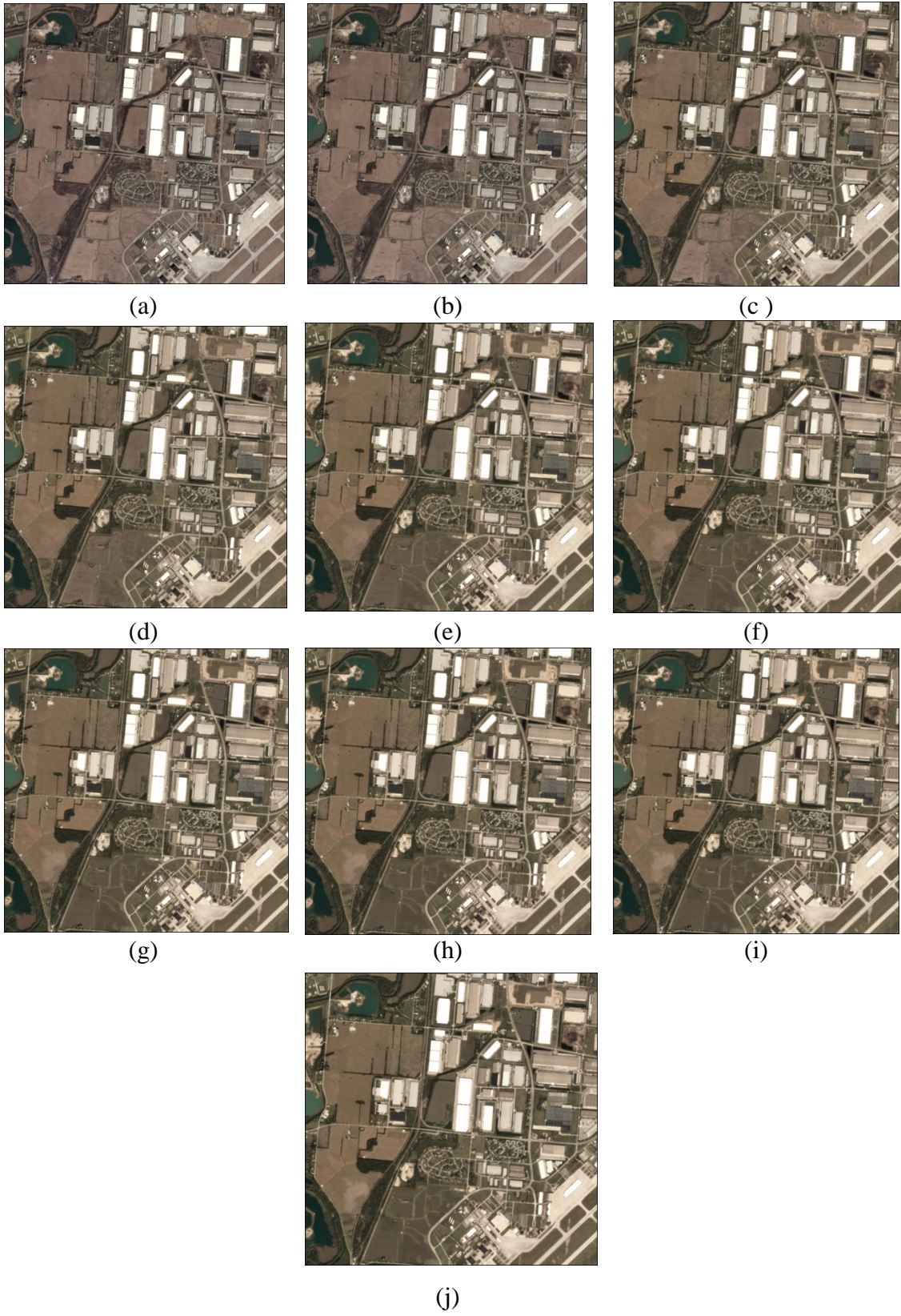

*Figure 18. Experiment III Filtering Results of Image 1. Filtering Results of Parameters σt from 0 to 0.9 marked from a to j*

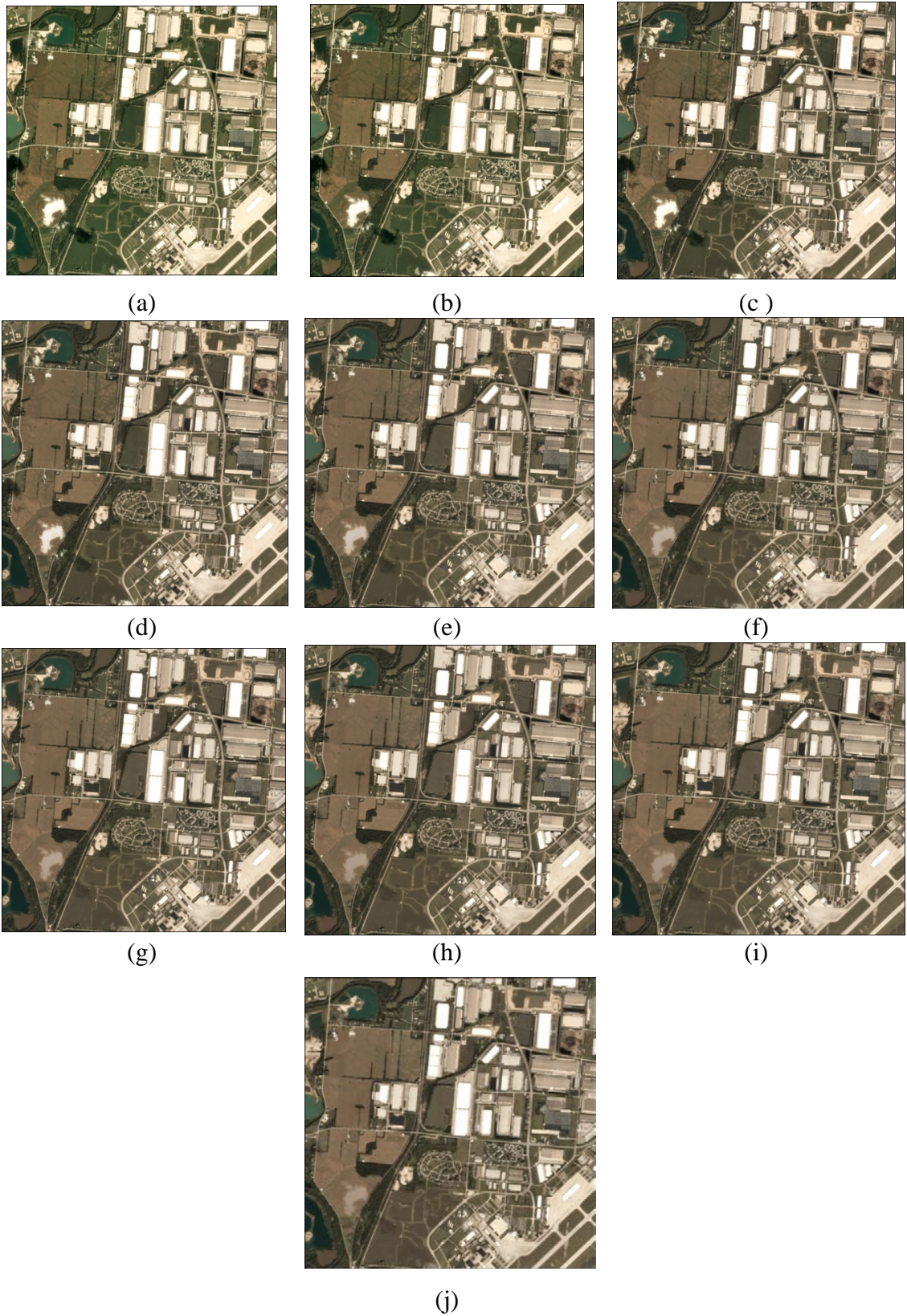

*Figure 19. Experiment III Filtering Results of Image 2. Filtering Results of Parameters σt from 0 to 0.9 marked from a to j*

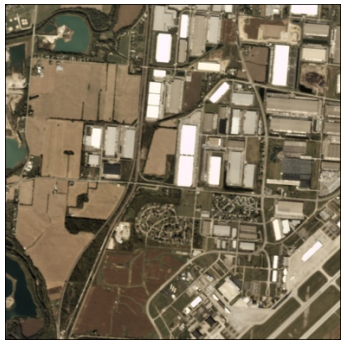
(a)
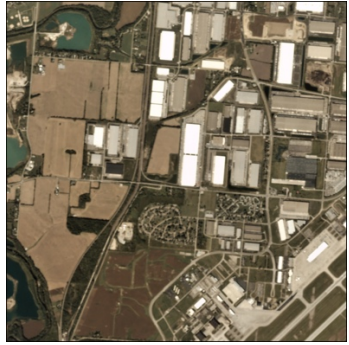
(b)
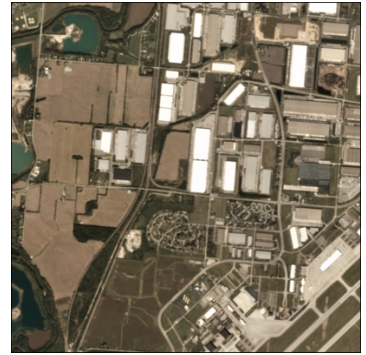
(c )

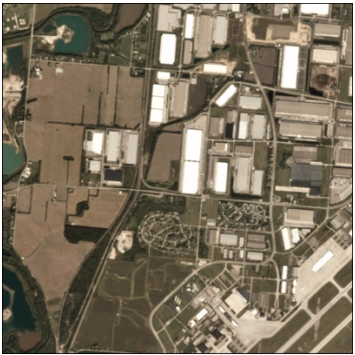
(d)
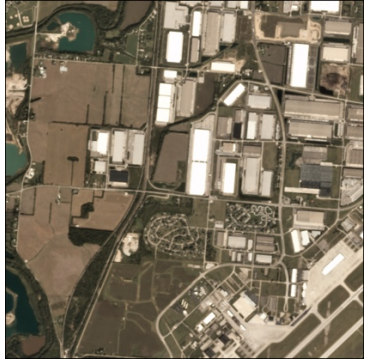
(e)
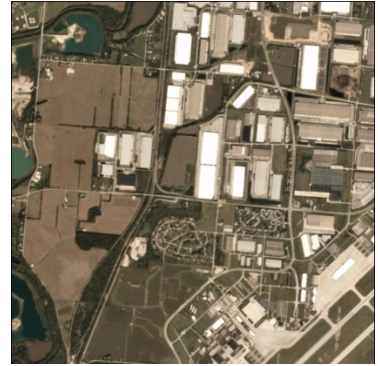
(f)

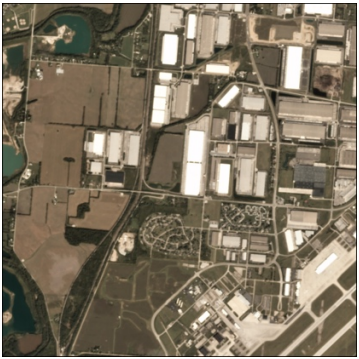
(g)
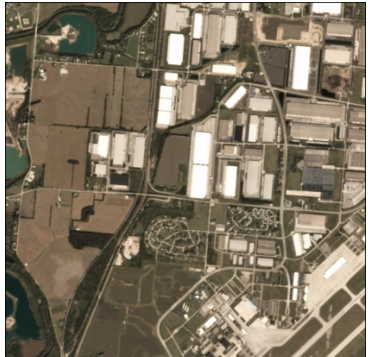
(h)
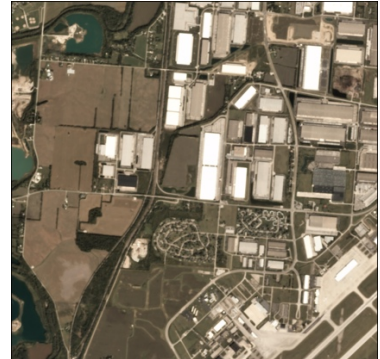
(i)

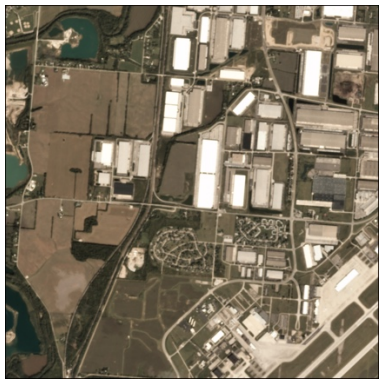
(j)

*Figure 20. Experiment III Filtering Results of Image 3. Filtering Results of Parameters σt from 0 to 0.9 marked from a to j*

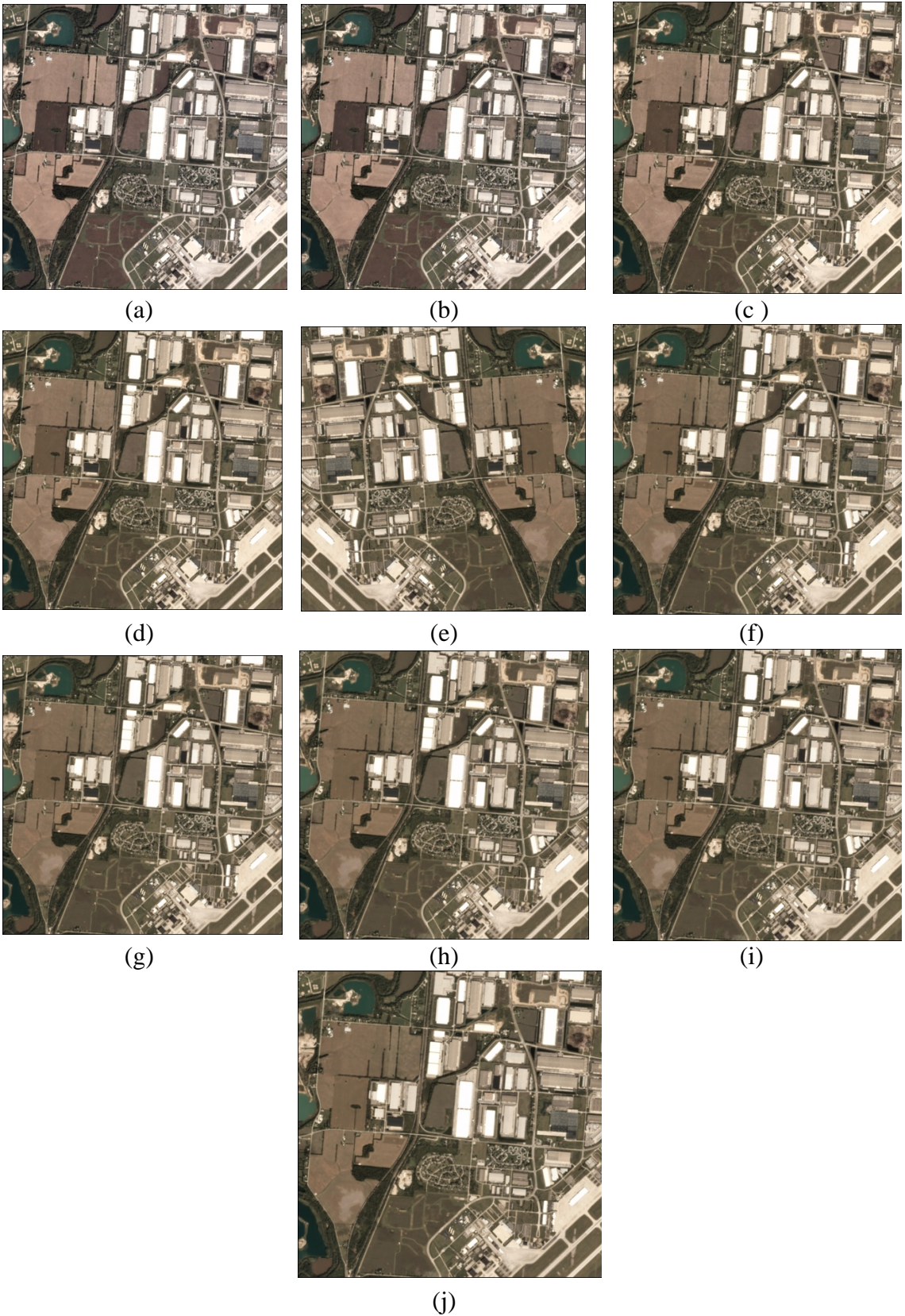

*Figure 21. Experiment III Filtering Results of Image 4. Filtering Results of Parameters σt from 0 to 0.9 marked from a to j*

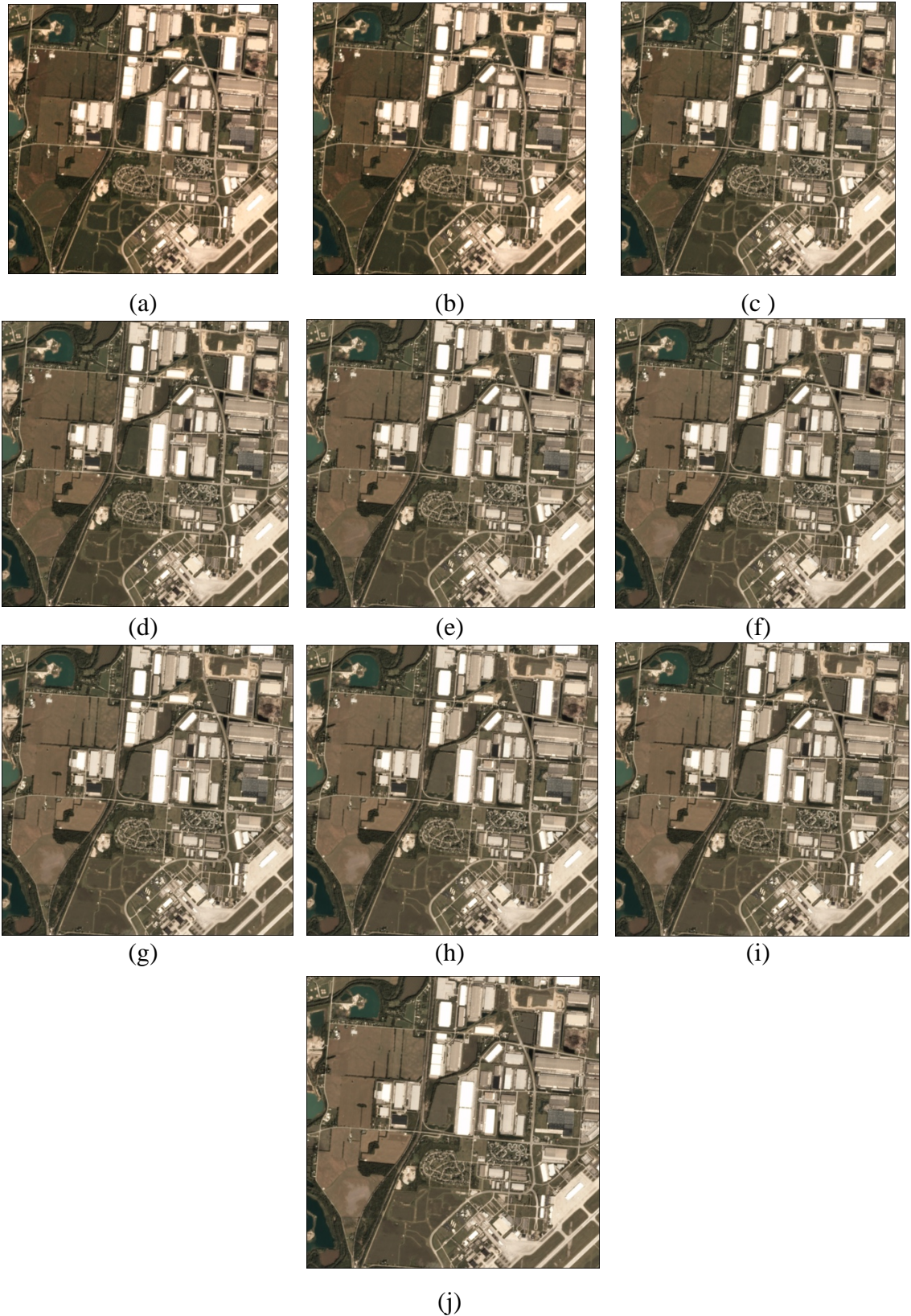

Figure 22. Experiment III Filtering Results of Image 5. Filtering Results of Parameters σt from 0 to 0.9 marked from a to j

### 4.3 Classification using SVM and Transfer Learning

The goal of this project is to enhance the classification results using transfer learning, and this is the main reason for developing the filter. The newly developed filter decreases the radiometric differences between the images, as well as the noises within them. Reducing the noises will improve classification in the classical way of learning (individually classified) or transfer learning (all images classified based on a single classifier/training set). However, decreasing the radiometric differences between the images will make them more uniform and will improve the transfer learning in particular. To illustrate how this work, remember that the filter forces the images to look alike as we increase the temporal parameter, and now that the images are more alike, the features in the target image are more recognizable by the training set. Further examples and explanations on how this filter enhanced classification using transfer learning are demonstrated in the next three sections.

In this experiment, one-against-all (multi-class) SVM was adopted to train all classes, in addition to radial basis function RBF as the kernel function. Furthermore, all bands were processed in the classification (the 8 and 3 bands from Landsat 8 and planet images respectively); it also has been shown in (Otukei & Blaschke, 2010) the handiness and effectiveness of using all bands in the classification. In the training phase, the features used were simple color values from the pixels in the images; the masks were created in a preceding step and were used to create the labels and locate features'. A sample number of the labels was defined to decide the number features going to be used in the training, while the rest will be used in the testing phase of the classification. The next step of the testing phase involves the unlabeled pixels that were not used in the training step. The testing step along with the known labels support the final stage, which is the validation, where a confusion matrix for all classes is constructed to evaluate the true and predicted results of classification.

Transfer learning is a time-saving approach that uses past experience of ready trained data set and apply it on a new unlabeled and untrained datasets, thus, reduces time spent on creating labels. In this work, transfer learning was used to classify images from different dates and types (the filtered and the original images); the classification steps are summarized in Figure 23. Despite the advantage of transfer (saves time and effort), it still reduces the accuracy of the classified data by about 10% in the first two experiments and 5% in the third experiment. This drop-in accuracy is noticed by comparing classification results of images classified individually (using their own labels and masks) with others using transfer learning (using single training data or classifier to classify the rest of the images). The reason for this drop in accuracy was discussed earlier in chapter 2 and 3.

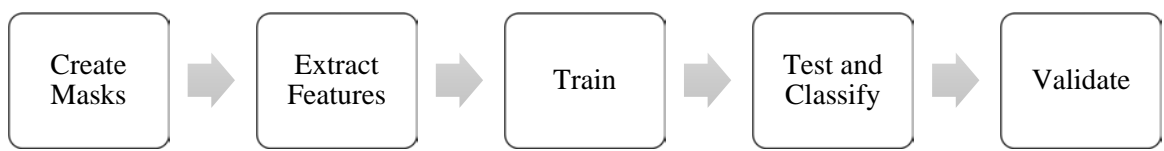

*Figure 23. Classification Steps*

### 4.3.1 Experiment I

To view any improvements in the classification results, an important comparison is made between classification results of the original and the filtered data, in addition to between the filtered data. Overall, the filter proved its ability to smooth the images and reduce the falsely detected features in classification (The classification results are depicted in appendix-A part(A)). There are two kinds of learning in this project, the conventional way (the individually classified) and the transfer learning, and the outcomes of the prior method are more accurate and close to the actual components in the images than the latter. However, transfer learning is used to reduce the computational complexity and time spent on creating labels and training datasets.

Choosing a reference image for the training and transfer learning is critical, thus, the reference image must be carefully chosen in a way that it encompass all possible colors of any feature. In this experiment, for example, image 3 has three colors for barren land that matches the same feature colors in the rest of the images. Unlike image 2 where barren land, in that case, has only one color and mostly disturbed by cloud shadows. A similar observation was noticed for water surfaces in the original image 4; it is very bright (white in some areas) and when training using this image, it might incorrectly classify some of the bright features in other images as water. Thus, having a variety of colors for one feature in the training is a demand since it supports and enhances classification. In this research, image 3 was chosen as a reference for training and transfer learning since it has various color options for many features such as vegetation and barren lands.

By monitoring the four classes in the classified images, it can be seen that most classes subject to change over this filter are the seasonal varying objects such as vegetation, barren land, and small lakes or ponds. This is mainly due to seasonal variations that influence the appearances of images, for example, in summer the image is mostly green (image 3 and 5), end of winter and early spring it has no vegetation, so mostly barren, etc. (see Figure 2). On the other hand, the impervious surfaces such as roads and buildings are more stable and do not experience any change over seasons. Hence, it gives an indication of the locations of the infrastructures or any urban sprawl. Looking closer at the smallest parameters of the filter $\sigma_t =0.1$ to $0.2$ and its effect on classification, these parameters preserve the individuality of the dataset while filtering. Thus, it makes an ideal filter to monitor any environmental and urban changes between the images during classification. Despite the fact that the individually classified results of the filtered images do not have higher accuracies than the original data (see Table 3), the transfer learning shows better accuracies of the filtered data at $\sigma_t \leq 0.3$ than the original (see Table 4). This is mainly because the filter adds more resemblance to the images, and provides common and close intensity values between them, thus better transfer learning. Generally, the filter has shown some remarkable enhancement to the classification outcomes in transfer learning. One of the challenging regions in classification is the boundaries of features, and according to a study by (Cusano, Ciocca, & Schettini, 2003) the most common misclassifications using SVM are around the boundaries where the pixel might be mistakenly interpreted to a different feature (pixels at the boundary might have mixed color of the two sides). Therefore, a spatiotemporal filter makes an ideal approach to reduce the misclassifications around the edges and boundaries, in addition to correct many misclassifications in the original results in several areas. Among these corrections of classification map (using transfer learning) due to the filter are:

1) Barren lands are more likely to be misclassified to impervious surfaces, due to the similar pixel values around their edges and if they experience any metrological conditions (such as wetness or shadows covering them). Nevertheless, as σt increases, the classification outcomes of barren lands are improved and corrected to their right classes (see Figure 27 and Figure 24).

2) Some parts in residential areas are interfered with barren lands as seen in the original image classification result see Figure 25, while in fact, it is nothing but roads and buildings with vegetation in between. Thus, applying this filter and increasing the values of σ$_t$, will correct these features and classify them to their true class (see Figure 25).

3) Greenish water surfaces (due to any vegetation or organic matter) sometimes are misclassified as vegetation. The filter gives these surfaces the normal or average color of water bodies, thus will be classified properly as water (see Figure *28*).

4) Images with haziness will experience less haze by this filter (see Figure 26).

5) Images with small clouds blocking the scene were corrected (see Figure 24).

6) The appearance of some features such as minor roads, narrow rivers, etc. (see Figure 24).

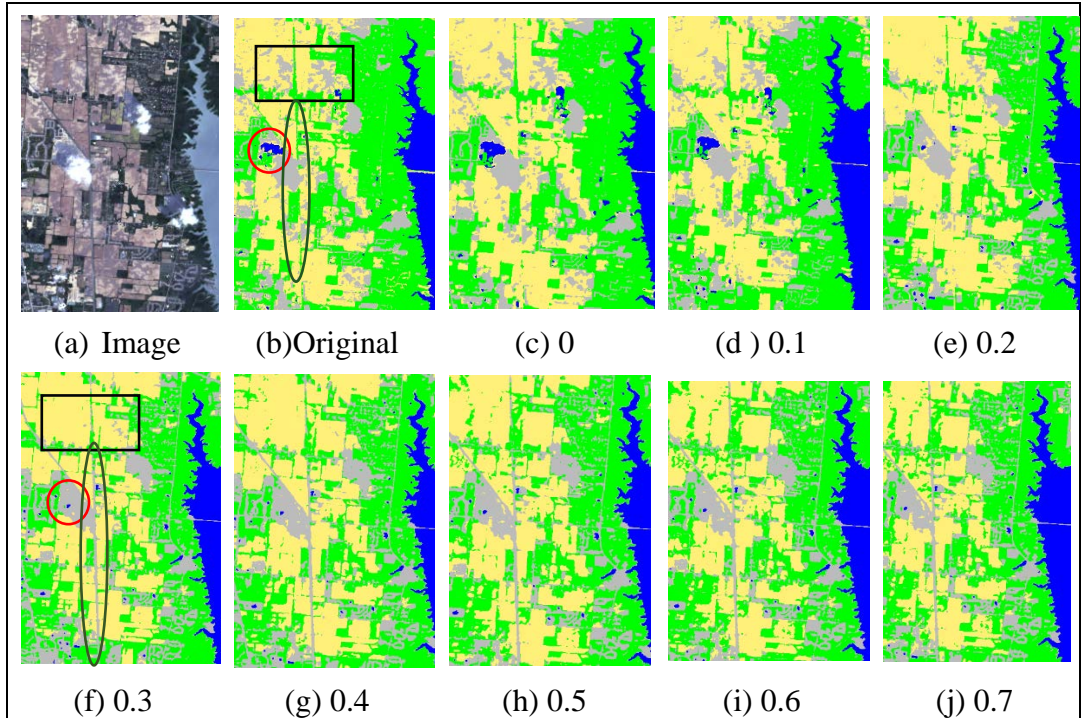

*Figure 24. Section of image 1. Enhancement of some classes by the spatiotemporal bilateral filter.*

*Note: The red circle: cloud disappearing over time, black rectangle: barren land covered by cloud shadows falsely detected as impervious surfaces, and the ellipse shows appearance of a feature (road).*

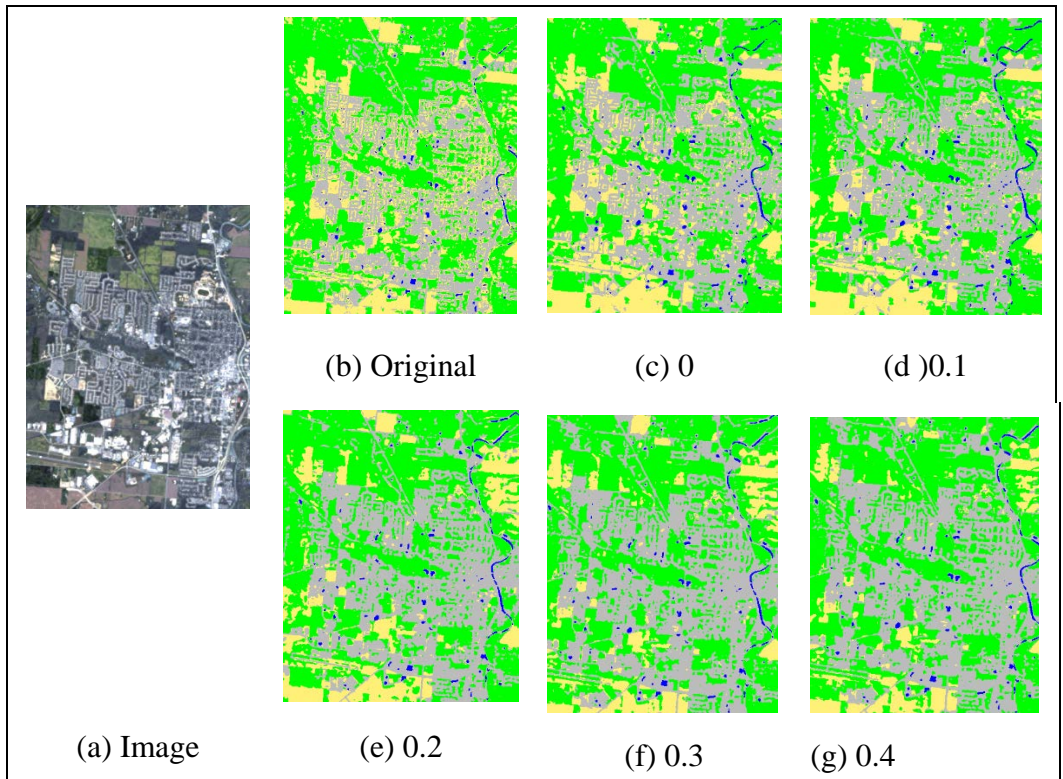

*Figure 25. Section of image 3. Residential areas corrected from false barren lands (transfer learning).*

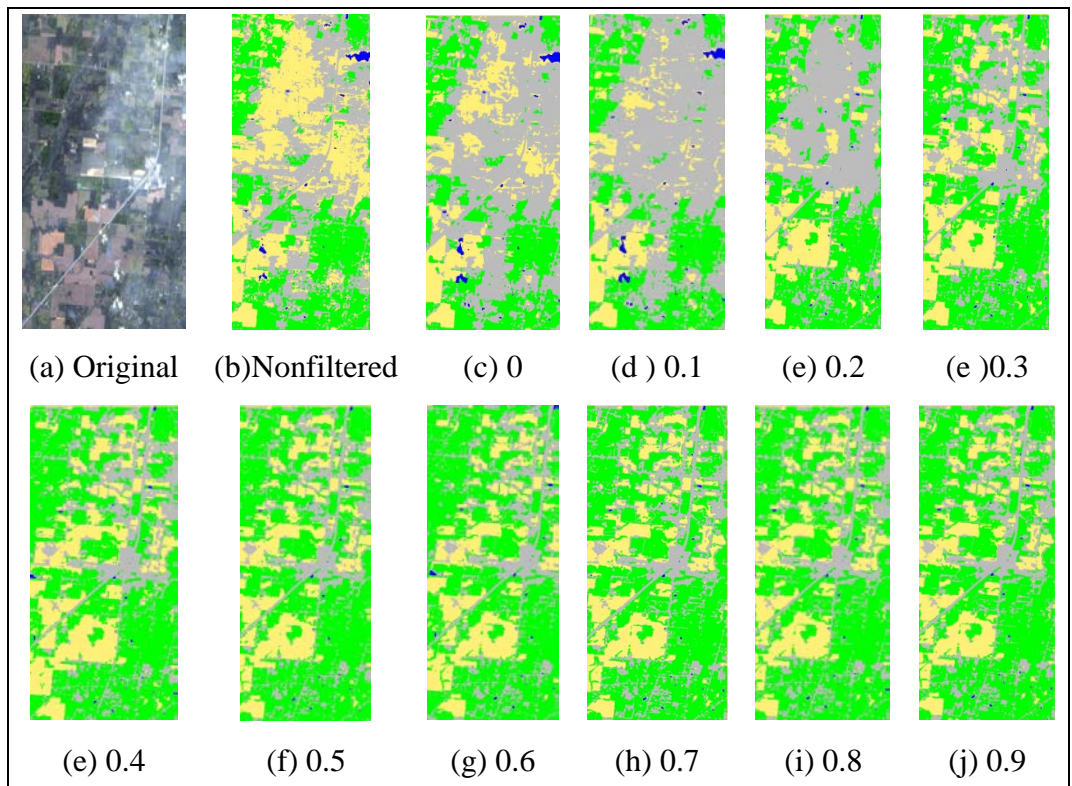

*Figure 26. Section of image 3. Correction of haziness by the spatiotemporal bilateral filter.*

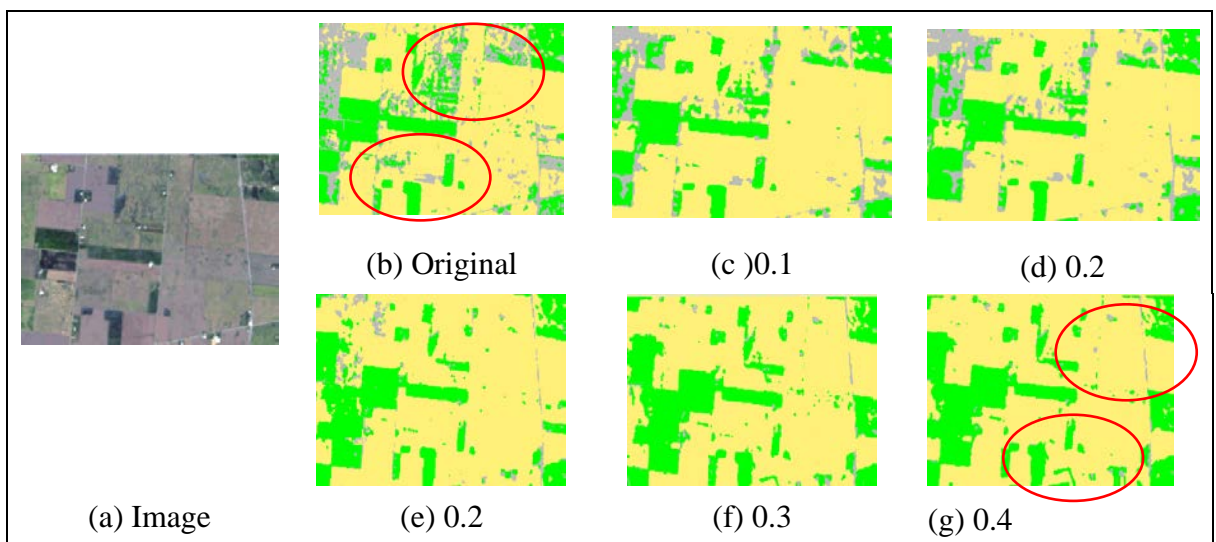

*Figure 27. A section in image 5. Boundaries are returned to their original components by the spatiotemporal filter.*

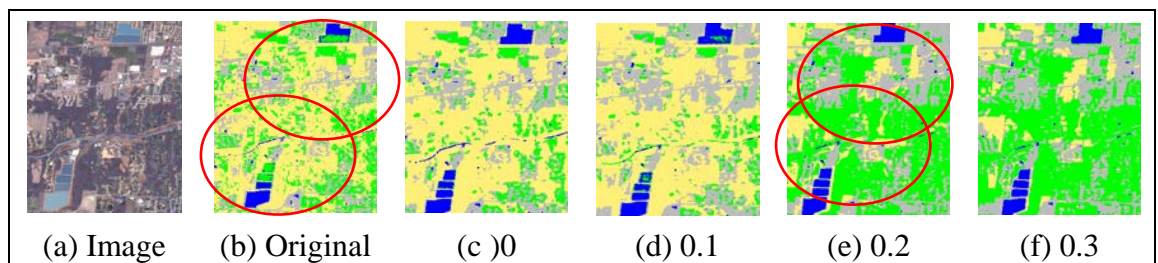

*Figure 28. Section of image 4. Water surfaces corrected by the spatiotemporal bilateral filter.*

### 4.3.2 Experiment II

After filtering the images from experiment II, the classification takes place to observe how the filter influences the outcomes. The classification results of the individually trained and classified images are better than the transfer learning and this is obvious in most of the figures (See appendix-A Part (B)). Image 5 has been selected as the reference classifier for classification in transfer learning, for the same reasons mentioned in experiment I. Since image 4 is the most variant image from the rest of the dataset, it is most likely to have a lot of features misclassified in transfer learning, this can be seen in appendix-A Part (B) for the original and filtered cases with $\sigma_t <= 0.1$. Filtering over the temporal parameter $\sigma_t$ will enhance the classification results using transfer learning. The filter reduces the differences between the images and facilitates the transformation of images in the transfer learning. Numerous areas in the classification outcomes were enhanced by this filter that includes borders, shadows, haziness, etc. Additionally, rising $\sigma_t$ will lead all classification results to look alike with slight differences, and the higher $\sigma_t$ the better the outcomes, but the results will be different that the original data. Since the dataset of experiment II is mostly urban structures and because of the large window size (11) used in the filtered, a lot of features of buildings and roads with similar intensities blended together and small details disappeared. Figure *29* through Figure 31 illustrate a couple of examples where the filter enhanced classification results.

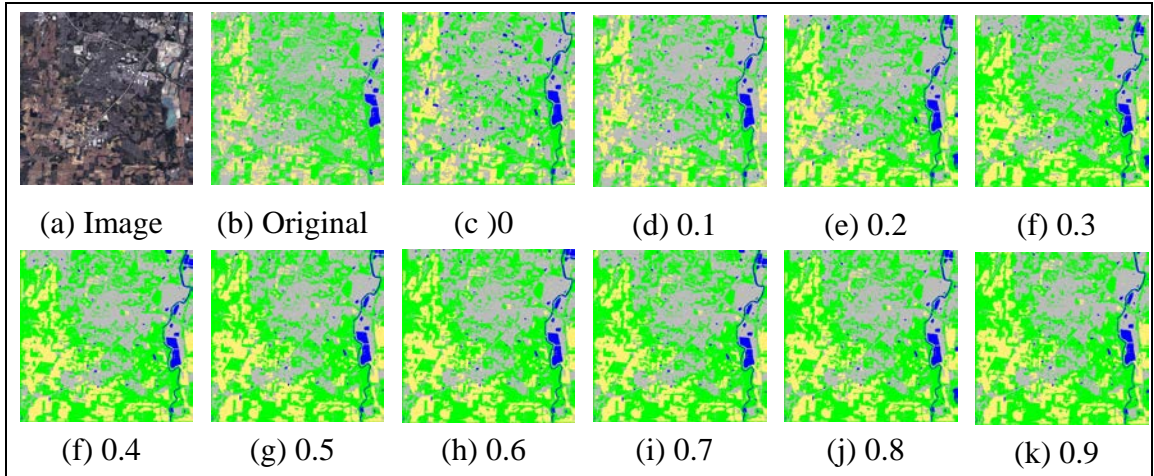

*Figure 29. Barren Lands falsely detected as impervious surfaces in image 1 was corrected by the filter.*

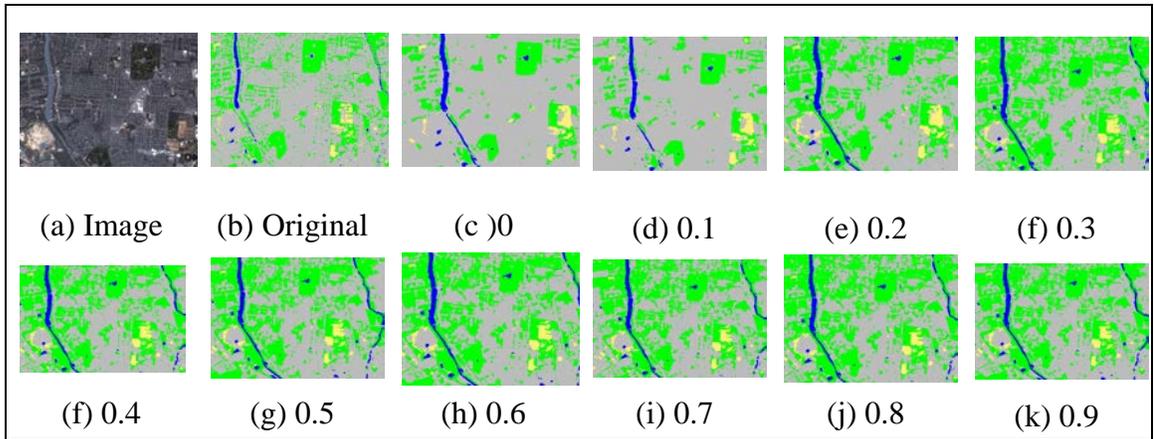

*Figure 30. Features like barren land and urban regions were improved by the filter*

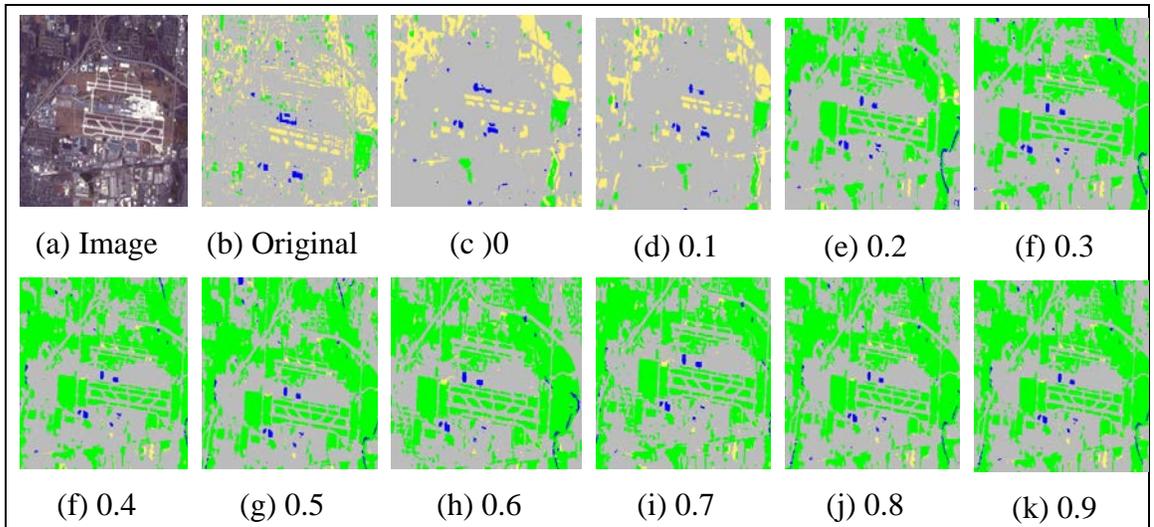

*Figure 31. Airport runaway improved by the spatiotemporal bilateral filter*

### 4.3.3       Experiment III

The high-resolution image allows detailed information to be retrieved from the image, therefore, in this experiment, five classes were used for classification: buildings, roads, water surfaces, barren land, and vegetation (see Table *2*). The filter and classification processes are applied in the same manner as in experiment I and II; the results are shown in Appendix-A Part (c ). It can be seen that the essence of the images is preserved as long as $\sigma_t$ has low values of 0 to 0.3; the images will still maintain their original appearances. For larger values of $\sigma_t$, the images will differ in appearance from what they used to be, but the differences between them will decrease and they will almost look alike with slight dissimilarities. In some cases, the filter enhances classification like the case in image 1, where, part of the barren land was covered with a gray color representing the roads (lower left corner), however, this misclassification will disappear gradually and will fade around $\sigma_t = 0.3$ (see Figure 34).

    Since this project uses color factor to classify, it is important to look for an image that has several of colors for each feature. Images with few colors of features will definitely lead to misclassifications. This is the case of image 1, since it does not have a diversity of colors in the input data, thus had many dark locations misclassified as water bodies. Similarly, some dark buildings like in image 3 are mistaken as water surfaces, and this is because there are not enough labels to train buildings with this specific color. Several iterations on classification were performed to decide which image is best used as a classifier for transfer learning; in this experiment image 4 was found to be the best reference to train and classify the rest of the data.

Image quality is very important and it affects the quality of the filter and classification outcomes. As mentioned earlier in the discussion of the filter, urban reflectance influences the classification results; many bright surfaces such roads and barren lands are mistakenly interpreted as buildings (see Figure 33). The existence of any atmospheric affects such as haziness and low clouds could also adversely affect the classification results of high-resolution images. This filter carries these impurities in the image to the rest of the dataset; hence effect the final results of classification. For example, image 2 had a cloud that is very bright and large considering the size of the image, this cloud imprinted on the rest of the images in the filtering process and can be obvious in all filtered images after the parameter $\sigma_t >= 0.4$. Hence, classification results will be affected by the cloud, where it is misclassified as roads or buildings. Overall, the filter managed to correct some misclassifications in transfer learning and improved the outcomes as can be seen in the examples provided by Figure 32 through Figure 34.

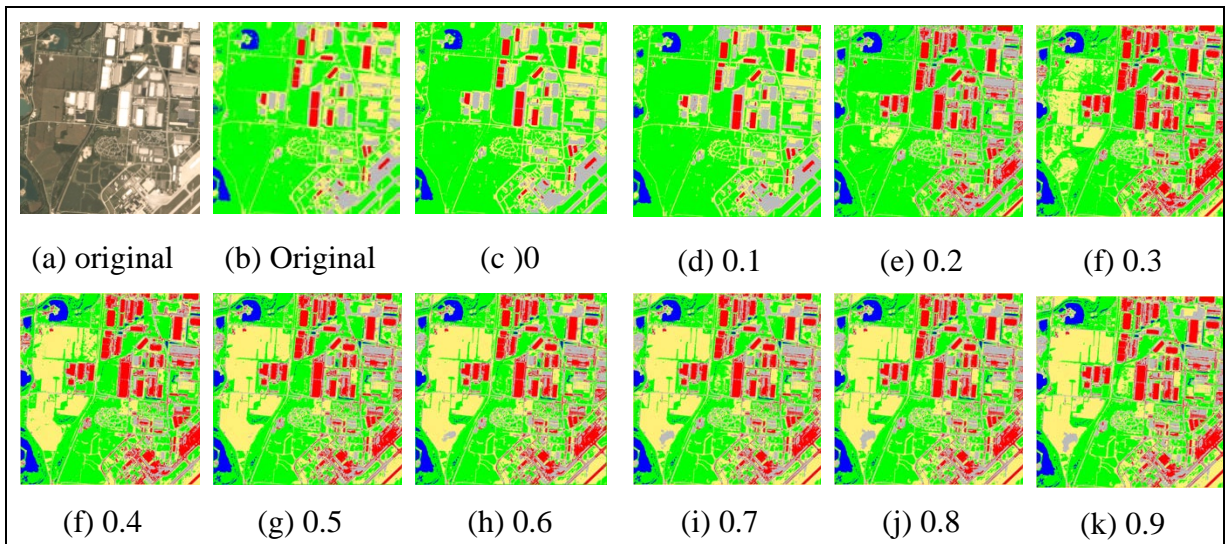

*Figure 32. Section image 5 where buildings and barren lands were corrected by the filter in transfer learning.*

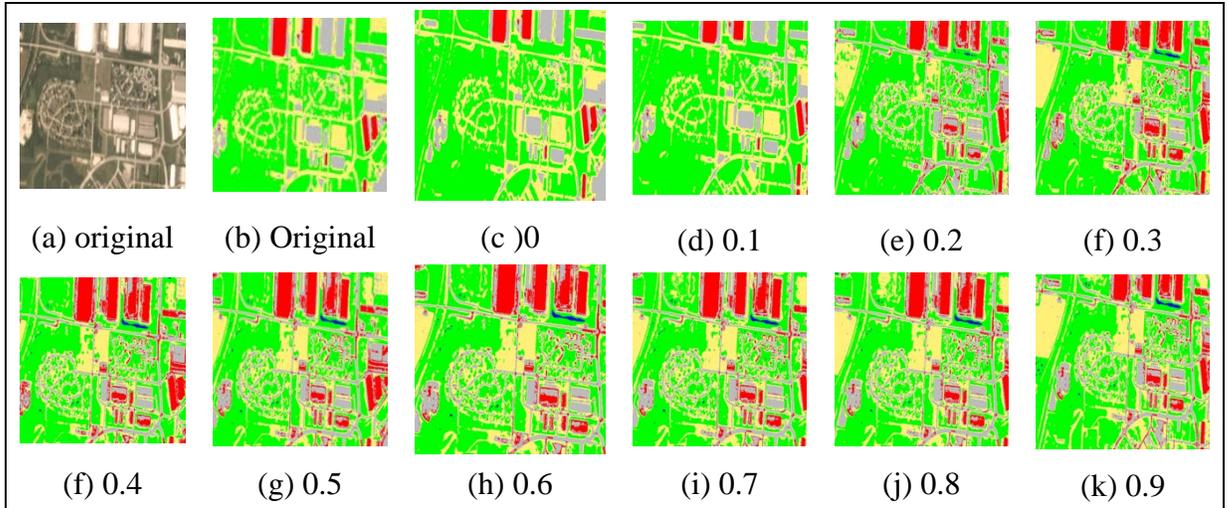

*Figure 33. Roads, buildings and barren lands are corrected to their original appearances.*

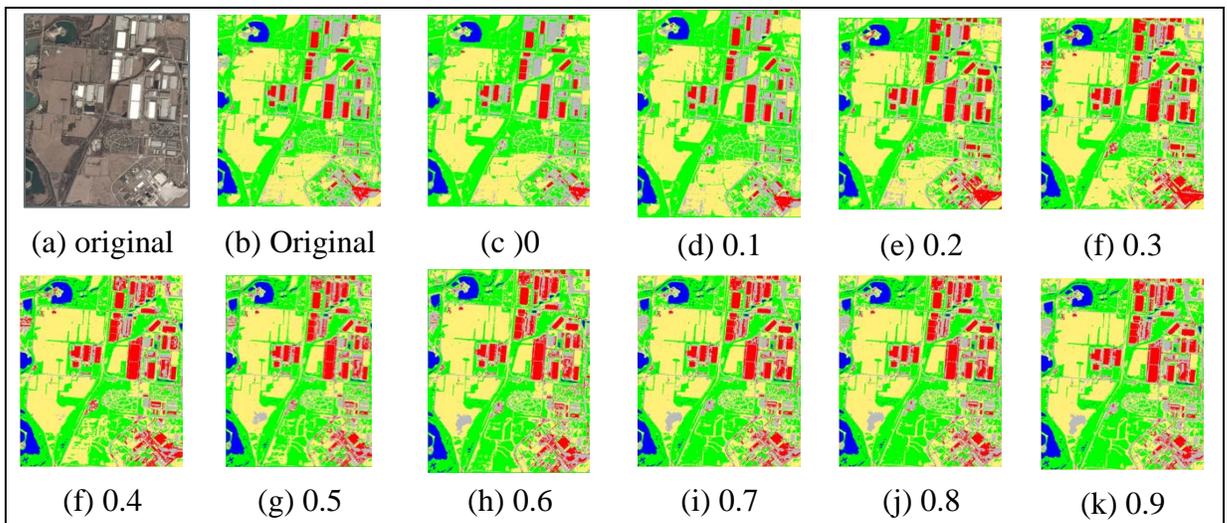

*Figure 34. Section in image 1, barren land covered by roads and buildings classified to roads is improved over this filtered.*

### 4.4 Measure of performance

A measure of performance is a powerful tool to assess the technique and methodology followed in any process; It is also used to evaluate if the desired goal is reached or satisfied. Therefore, in this research, the accuracy of classification is reported using basic confusion matrix and kappa measurements. The confusion matrix evaluates the performance of classification method by counting how many features were classified correctly and how many did not. This evaluation is defined by terms known as true positives (TP), true negatives (TN), false positives (FP) and false negatives (FN). The overall accuracy is measured and reported from these parameters using the following equation:

$$\text{Overall Accuracy} = \frac{TP + TN}{\text{Total}} \qquad (9)$$

The accuracy evaluation was made between the original and filtered classification results, the filtered results of several temporal parameters in the individually classified and transfer learning. Overall, the filtering results do not show that much of an improve in the accuracy of the individually classified data, but it does in the transfer learning. As a result, several areas show enhancement in the appearances and classification outcomes as was mentioned in the previous section. The classification accuracies were tested and measured for each temporal parameter ($\sigma_t$) with varying values ranging from 0 to 0.9, where $\sigma_t = 0$ is simply the conventional bilateral filter.

### 4.4.1 Experiment I

The major finding in this experiment and in the classification process is the improve in the accuracies of filtered data in the transfer learning. This is because the spatiotemporal bilateral filter forces the images to have common or similar values, thus can be easily recognized by the classifier in the transfer learning process. The improve in the accuracy of transfer learning due to the filter was found to be around 4.88% ≈5% better than the original data. This can be computed by subtracting the accuracy of the filtered results at $\sigma_t = 0.2$ from Table 4 and the original results. Moreover, the conventional bilateral filter shows that it has lower accuracies in both the individually classified and transfer learning methods than the filtered data at parameter 0.1 - 0.3 (Figure 35 and Table 4). This indicates that the spatiotemporal bilateral filter exceeds the performance of the conventional one. Moreover, at low parameter as $\sigma_t$ =0.1-0.3, the classification result seems to capture the difference between the images and maintain them to some extent. However, at high parameters larger than $\sigma_t$ =0.4, the classification results will appear almost the same but still will have minor differences. This is because low values of $\sigma_t$ (0.1-0.3) does not apply a lot of weight on images, thus, it filters and preserves the original structure of the image, and this explains why the accuracies keep increasing or decreasing at that range. On the other hand, $\sigma_t$ greater than or equal to 0.4 influences the classified results and the appearance as well, this is because the filter distributes higher weights on images and force them to have close pixel values and that by default affects the color of the images, and justifies the similar appearances. As a result, after $\sigma_t$ =0.4, the drop in accuracy is gradual and it keeps decreasing as $\sigma_t$ increases. All images whether its transfer learning or not will show a drop in accuracy as $\sigma_t$ rises and that can be observed in Figure 35. Looking at image 4 for instance, the original image is mostly covered by barren lands and as $\sigma_t$ increases, more vegetation is added and the more variant it would be from its original. The filtering results are affected by the available images, their contents and the season, if most of the images have vegetation in a certain area, the filter will force other images to have vegetation in the same location by adding more weight to it, the same for the other features. In this project, images (1,2,3,5) all have vegetation at the lower half of them, however, image 4 has mostly barren land, so the result of the filter as $\sigma_t$ increases forced it to have vegetation in the lower half. The same conclusion applies to barren lands, in all five images it is concentrated in the mid-upper part of the images and that explain why it is positioned at that specific location. This can explain the appearances of the classified images, as well as, the decrease in the accuracy values. Because the filter transforms all or part of the images to something other than its original components, it no longer matches the masks and labels initially created, thus, different data will be used in the training, testing and validation phases.

Table 3. Experiment I Original and Filtered Images Classification Accuracy Results of Individually Classified Images.

| Images $\sigma_t$ | 1 | 2 | 3 | 4 | 5 |
|---|---|---|---|---|---|
| **Original** | 97.11% | 97.13% | 93.30% | 90.59% | 97.05% |
| **0** | 96.58% | 96.66% | 92.42% | 89.17% | 96.82% |
| **0.1** | 96.63% | 96.29% | 93.21% | **92.48%** | 96.80% |
| **0.2** | 96.04% | 96.72% | **93.95%** | **91.53%** | 96.76% |
| **0.3** | 95.82% | **97.16%** | 93.22% | **91.46%** | 94.35% |
| **0.4** | 95.82% | 96.92% | 92.48% | 90.28% | 93.10% |
| **0.5** | 95.82% | 96.71% | 91.86% | **90.66%** | 92.22% |
| **0.6** | 95.75% | 96.57% | 91.12% | 90.56% | 91.56% |
| **0.7** | 95.71% | 96.54% | 90.66% | 90.22% | 91.16% |
| **0.8** | 95.70% | 96.42% | 90.35% | 90.14% | 90.95% |
| **0.9** | 95.70% | 96.42% | 90.11% | 90.11% | 90.72% |

Table 4. Experiment I Original and Filtered Images Classification Accuracy Results of Transfer Learning Based on Image 3.

| Image<br>σt | 1 | 2 | 3 | 4 | 5 |
|---|---|---|---|---|---|
| **Original** | 80.88% | 74.15% | 93.30% | 93.59% | 91.97% |
| **0** | **82.21%** | **74.68%** | 92.42% | **96.57%** | **94.03%** |
| **0.1** | **83.08%** | **79.95%** | 93.21% | **96.36%** | **94.01%** |
| **0.2** | **91.99%** | **91.96%** | **93.95%** | 86.08% | **94.30%** |
| **0.3** | **92.93%** | **93.34%** | 93.22% | 83.74% | 91.73% |
| **0.4** | **87.43%** | **87.61%** | 92.48% | 76.47% | 89.96% |
| **0.5** | **80.97%** | **81.10%** | 91.86% | 68.94% | 88.28% |
| **0.6** | 76.79% | **76.88%** | 91.12% | 64.21% | 86.78% |
| **0.7** | 76.16% | **76.23%** | 90.66% | 63.56% | 86.12% |
| **0.8** | 75.88% | **75.91%** | 90.35% | 63.20% | 85.67% |
| **0.9** | 75.21% | **75.24%** | 90.11% | 62.47% | 85.25% |

From the four graphs presented in Figure 35, it can be seen that it is true that the transfer learning has less accuracy than individually classified. Besides, the highest accuracies of individually classified and transfer learning are within the range of 0.2-0.3 temporal parameters. However, in the individually classified data, it is not noticeable since any increase or decrease is very slight. Furthermore, it can be noticed that since image 3 and image 5 are more alike (the images are taken on the same season), the accuracies in transfer learning for image 5 shows the highest range (from as low as 85% - 95%) unlike the rest of the images (with lowest accuracies from 62%- 75%). This provide a clue that using images with the same season can maintain the original appearances of images and increase their accuracies in classification using transfer learning. Image 4 on the other hand, shows three parts where transfer learning shows higher accuracy than individually classified these points are at original and filtered at parameters (0, 0.1) (see Table 3 and Table 4). This can be justified by the fact that image 4 has areas where it is hard to extract information from in the labeling process especially when it is done manually and at low-resolution images. For example, masking vegetation in image 4 is difficult to find because there are few vegetation spots and at narrow areas, and can be missed during masking. Thus, when user fail to label and mask all possible different features in the image, they will not be classified in the individually classified approach. On the other hand, the program manages to get these features classified correctly in the transfer learning since it uses the training of an image that recognize these features. To demonstrate this point more, remember that images 3 has variety of colors for most of the features, while image 4 has few colors for all features, during training these unique few features might not be considered in the individually classified, but are considered during transfer learning.

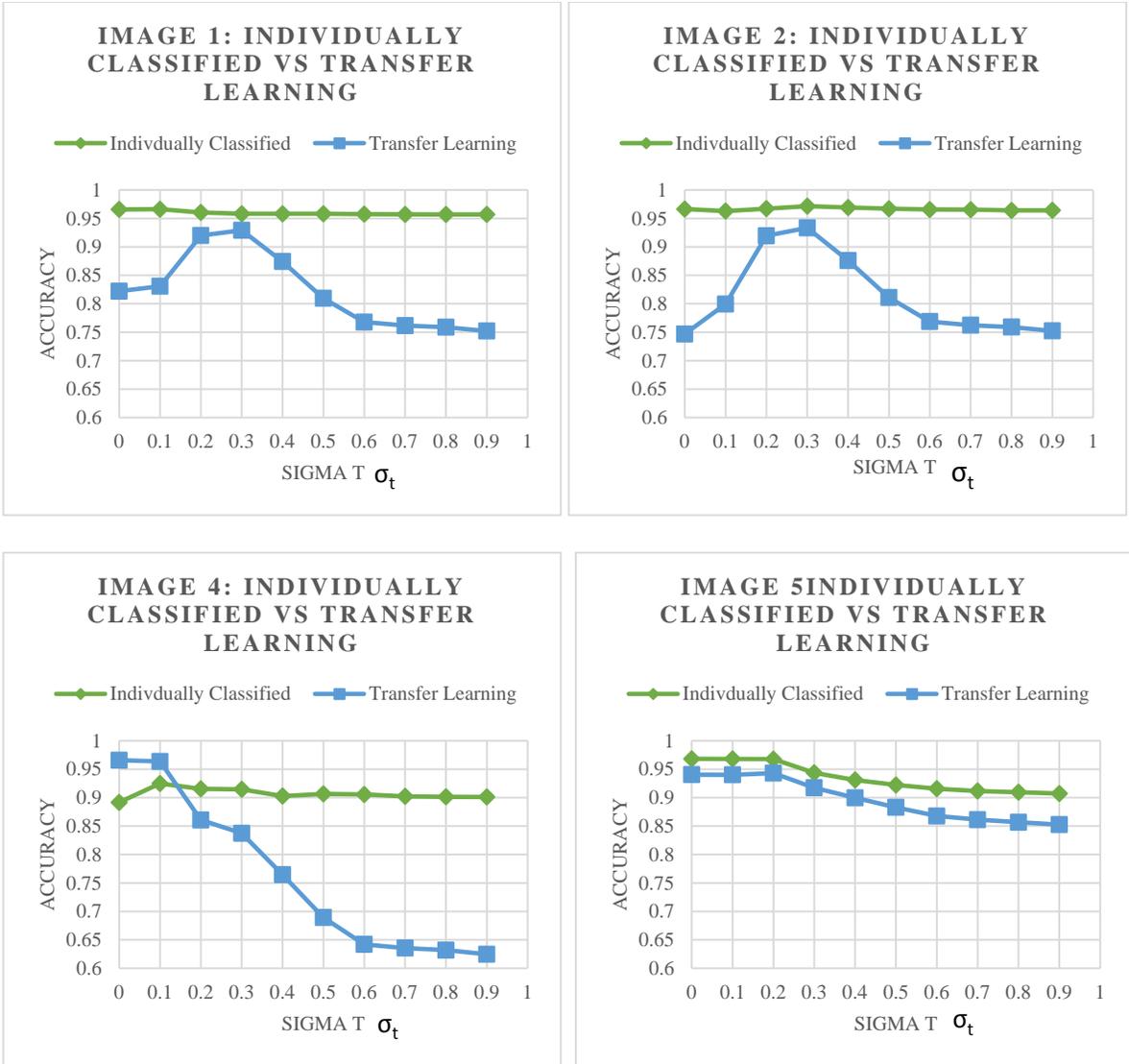

*Figure 35. Classification Accuracies of Filtered Images (Individually classified vs. Transfer Learning) of Experiment I.*

*Note: Image 3 is excluded because the Transfer Learning is the same as individually classified classification*

Figure 36 and Figure 37 show the accuracy of all classes within the individually classified or a transfer learning classification approaches. In general, large water surfaces have almost constant to steady accuracies over any parameter of the filter, and that is true in both cases (see Figure 36 and Figure 37). Nevertheless, in transfer learning, barren land has the lowest accuracies among all other classes, and as the $\sigma_t$ increases the lower accuracies it will have. Figure 36 and Figure 37 demonstrate how the accuracy of Barren land and vegetation drops. This is mostly due to the metrological conditions that affect the appearance and intensities of the barren land such as rain that causes wetness or existence of shadows, which result in similar intensities to those of impervious surfaces (buildings in particular), thus, might be misclassified as impervious surfaces. Moreover, the only change between the images are the natural classes (vegetation and barren lands), and this due to the seasonal change; manmade structures and large water surfaces, on the other hand, do not experience any change in their locations over times or seasons. This explains s why there is a drop in the accuracy of the first two classes and a stability in the others. Another observation is that in the individually classified data the lowest accuracies usually represented by impervious surfaces and barren land. However, impervious surfaces are more often to have lower accuracies than barren land like in image 2 and 4, and this can be justified that many bright barren lands have similar spectral values as buildings, hence, mistaken as manmade structures Figure 36.

Kappa measures the agreement level in classification, by comparing the actual and expected agreements. These values of kappa were reported in this research for all classified images with all parameters, and it can be seen in the appendix. Kappa's ranges for transfer learning were averaged for each filtering parameters; the range of kappa was found between ($\approx 0.77$- $\approx 0.913$), which represents good agreement. The individually classified, on the other hand, have kappa values between ($\approx 0.85 - \approx 0.95$), which represent better agreement than transfer learning.

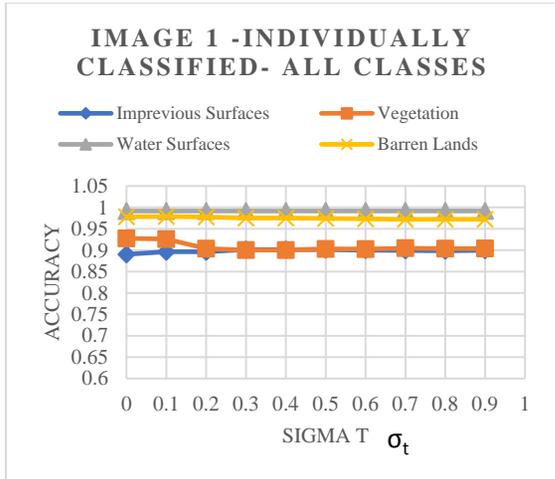
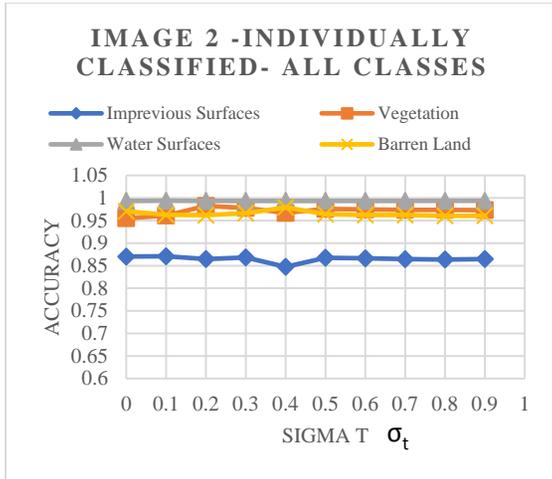
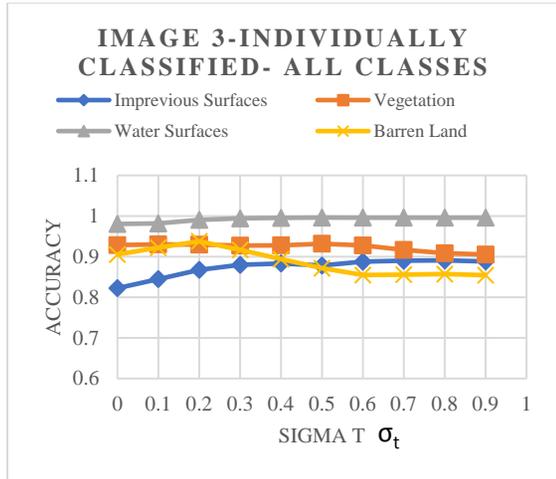
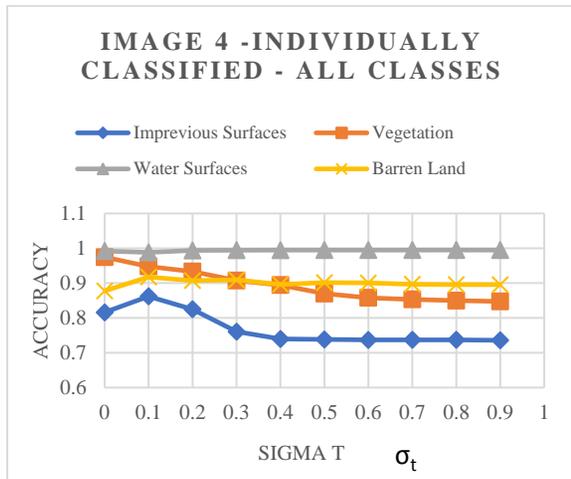
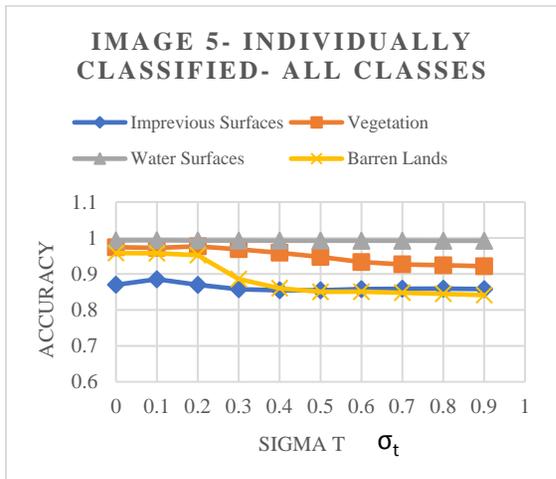

*Figure 36. Experiment I Classes Comparison for Images (1 to 5) in the Individually Classified*

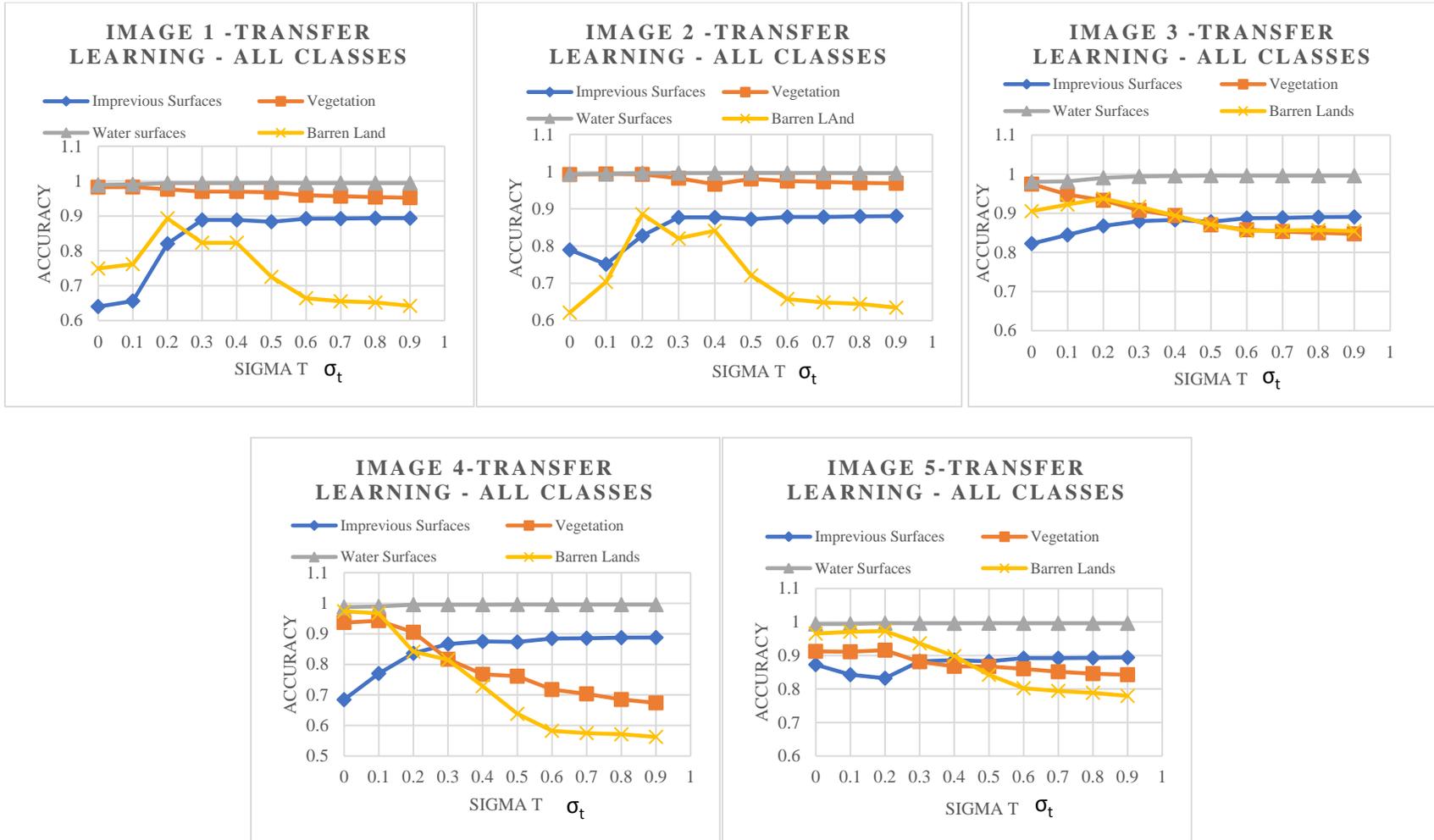

Figure 37. Experiment I Classes Comparison for Images (1 to 5) in the Transfer Learning

### 4.4.2 Experiment II

The accuracies of the individually classified and transfer learning are shown in Table 5 and Table 6. Of course, all values after $\sigma_t$ equals to 0.3 will be decreasing in the individually classified and transfer learning, because the filtered data will no longer match the labels that were created for the original images. However, at $\sigma_t$ from 0.1 to 0.3 the accuracies are fluctuating increasing and decreasing depending on the filtering outcomes, and the maximal values were found in that region.

Table 6 shows that filtering produce higher accuracies than original results in the transfer learning and it shows an improve in the accuracy by about 15%. This can be found by differencing filtered classification results at $\sigma_t=0.2$ from the original results in Table 6. The drop in accuracy between the two approaches (individually classified and transfer learning) is about 10%. Figure 38 shows the highest accuracies due to filtering process, which are captured at $\sigma_t=0.2$, in addition, it shows that transfer learning has less accuracies than individually classified. Kappa measurement was also reported in it shows that in transfer learning most of the values were between ($\approx 0.6$-$\approx 0.8$), and this indicates moderate agreement. Nevertheless, the individually classified have higher agreement levels and values of kappa that is between ($\approx 0.78$-$\approx 0.87$). The comparison between all classes in the images is shown in Figure 39 and Figure 40. In most the images and in both types of learning barren lands shows the lowest accuracies among all other classes, and as in experiment I, water surfaces are mostly the most stable class with the highest accuracies. In urban regions, many structures would have similar pixel values with barren lands especially in areas with high reflectance, which may cause the barren land to be mistaken as buildings and cause misclassifications.

Table 5. Experiment II Original and Filtered Images Classification Accuracy Results of Individually Classified Images.

| Images $\sigma_t$ | 1 | 2 | 3 | 4 | 5 |
|---|---|---|---|---|---|
| Original | 92.62% | 93.39% | 92.69% | 90.80% | 93.75% |
| 0 | 92.04% | 93.32% | 88.67% | **93.01%** | **94.64%** |
| 0.1 | 92.37% | **93.81%** | 89.47% | **93.27%** | **94.70%** |
| 0.2 | 92.16% | **94.03%** | 91.72% | 85.75% | 93.10% |
| 0.3 | 92.25% | **94.03%** | 85.94% | 83.99% | 90.70% |
| 0.4 | 92.09% | 93.90% | 81.58% | 83.19% | 89.16% |
| 0.5 | 91.97% | 93.71% | 81.23% | 83.03% | 88.26% |
| 0.6 | 91.92% | 93.60% | 81.06% | 82.96% | 87.78% |
| 0.7 | 91.91% | 93.55% | 83.34% | 82.91% | 87.48% |
| 0.8 | 91.88% | 93.50% | 83.27% | 82.89% | 87.27% |
| 0.9 | 91.86% | 93.47% | 80.88% | 82.87% | 87.14% |

Table 6. Experiment II Nonfiltered and Filtered Images Classification Accuracy Results of Transfer Learning Based on Image 5.

| Image $\sigma t$ | 1 | 2 | 3 | 4 | 5 |
|---|---|---|---|---|---|
| Original | 72.61% | 67.91% | 81.01% | 50.21% | 93.75% |
| 0 | **76.12%** | **71.13%** | 80.58% | 47.48% | **94.64%** |
| 0.1 | **77.24%** | **73.13%** | 78.74% | **53.60%** | **94.70%** |
| 0.2 | **89.29%** | **90.60%** | **91.35%** | **77.82%** | 93.10% |
| 0.3 | **90.52%** | **90.75%** | **90.15%** | **76.22%** | 90.70% |
| 0.4 | **90.33%** | **90.06%** | **88.01%** | **74.92%** | 89.16% |
| 0.5 | **89.94%** | **89.55%** | **87.10%** | **74.14%** | 88.26% |
| 0.6 | **89.64%** | **89.19%** | **86.59%** | **73.64%** | 87.78% |
| 0.7 | **89.39%** | **88.93%** | **86.36%** | **73.27%** | 87.48% |
| 0.8 | **89.26%** | **88.80%** | **86.18%** | **73.05%** | 87.27% |
| 0.9 | **89.16%** | **88.70%** | **86.08%** | **72.94%** | 87.14% |

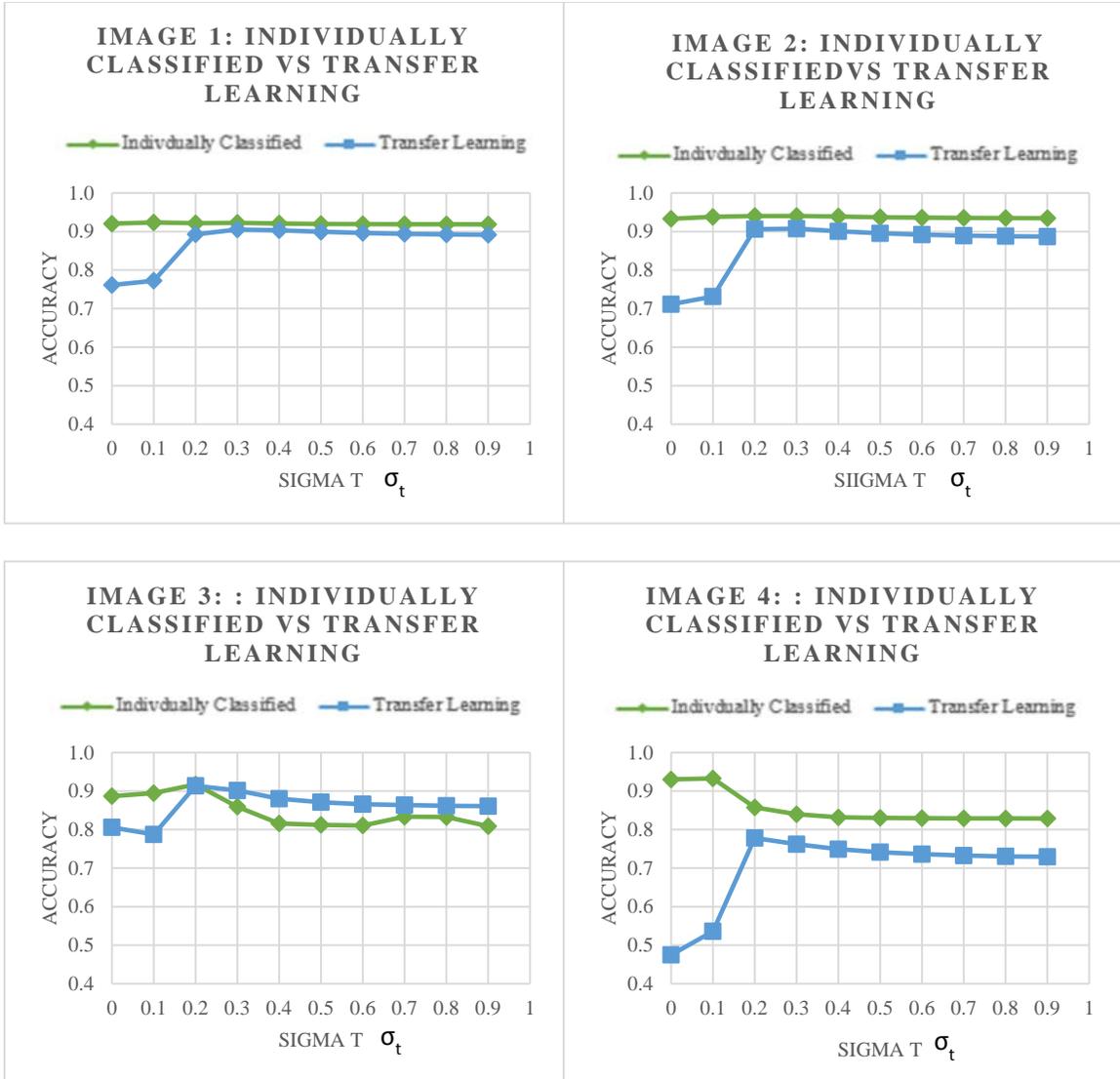

*Figure 38. Classification Accuracies of Filtered Images (Individually classified vs. Transfer Learning) of Experiment II.*

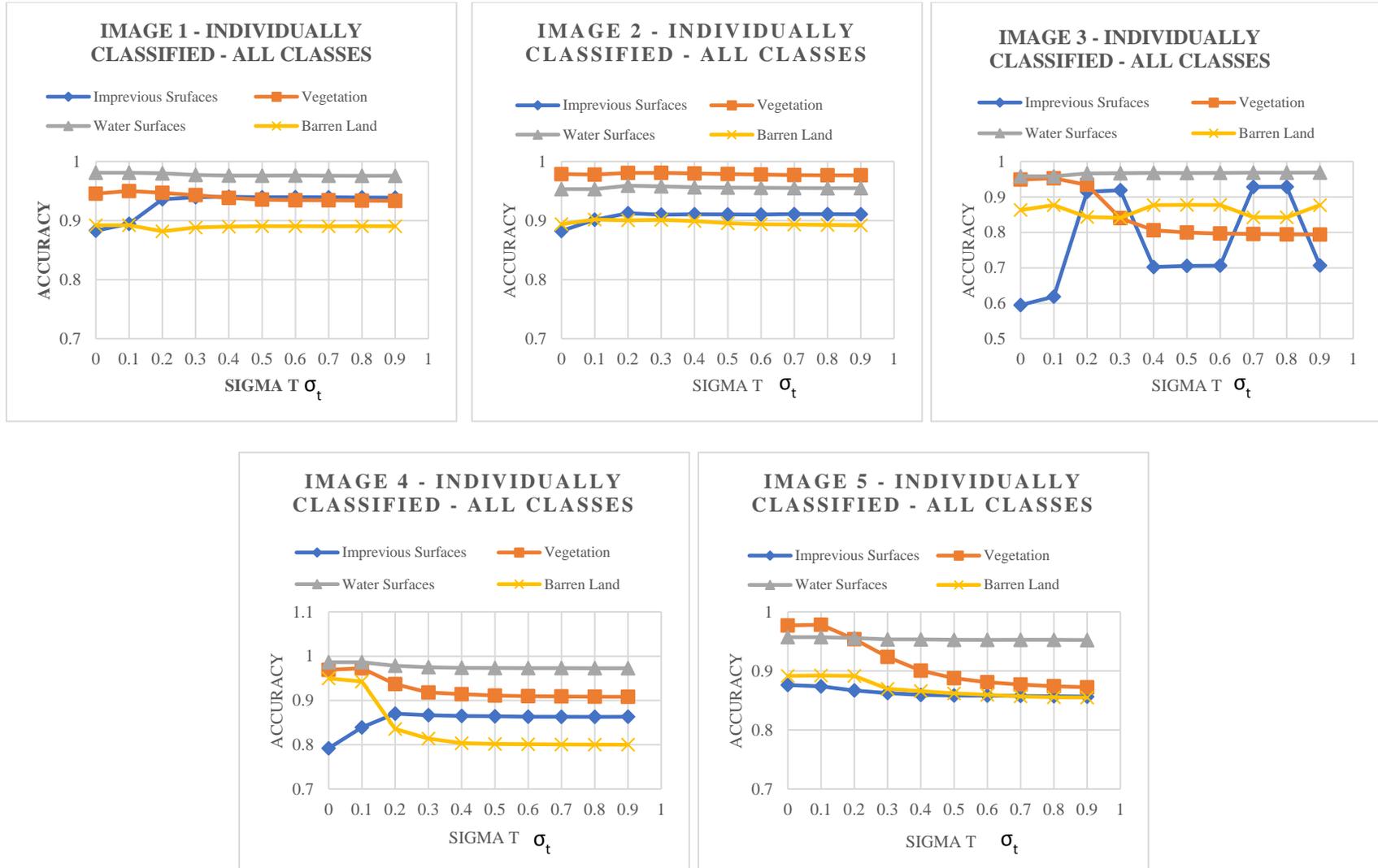

*Figure 39. Experiment II Classes Comparison for Images (1 to 5) in the individually trained.*

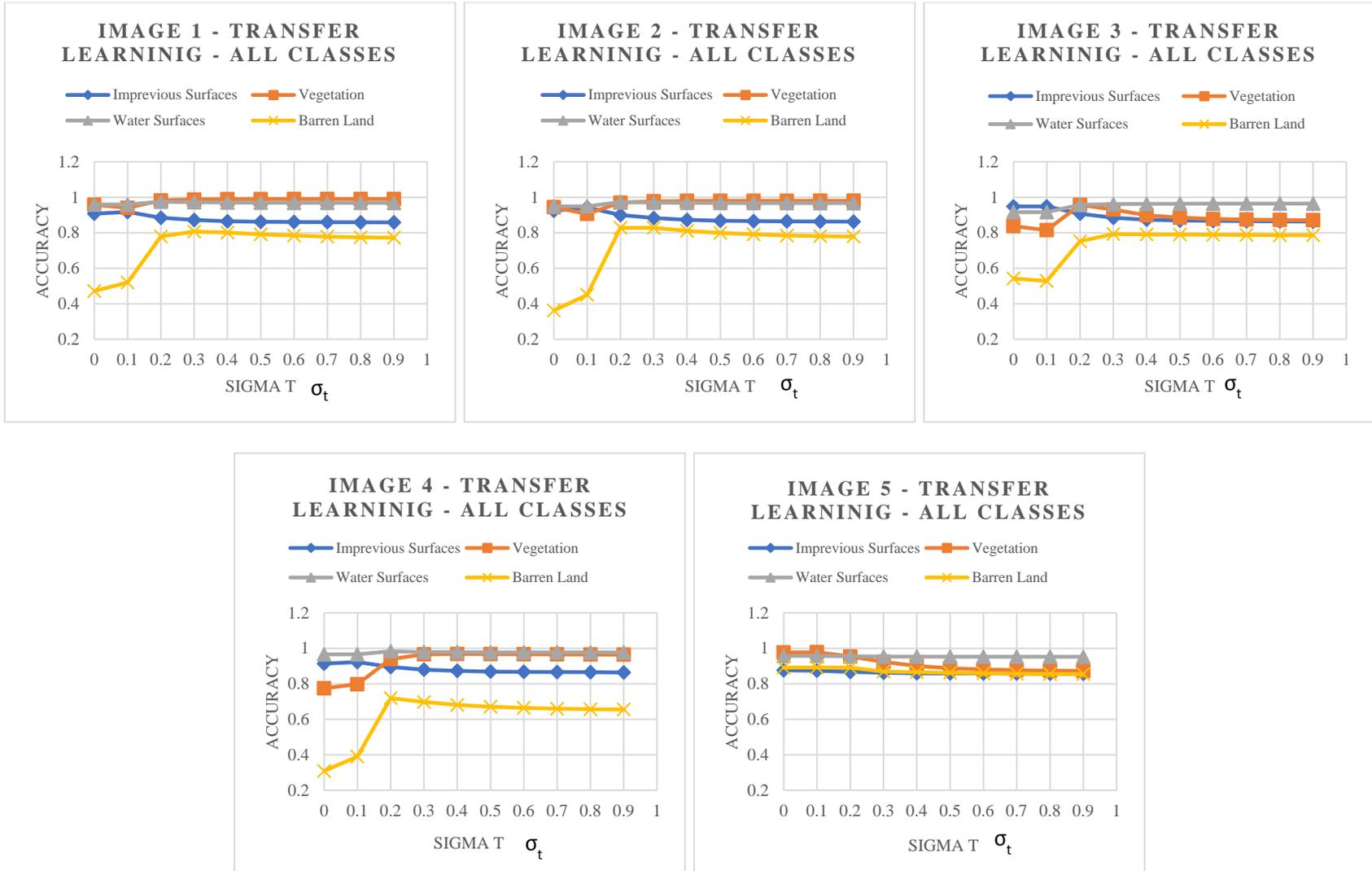

*Figure 40. Experiment II Classes Comparison for Images (1 to 5) in the Transfer Learning*

### 4.4.3 Experiment III

The findings in this experiment are similar to experiment II and I. The most important observation here is the improvement of the accuracies of classification using transfer learning of filtered data at $\sigma_t = 0.2$, where it exceeds the accuracy of the original data by 1.84≈ 2%. Because of the change in appearance due to the filter, the images no longer meet or match the original images and so are the masks, and that will cause a drop in accuracy after $\sigma_t = 0.3$. Thus, as the temporal filtering parameter increase, the classification accuracies decrease. Furthermore, the transfer learning has accuracies less than the individually classified data, and this can be shown in Figure 41 and in

Table *8* where the drop in the accuracy is less than 6%. Image 1 however, shows that transfer learning accuracy better than the individually classified. By looking at the original image 1 it can be noticed that it do not have any vegetation, thus, the mask has no labels to identify this class. Because image 1 does not have a variety of colors for this particular feature, and as the filter impose vegetation to all the images, it will not be identified in the individually classified approach. Thus, vegetation in image 1 will be classified to the next close feature, which is water surfaces in the individually classified data process; this can be noticed after $\sigma_t=0.2$. Nevertheless, transfer learning uses the training data of image 4, which has a wide range of feature colors and can simply distinguish between similar or close colors of features. Therefore, in the case of image 1 (with few training data for a specific class), it is better to use transfer learning. The highest classification accuracies in the transfer learning are around 0.2 to 0.3 $\sigma_t$, beyond this range the accuracies keep decreasing since the images lose their original structures and look a lot different from what they used to be, and more similar to each other.

Table 7. Experiment III. Original and Filtered Images Classification Accuracy Results of Individually Classified Images.

| t | 1 | 2 | 3 | 4 | 5 |
|---|---|---|---|---|---|
| **Original** | 68.12% | 71.64% | 70.08% | 67.17% | 75.29% |
| **0** | 66.64% | **73.48%** | **70.72%** | 66.49% | **77.04%** |
| **0.1** | 66.03% | **73.43%** | 69.76% | 65.82% | **77.19%** |
| **0.2** | 66.17% | **74.56%** | 70.34% | 65.10% | 73.87% |
| **0.3** | 62.21% | **74.19%** | 68.56% | 64.33% | 72.27% |
| **0.4** | 54.52% | **74.72%** | 67.62% | 65.21% | 71.26% |
| **0.5** | 51.64% | **75.05%** | 66.68% | 65.39% | 70.58% |
| **0.6** | 49.72% | **74.68%** | 66.23% | 64.85% | 69.68% |
| **0.7** | 47.37% | **74.02%** | 65.64% | 65.44% | 68.00% |
| **0.8** | 47.58% | **74.42%** | 65.37% | 65.47% | 67.88% |
| **0.9** | 47.55% | **74.50%** | 64.99% | 65.00% | 67.47% |

Table 8. Experiment III Original and Filtered Images Classification Accuracy Results of Transfer Learning Based on Image 4.

| Image t | 1 | 2 | 3 | 4 | 5 |
|---|---|---|---|---|---|
| **Original** | 66.48% | 74.14% | 68.35% | 67.17% | 73.30% |
| **0** | **66.64%** | 74.13% | 68.33% | 66.49% | 73.10% |
| **0.1** | 65.88% | **74.67%** | **69.80%** | 65.82% | **74.16%** |
| **0.2** | **66.75%** | **76.20%** | **72.06%** | 65.10% | **78.54%** |
| **0.3** | **68.01%** | **75.71%** | 67.72% | 64.33% | 70.20% |
| **0.4** | 64.16% | 74.43% | 67.08% | 65.21% | 65.49% |
| **0.5** | 62.18% | 72.34% | 66.00% | 65.39% | 63.19% |
| **0.6** | 58.46% | 70.53% | 64.90% | 64.85% | 61.32% |
| **0.7** | 58.36% | 69.91% | 65.10% | 65.44% | 60.22% |
| **0.8** | 57.06% | 69.86% | 65.01% | 65.47% | 59.95% |
| **0.9** | 57.04% | 68.82% | 64.55% | 65.00% | 58.87% |

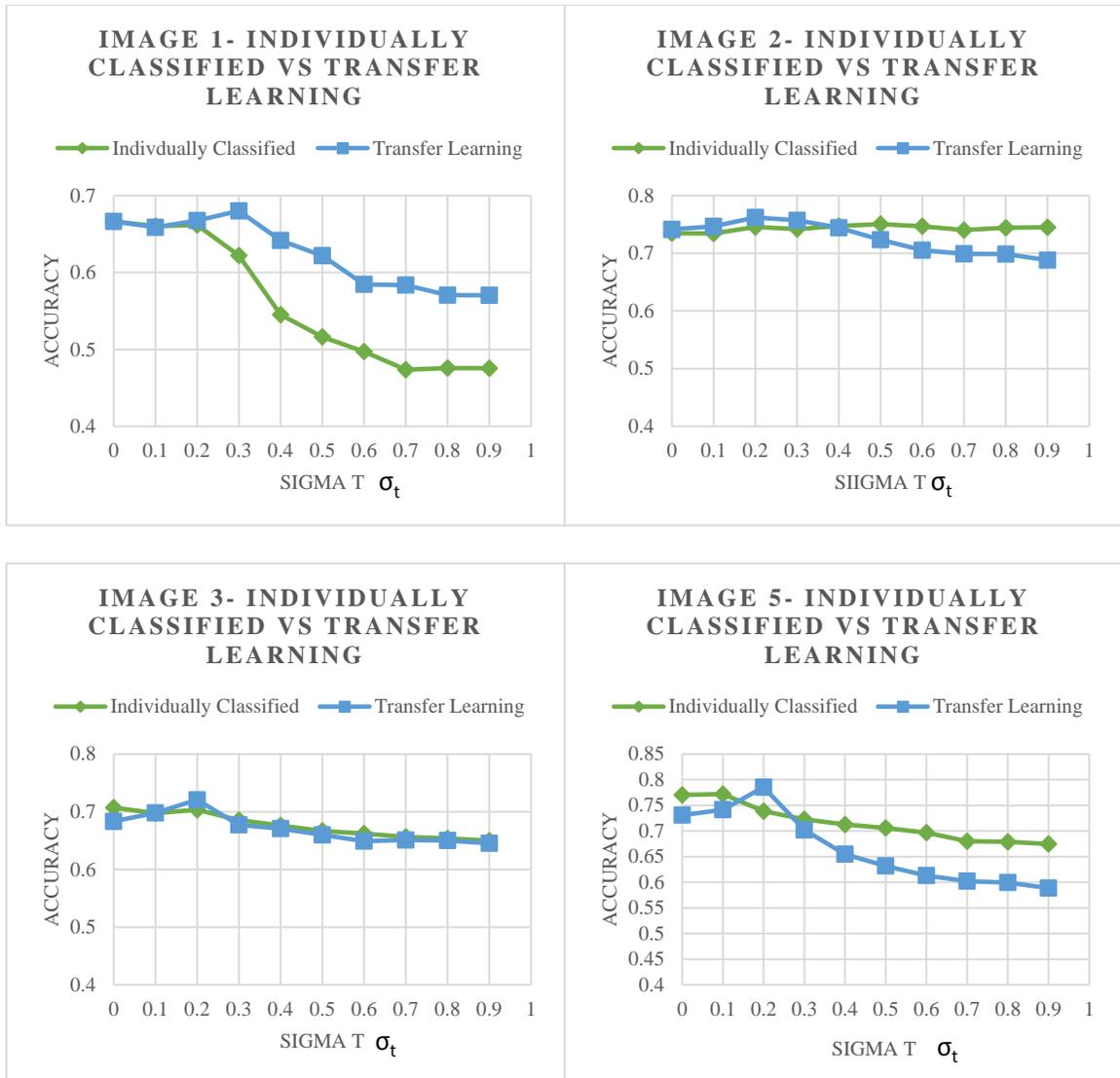

*Figure 41. Classification Accuracies of Filtered Images (Individually classified vs. Transfer Learning) of Experiment III.*

It is important to understand which class is most sensitive to change over the filter and which is the most stable. This sensitivity test is measured by confusion matrix and used later to compare between the accuracies of all classes in a single image. In both types of learning, the most sensitive class is the buildings and the least sensitive class is water surfaces, this can be seen in Figure 42 and Figure 43, where these two classes have the lowest and highest accuracies. Therefore, we can see that unlike Landsat images, where environmental elements are the classes with the lowest accuracies in transfer learning, the high-resolution images have the buildings as lowest accuracies of classes. The second lowest accuracy of features in classification are barren lands and roads, these two classes along with the buildings are easily mixed up due to their similar intensities at some locations. Another reason for misinterpretation of these classes is that empty lands can be very bright, therefore, misclassified as buildings; bright shinny roads behave in a similar manner. Kappa coefficients were also reported in the appendix. These kappa values were found to be between ($\approx 0.49 - \approx 0.56$) in individually classified, and ($\approx 0.47 - \approx 0.58$) in the transfer learning. These intervals show much lower agreement levels than those resulted from Landsat 8. It also shows that the transfer learning and individually classified in high-resolution images almost have the same level of agreement.

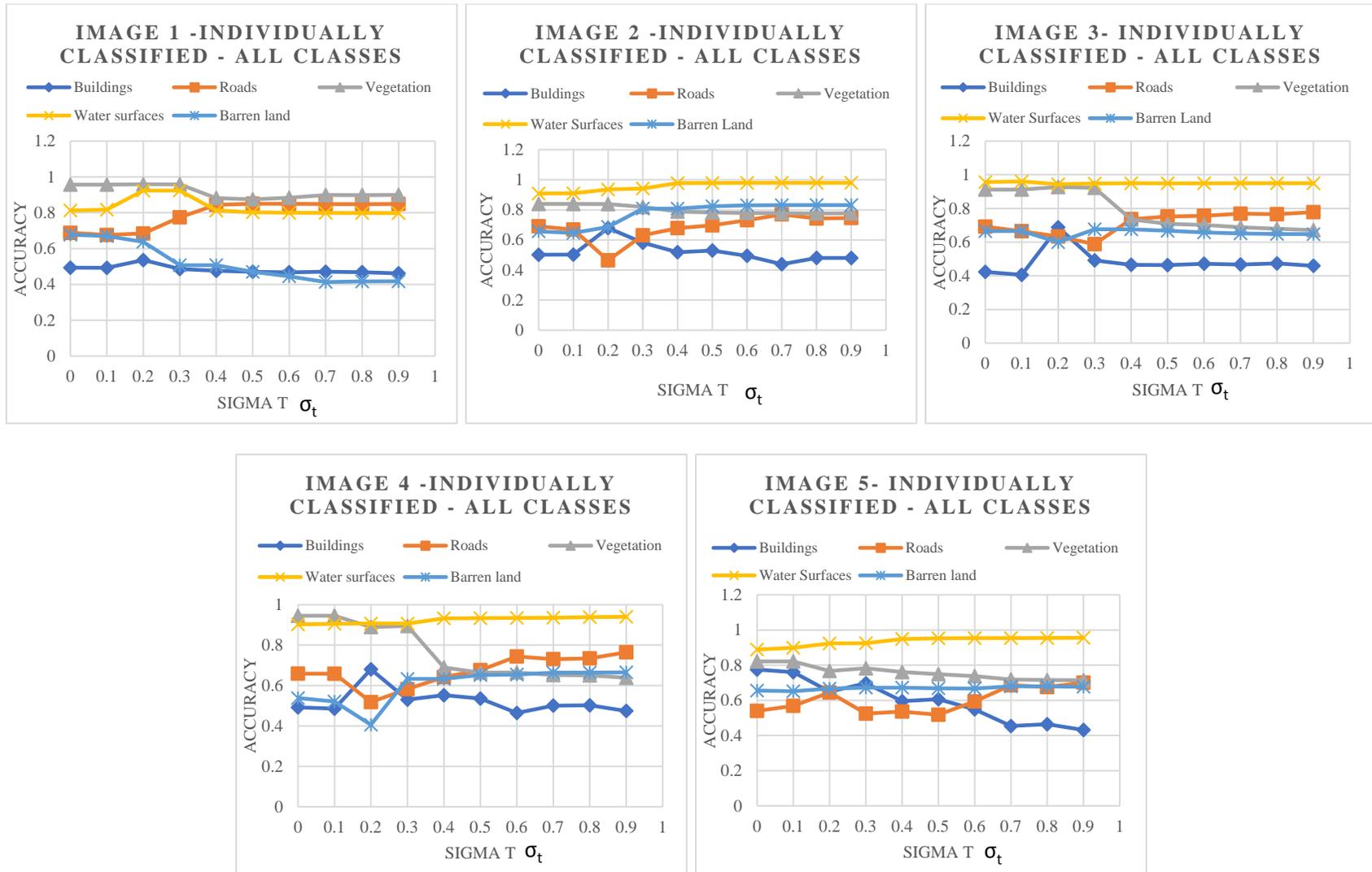

*Figure 42. Experiment III Classes Comparison for Images (1 to 5) in the Individually Classified.*

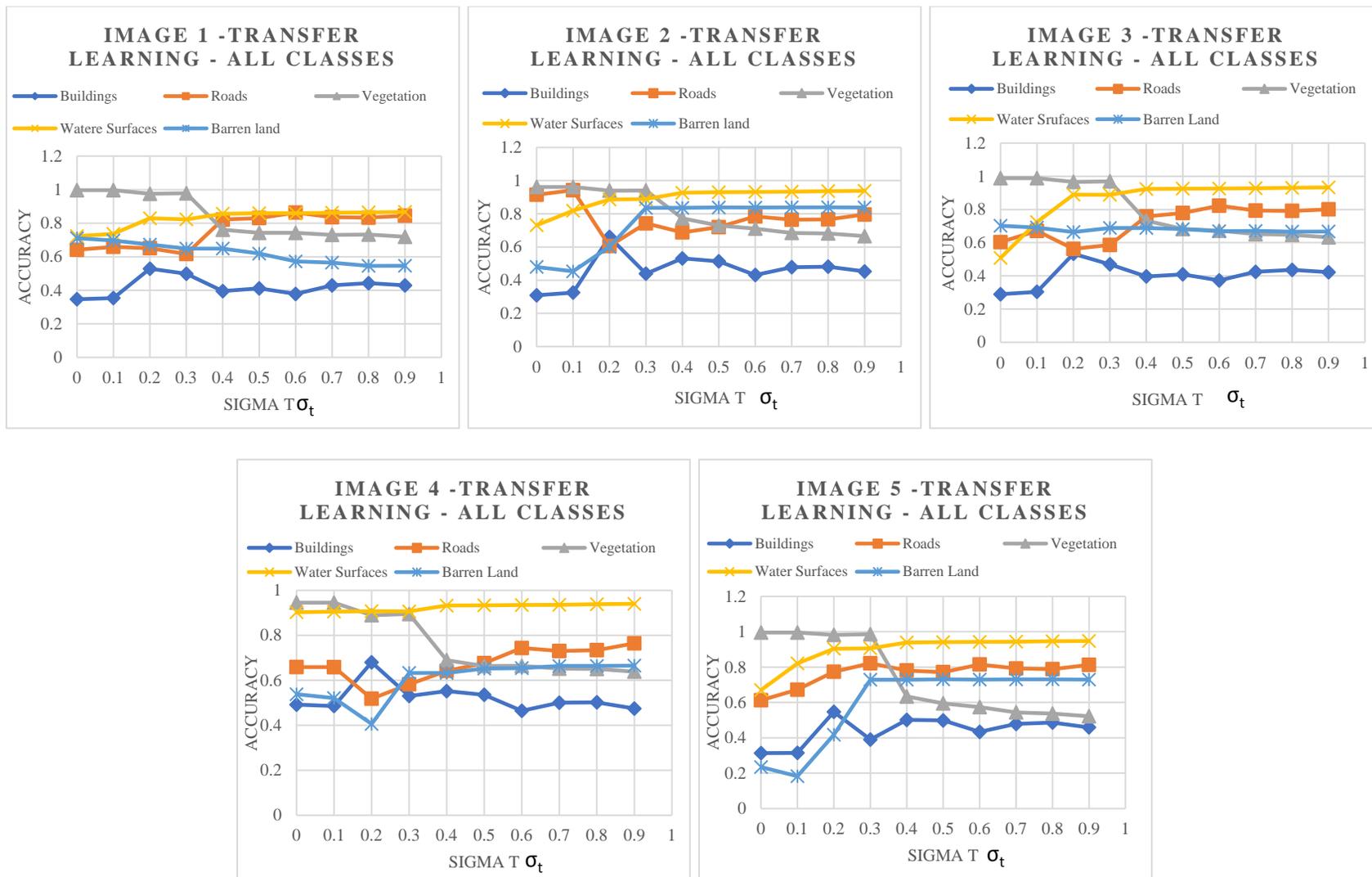

Figure 43. Experiment III Classes Comparison for Images (1 to 5) in the Transfer Learning

Chapter 5: Conclusion

In this research, it was proposed to use the temporal information within a sequence of satellite images to enhance the quality of images and the classification using transfer learning by developing a spatiotemporal bilateral filter. The influence of the filter on the quality of satellite images was observed, investigated and reported in this work. This newly developed filter is derived from the conventional bilateral filter, which is on its own suits many applications due to its simple formation, mechanism, ability to smooth while maintaining the edges, and finally, its speed since it is not an iterative method. Hence, modifying it by incorporating the temporal domain makes it stronger and more effective. The new filter proved to be better than the bilateral filter in terms of improving classification accuracy in transfer learning as it was shown in section 4.4. Most attempts to modify the bilateral filter in the temporal domain used two images taken by a camera or video scenes such as in dual bilateral filter and joint bilateral filter (Eisemann & Durand, 2004; Bennett, Mason, & McMill, 2007; Petschnigg et al., 2004). Only a few attempts were made on applying this filter on satellite images, therefore, the new contribution to this filter by considering the temporal domain of satellite images and evaluating its influence is considered valuable. On the contrary to what (Bennett & McMillan, 2005) stated about simple averaging of static scenes being sufficient, the developed filter proved that differencing and measuring Euclidian distances among the images and bands was better in terms of preserving the individuality of each image at certain temporal parameters. As a result, the new filtering addition uses images of different dates, filters in the spectral, spatial and temporal domains. Choosing the right window size, $\sigma_s$, $\sigma_r$ and $\sigma_t$ is critical and control the amount of filtering going over an image; therefore, choosing the right parameter involves trying several values to come up with the optimum outcome. Choosing the temporal parameter depend on the desired outcome, if an approximation of the map is required, high $\sigma_t$ should be used (larger weights will be assigned to the pixels and will lead to all images to be similar in appearance). Otherwise, if an image is to be enhanced visually with the desire to preserve its original appearance, a small value of $\sigma_t$ are to be used (0-0.3). In general, the filter works on reducing the differences between the images, as well as any possible noises such as haziness and high reflectance of surfaces.

Reducing the radiometric differences between the images will facilitate the transformation of the data from a reference classifier to the rest of the images. Thus, normalizing the images and forcing the images to have a similar appearance to improve the classification results using transfer learning (the examples were shown in section 4.3). Generally, the filter reduced a lot of misclassifications and provided more smoothed classification outcomes. The highest accuracies of classification (using transfer learning) were found around the values of $\sigma_t$ 0.1 to 0.3 and these are the values where the appearances are still preserved. At larger values of $\sigma_t$ ($\geq 0.4$), the classification results will be better in terms of distinguishing between different features, but they will no longer match the original data. Although the classification accuracies were not enhanced by the filtering process in the individually classified approach, but it was enhanced in the transfer learning where it exceeds the original classified data by ($\approx 5\%$, $\approx 15\%$, and $\approx 2\%$) in the three experiments respectively. This improve in the transfer learning accuracy due to the spatiotemporal bilateral filters is the major contribution to this research. Overall, the classification results of Landsat 8 were better than Planet in terms of accuracy ranges, and this is because high-resolution images are more sensitive to noise (could convey strong noises to the rest of the data in the filtering process).

Considering that one of the goals of this project is to use a minimum source of information (color and single classifier) for filtering and classification, the results were satisfactory. A rough estimate of ground components and their locations could be obtained, in addition to the ability to monitor environmental and urban changes. The filter provides smooth, better quality images and classification results, and manages to enhance a lot of regions with atmospheric effects or blockage in the scene that might cause some misinterpretations.

## 5.1 Future works

There are many areas in this research that can be further studied and enhanced. The classification results for the high-resolution images is one area that needs more enhancement. The use of an external source of data such as height information can be beneficial to support and enhance classification. Another direct is to use more complicated feature descriptor such as the spatial and height features. Transfer learning also can be improved in future researches, using more than one source of training data can be helpful to enhance transfer learning.

Appendix A: Classification results

# Classification Results

The following tables show the classification results of the three experiments. The comparison in the tables is made between the original and filtered classification data and between the filtered data, in addition to between classifications performed in the classic way (individually classified denoted as I) or using transfer learning (denoted T). Since it is important to visually inspect the classification results, we provide tables that help the reader to see how images change over the temporal parameter ($\sigma_t$) from 0 to 0.9 with increment 0.1 in the individually classified and transfer learning approaches.

**Part (A):** This section cover the Classification Outcomes of Experiment I; the legend of the classes is provided on the top of the table.

| Classes | ● Vegetation | ● Water Surfaces | ● Barren Land | ● Impervious Surfaces |
|---|---|---|---|---|

| | | | Image | | |
|---|---|---|---|---|---|
| | 1 | 2 | 3 | 4 | 5 |
| | colspan: Original Nonfiltered | | | | |
| I | 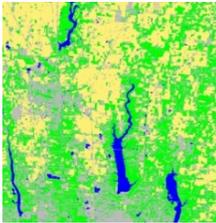 | 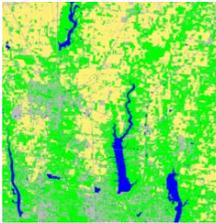 | 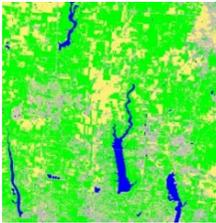 | 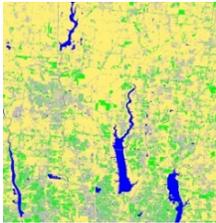 | 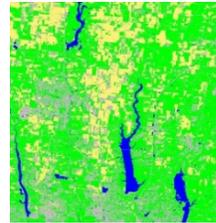 |
| T | 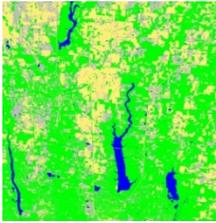 | 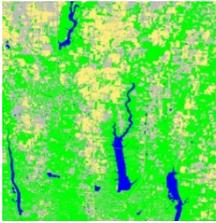 | 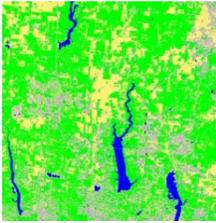 | 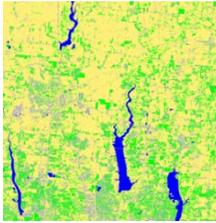 | 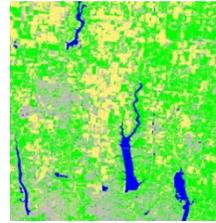 |
| $\sigma_t$ | colspan: **Filtered** | | | | Continued |

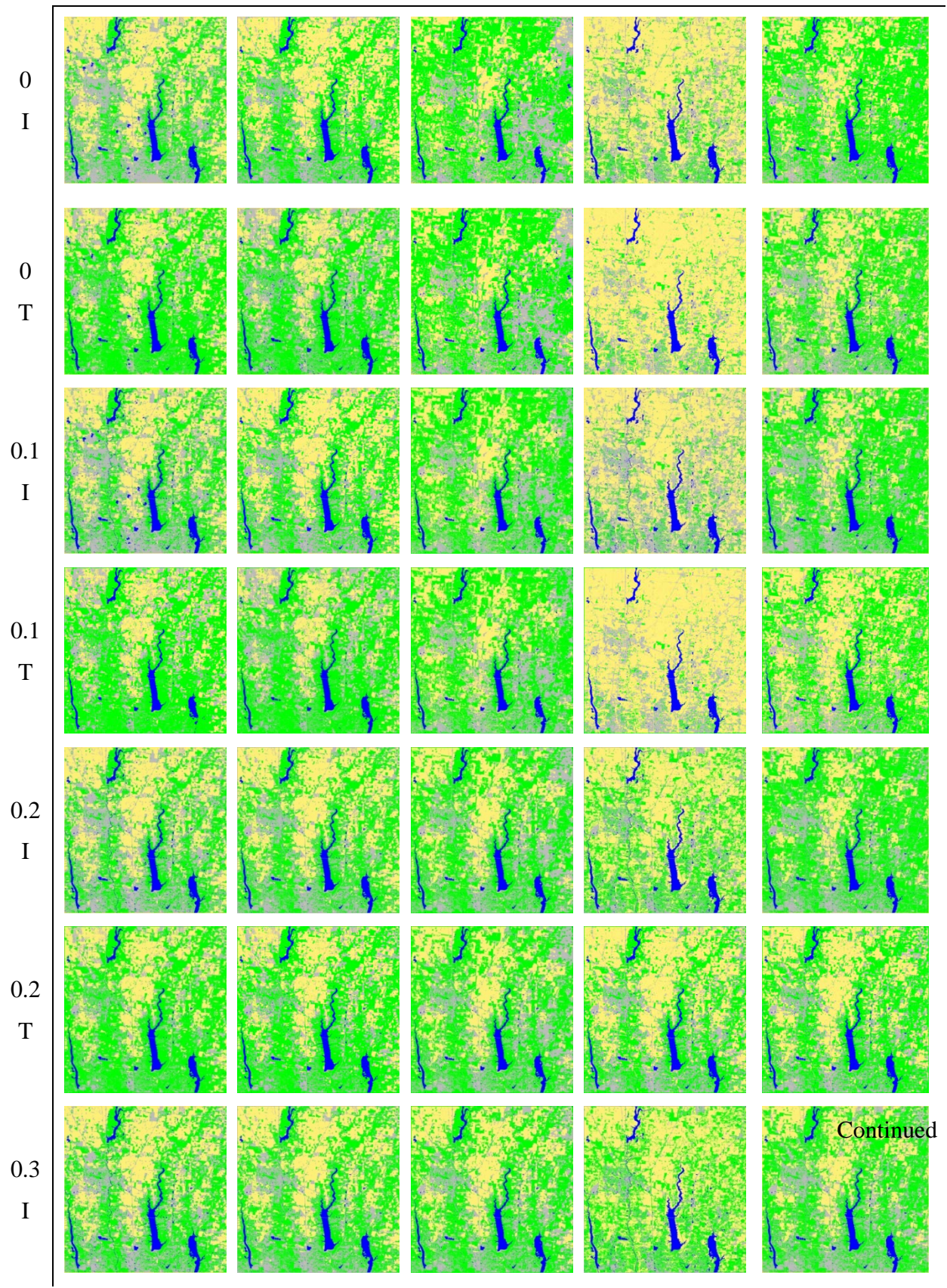

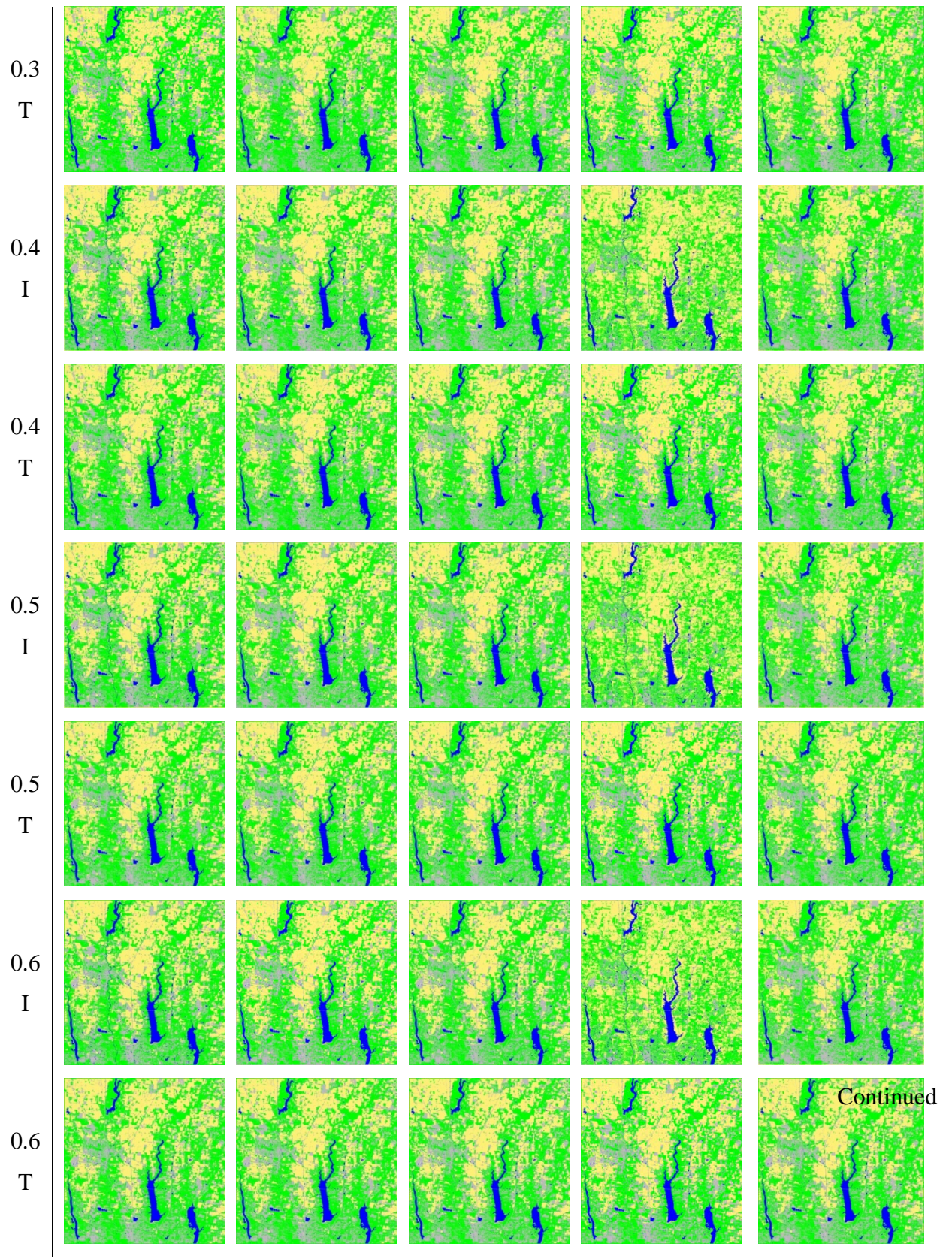

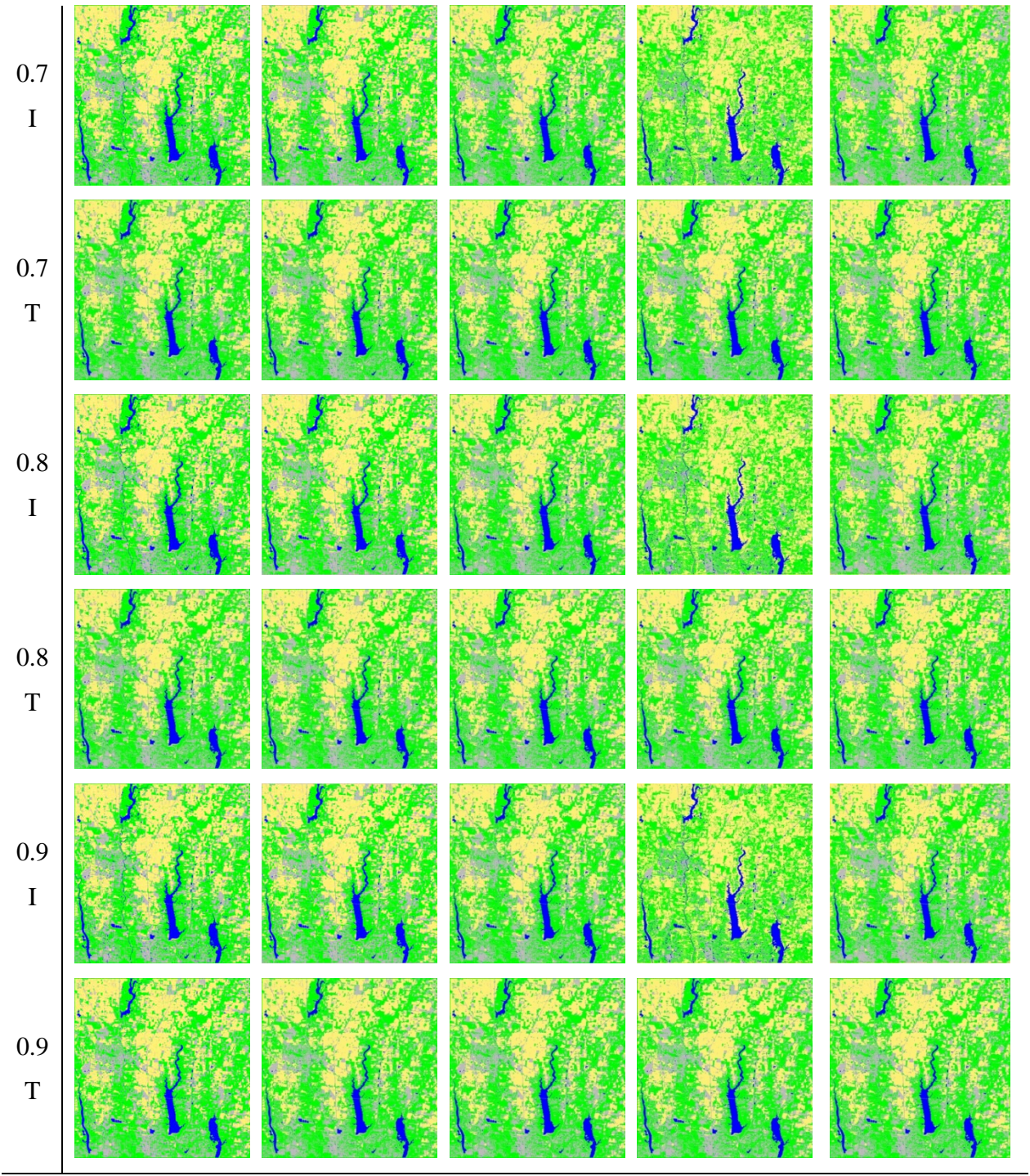

**Part (B):** This section covers the Classification Outcomes of Experiment II, the legend of the classes is provided on the top of the table.

| Classes | ● Vegetation | ● Water Surfaces | ● Barren Land | ○ Impervious Surfaces |

| | | | Image | | |
|---|---|---|---|---|---|
| | 1 | 2 | 3 | 4 | 5 |
| | | | **Original Nonfiltered** | | |
| I |  |  |  |  | 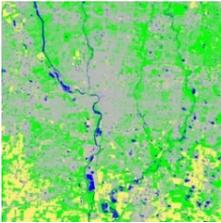 |
| T | 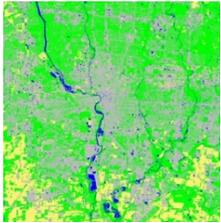 | 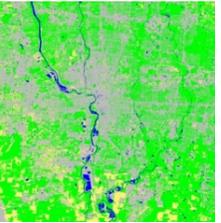 | 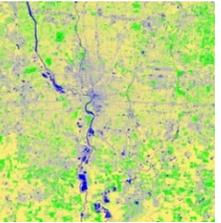 | 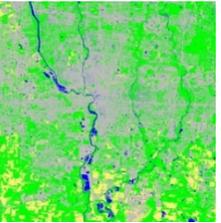 | 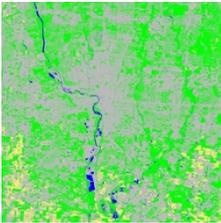 |
| $\sigma_t$ | | | **Filtered** | | |
| 0 I | 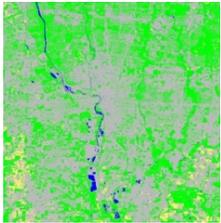 | 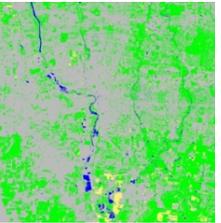 | 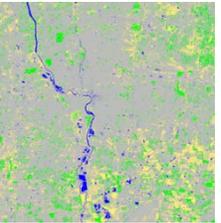 | 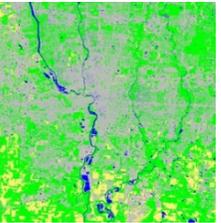 | 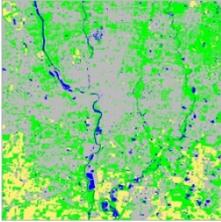 |
| 0 T | 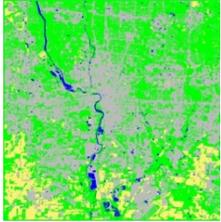 | 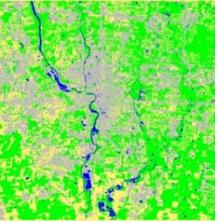 | 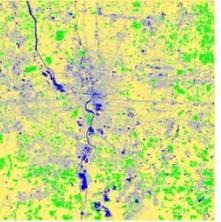 | 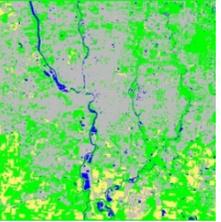 | 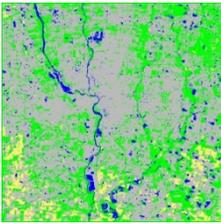 |

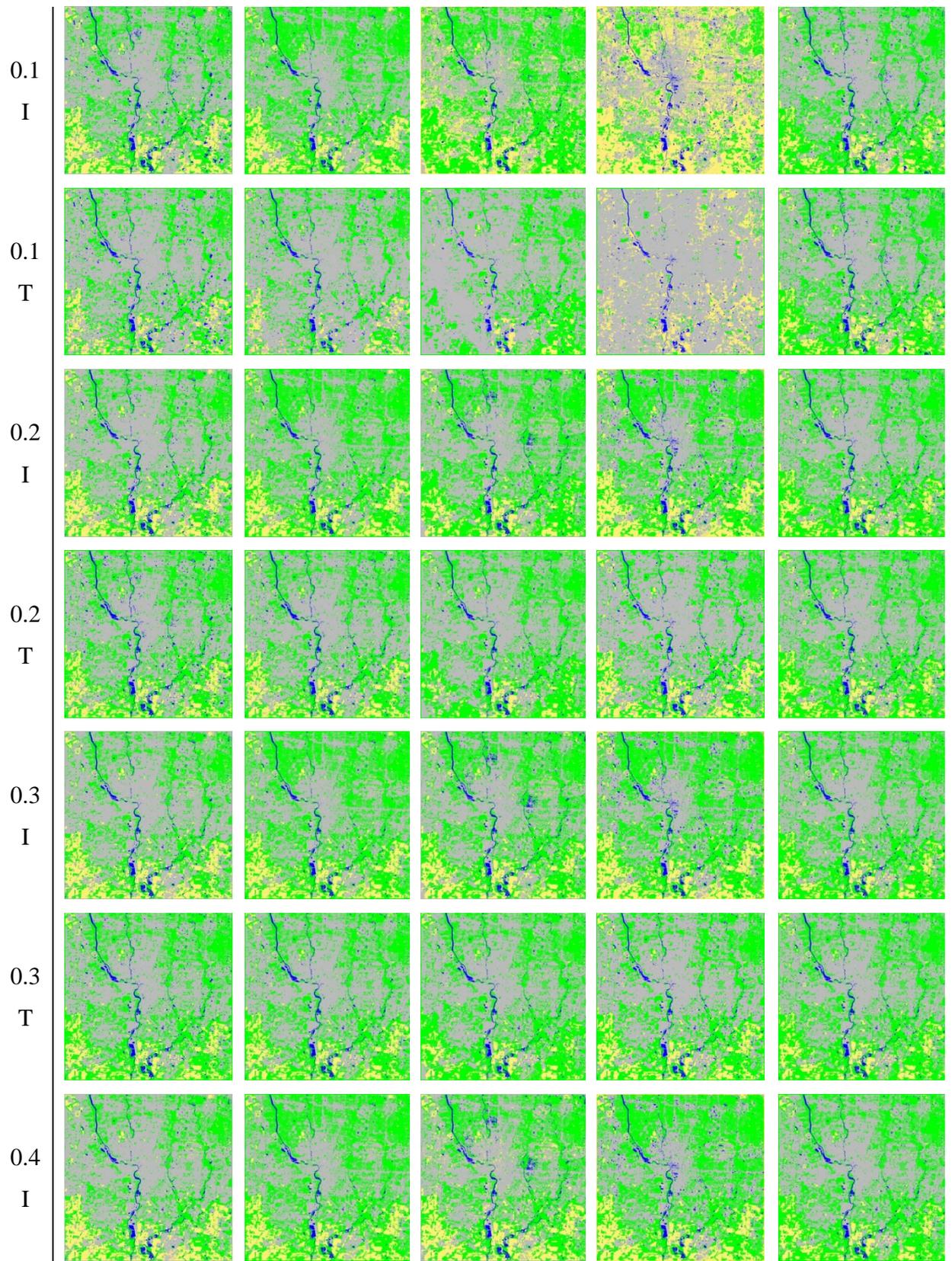

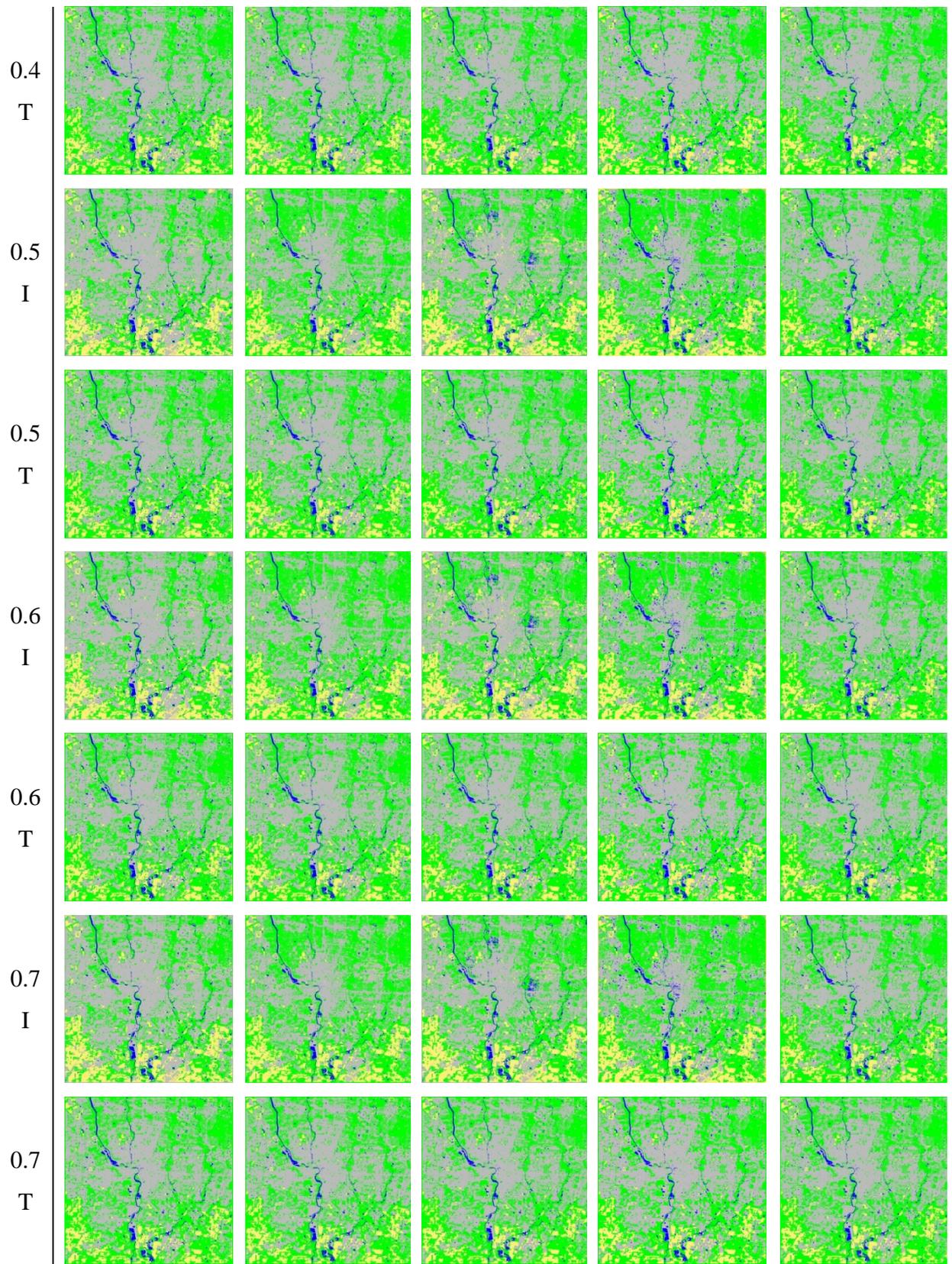

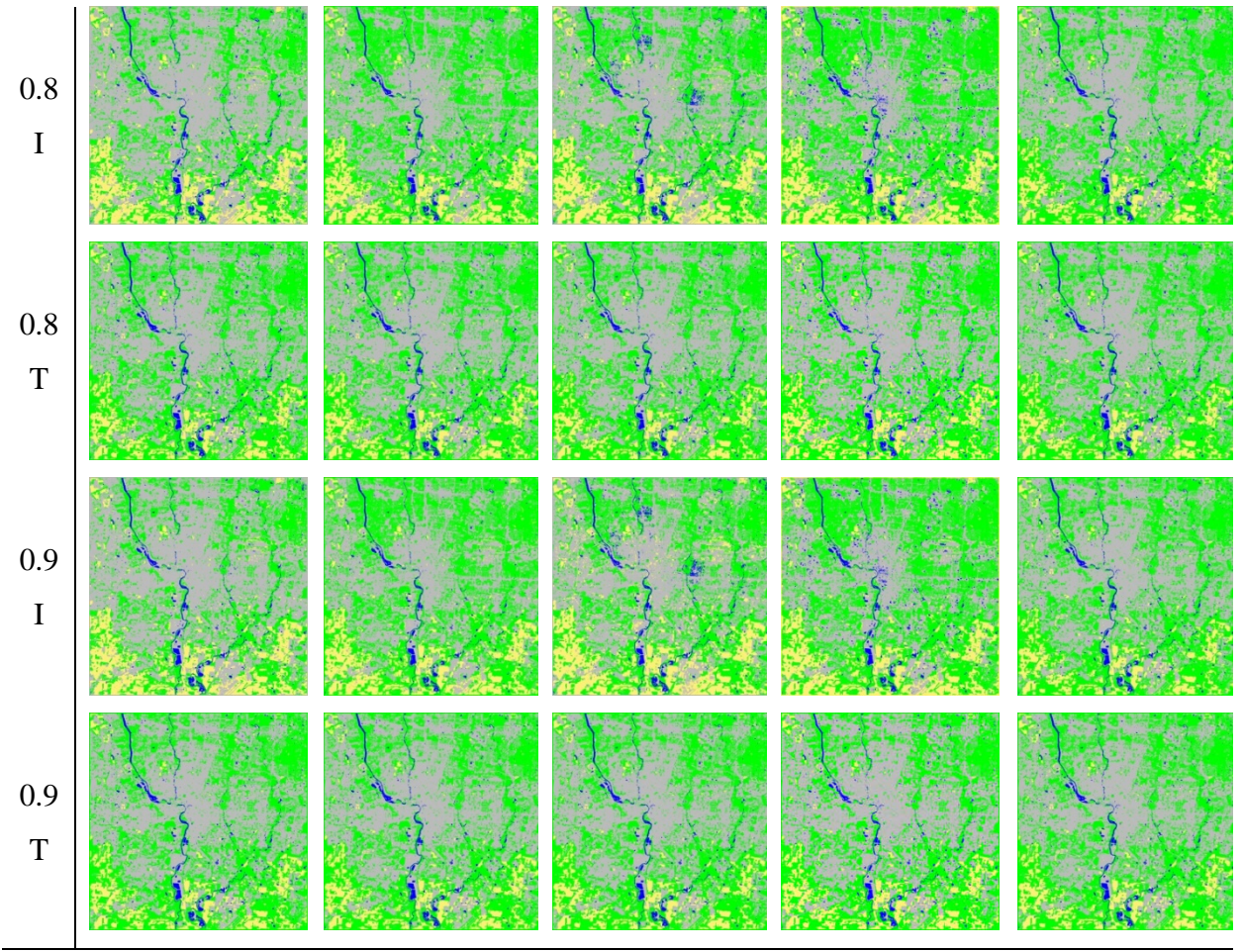

**Part (C):** This section covers the Classification Outcomes of Experiment III, the legend of the classes is provided on the top of the table.

| Classes | 🟢 Vegetation | 🔵 Water Surfaces | 🟡 Barren land | ⚪ Roads | 🔴 Buildings |
|---|---|---|---|---|---|
| | | | **Image** | | |
| $\sigma_t$ | 1 | 2 | 3 | 4 | 5 |
| | | | **Original Nonfiltered** | | |
| I |  |  |  |  |  |
| T | 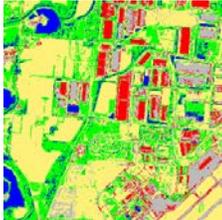 | 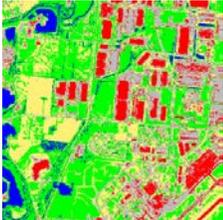 | 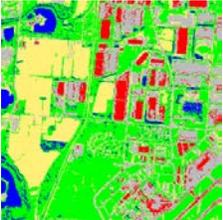 | 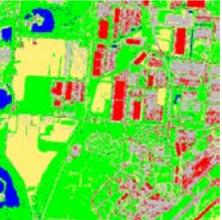 | 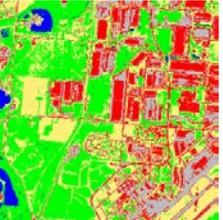 |
| | | | **Filtered** | | |
| 0 I | 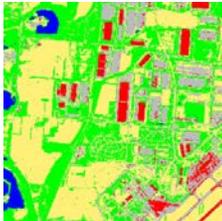 | 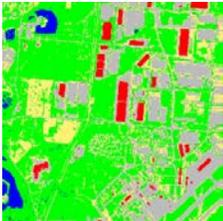 | 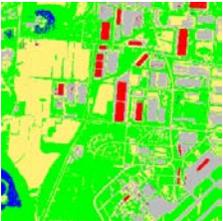 | 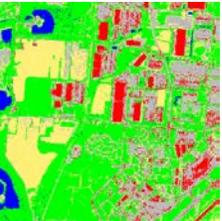 | 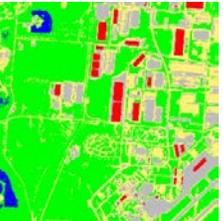 |
| 0 T | 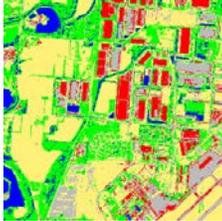 | 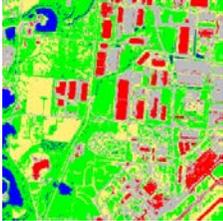 | 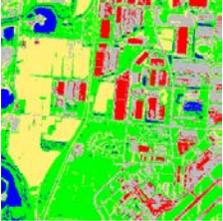 | 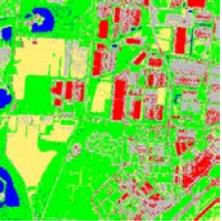 | 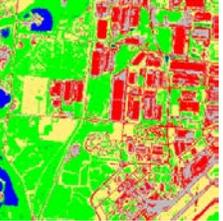 |

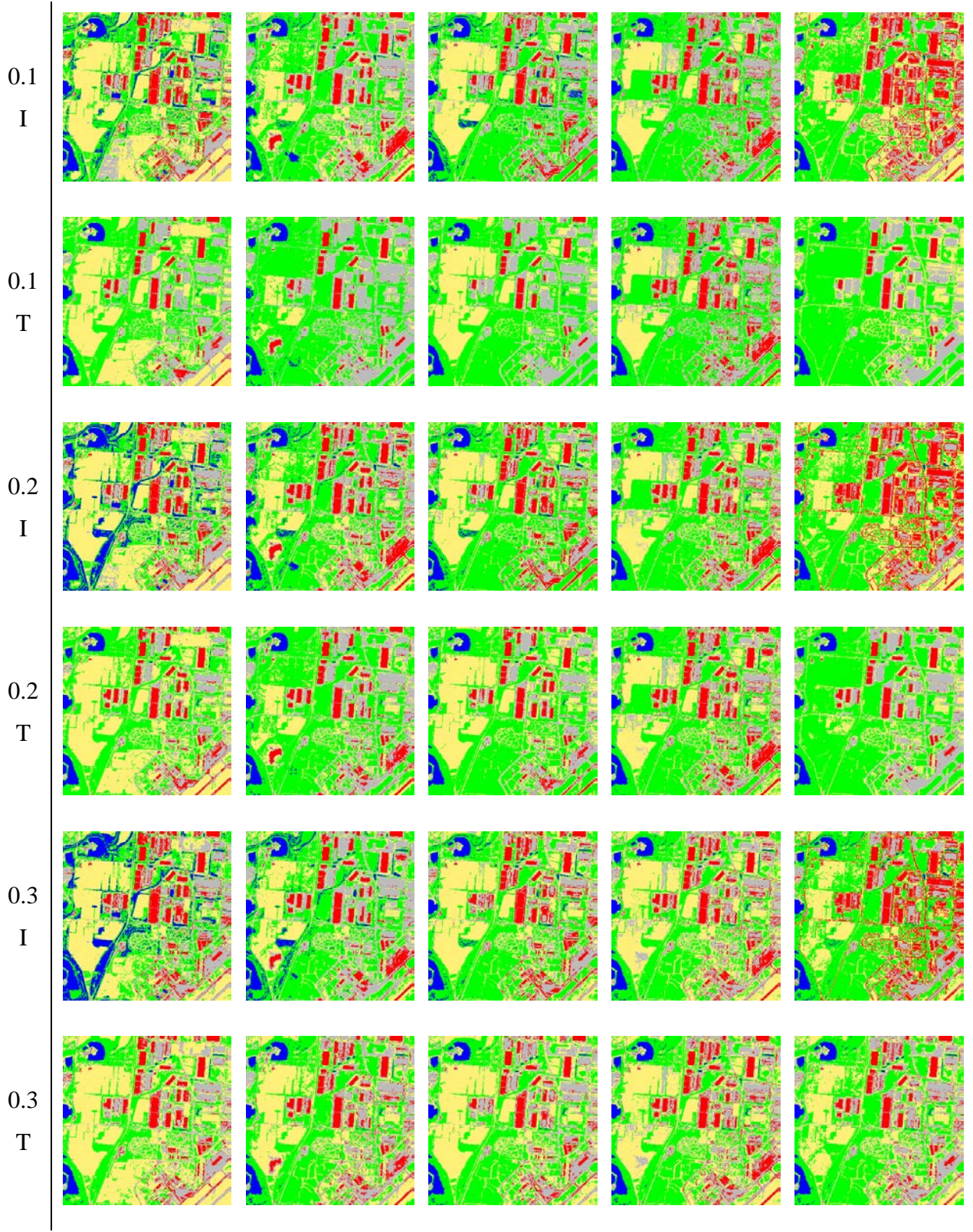

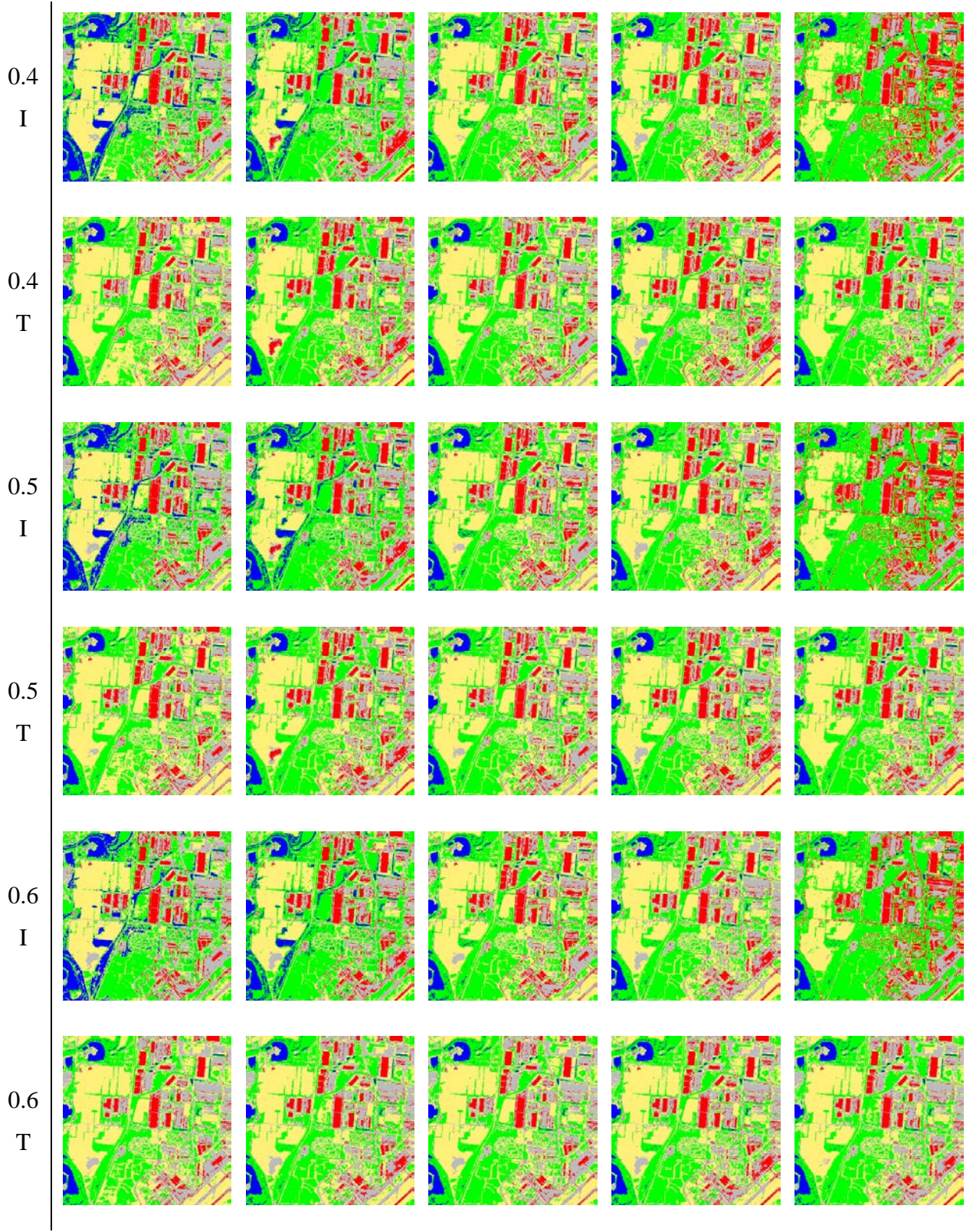

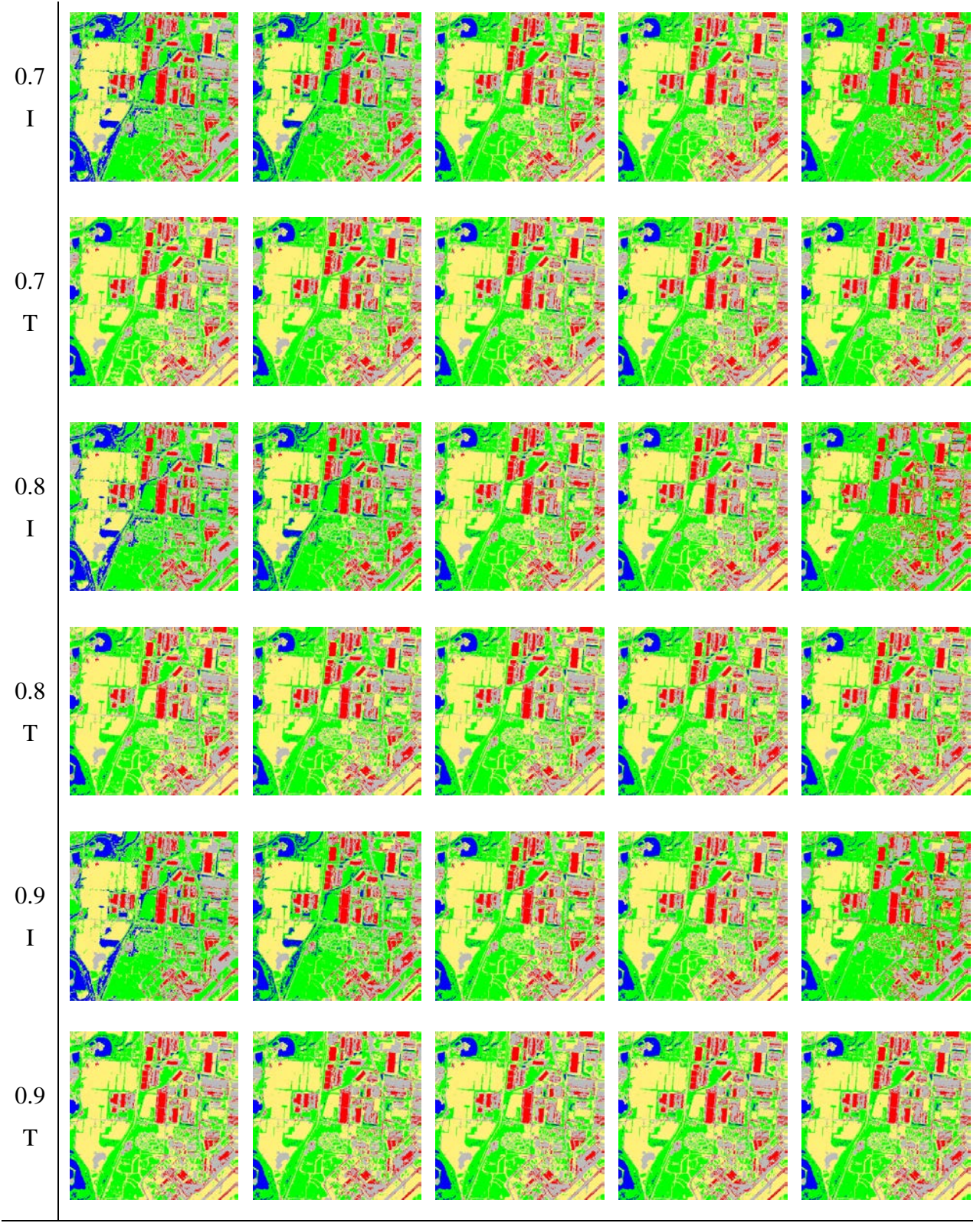

Appendix B: Kappa accuracy assessment

# Kappa accuracy assessment

An accuracy assessment using Kappa Cohen's is reported in the tables below. These values are very important to measure the level of agreement between the images and the classification results. Kappa is measured for all three experiments, and the average of all types of images whether it is filtered on not is calculated.

**Part (A):** Kappa coefficients for Experiment I.

| Individually classified, classification results of Kappa | | | | | | | | | | | |
|---|---|---|---|---|---|---|---|---|---|---|---|
| Type | Original | 0 | 0.1 | 0.2 | 0.3 | 0.4 | 0.5 | 0.6 | 0.7 | 0.8 | 0.9 |
| 1 | 0.94 | 0.93 | 0.94 | 0.92 | 0.92 | 0.92 | 0.92 | 0.92 | 0.92 | 0.92 | 0.92 |
| 2 | 0.95 | 0.94 | 0.93 | 0.94 | 0.95 | 0.94 | 0.94 | 0.94 | 0.93 | 0.93 | 0.93 |
| 3 | 0.90 | 0.88 | 0.90 | 0.91 | 0.90 | 0.88 | 0.87 | 0.86 | 0.86 | 0.85 | 0.85 |
| 4 | 0.74 | 0.72 | 0.78 | 0.76 | 0.76 | 0.73 | 0.74 | 0.74 | 0.73 | 0.73 | 0.73 |
| 5 | 0.95 | 0.95 | 0.95 | 0.95 | 0.91 | 0.89 | 0.88 | 0.87 | 0.86 | 0.86 | 0.86 |
| average | 0.90 | 0.88 | 0.90 | 0.90 | 0.89 | 0.87 | 0.87 | 0.86 | 0.86 | 0.86 | 0.86 |

| Transfer Learning, classification results of Kappa | | | | | | | | | | | |
|---|---|---|---|---|---|---|---|---|---|---|---|
| Type | Original | 0 | 0.1 | 0.2 | 0.3 | 0.4 | 0.5 | 0.6 | 0.7 | 0.8 | 0.9 |
| 1 | 0.69 | 0.71 | 0.72 | 0.86 | 0.87 | 0.78 | 0.68 | 0.62 | 0.61 | 0.61 | 0.60 |
| 2 | 0.61 | 0.62 | 0.68 | 0.86 | 0.88 | 0.78 | 0.68 | 0.62 | 0.61 | 0.61 | 0.60 |
| 3 | 0.90 | 0.88 | 0.90 | 0.91 | 0.90 | 0.88 | 0.87 | 0.86 | 0.86 | 0.85 | 0.85 |
| 4 | 0.80 | 0.89 | 0.88 | 0.65 | 0.61 | 0.50 | 0.42 | 0.37 | 0.37 | 0.36 | 0.36 |
| 5 | 0.88 | 0.91 | 0.91 | 0.91 | 0.87 | 0.85 | 0.82 | 0.80 | 0.79 | 0.78 | 0.77 |
| average | 0.78 | 0.80 | 0.82 | 0.84 | 0.83 | 0.76 | 0.70 | 0.66 | 0.65 | 0.64 | 0.64 |

**Part (B):** Kappa coefficients for Experiment II.

| Individually classified, classification results of Kappa | | | | | | | | | | | |
|---|---|---|---|---|---|---|---|---|---|---|---|
| Type | Original | 0 | 0.1 | 0.2 | 0.3 | 0.4 | 0.5 | 0.6 | 0.7 | 0.8 | 0.9 |
| 1 | 0.88 | 0.87 | 0.88 | 0.88 | 0.88 | 0.87 | 0.87 | 0.87 | 0.87 | 0.87 | 0.87 |
| 2 | 0.90 | 0.90 | 0.90 | 0.91 | 0.91 | 0.90 | 0.90 | 0.90 | 0.90 | 0.90 | 0.90 |
| 3 | 0.87 | 0.80 | 0.82 | 0.86 | 0.77 | 0.70 | 0.69 | 0.69 | 0.73 | 0.73 | 0.69 |
| 4 | 0.81 | 0.85 | 0.86 | 0.73 | 0.70 | 0.69 | 0.69 | 0.69 | 0.69 | 0.69 | 0.68 |
| 5 | 0.89 | 0.90 | 0.91 | 0.88 | 0.84 | 0.81 | 0.80 | 0.79 | 0.78 | 0.78 | 0.78 |
| average | 0.87 | 0.86 | 0.87 | 0.85 | 0.82 | 0.80 | 0.79 | 0.79 | 0.79 | 0.79 | 0.78 |

| Transfer Learning, classification results of Kappa |||||||||||
|---|---|---|---|---|---|---|---|---|---|---|
| Type | Original | 0 | 0.1 | 0.2 | 0.3 | 0.4 | 0.5 | 0.6 | 0.7 | 0.8 | 0.9 |
| 1 | 0.60 | 0.65 | 0.66 | 0.83 | 0.85 | 0.84 | 0.84 | 0.83 | 0.83 | 0.82 | 0.82 |
| 2 | 0.55 | 0.59 | 0.62 | 0.86 | 0.86 | 0.84 | 0.84 | 0.83 | 0.83 | 0.82 | 0.82 |
| 3 | 0.68 | 0.68 | 0.66 | 0.85 | 0.83 | 0.79 | 0.78 | 0.77 | 0.77 | 0.77 | 0.76 |
| 4 | 0.31 | 0.29 | 0.34 | 0.62 | 0.60 | 0.58 | 0.57 | 0.56 | 0.56 | 0.56 | 0.56 |
| 5 | 0.89 | 0.90 | 0.91 | 0.88 | 0.84 | 0.81 | 0.80 | 0.79 | 0.78 | 0.78 | 0.78 |
| average | 0.61 | 0.62 | 0.64 | 0.81 | 0.79 | 0.77 | 0.76 | 0.76 | 0.75 | 0.75 | 0.75 |

**Part (C):** Kappa coefficients for Experiment III.

| Individually classified, classification results |||||||||||
|---|---|---|---|---|---|---|---|---|---|---|
| Type | Original | 0 | 0.1 | 0.2 | 0.3 | 0.4 | 0.5 | 0.6 | 0.7 | 0.8 | 0.9 |
| 1 | 0.43 | 0.42 | 0.42 | 0.40 | 0.43 | 0.33 | 0.31 | 0.30 | 0.28 | 0.28 | 0.28 |
| 2 | 0.60 | 0.62 | 0.62 | 0.65 | 0.64 | 0.65 | 0.65 | 0.65 | 0.64 | 0.65 | 0.65 |
| 3 | 0.59 | 0.60 | 0.59 | 0.61 | 0.59 | 0.55 | 0.53 | 0.53 | 0.52 | 0.52 | 0.51 |
| 4 | 0.55 | 0.55 | 0.54 | 0.48 | 0.52 | 0.51 | 0.51 | 0.51 | 0.51 | 0.51 | 0.51 |
| 5 | 0.63 | 0.66 | 0.66 | 0.61 | 0.61 | 0.58 | 0.57 | 0.56 | 0.54 | 0.54 | 0.53 |
| average | 0.56 | 0.57 | 0.56 | 0.55 | 0.56 | 0.52 | 0.52 | 0.51 | 0.50 | 0.50 | 0.50 |

| Transfer Learning, classification results |||||||||||
|---|---|---|---|---|---|---|---|---|---|---|
| Type | Original | 0 | 0.1 | 0.2 | 0.3 | 0.4 | 0.5 | 0.6 | 0.7 | 0.8 | 0.9 |
| 1 | 0.40 | 0.41 | 0.40 | 0.42 | 0.42 | 0.40 | 0.39 | 0.35 | 0.36 | 0.35 | 0.35 |
| 2 | 0.61 | 0.61 | 0.62 | 0.70 | 0.64 | 0.64 | 0.62 | 0.59 | 0.59 | 0.59 | 0.57 |
| 3 | 0.55 | 0.55 | 0.58 | 0.61 | 0.61 | 0.54 | 0.52 | 0.51 | 0.51 | 0.51 | 0.50 |
| 4 | 0.55 | 0.55 | 0.54 | 0.48 | 0.52 | 0.51 | 0.51 | 0.51 | 0.51 | 0.51 | 0.51 |
| 5 | 0.55 | 0.55 | 0.57 | 0.70 | 0.65 | 0.52 | 0.50 | 0.47 | 0.46 | 0.46 | 0.45 |
| average | 0.53 | 0.53 | 0.54 | 0.58 | 0.57 | 0.52 | 0.51 | 0.49 | 0.49 | 0.48 | 0.48 |